\documentclass[nofootinbib,aps,11pt]{revtex4-1}
\usepackage{graphicx}
\usepackage{hyperref}
\usepackage{cancel}
\usepackage{amssymb}
\usepackage{textcomp}
\usepackage{amsmath}
\usepackage{bm}
\usepackage{times}
\usepackage{epsfig}
\usepackage{color}
\usepackage{ulem}
\usepackage{subfigure}

\newcommand{\BR}{{\rm BR}}

\newcommand{\GeV}{{\rm GeV}}
\newcommand{\MeV}{{\rm MeV}}
\newcommand{\TeV}{{\rm TeV}}
\newcommand{\eV}{{\rm eV}}

\newcommand{\fb}{{\rm fb}}
\newcommand{\pb}{{\rm pb}}

\newcommand{\mZ}{\mathbb{Z}}

\newcommand{\DM}{{\rm DM}}

\newcommand{\Znot}{\setminus\!\!\!\!Z}

\begin{document}
\title{\LARGE Radiative neutrino mass with $\mathbb{Z}_3$ Dark matter: From relic density to LHC signatures}
\bigskip
\author{Ran Ding~$^{1}$}
\email{dingran@mail.nankai.edu.cn}
\author{Zhi-Long Han~$^{2}$}
\email{hanzhilong@mail.nankai.edu.cn}
\author{Yi Liao~$^{3,2,1}$}
\email{liaoy@nankai.edu.cn}
\author{Wan-Peng Xie~$^{1}$}
\email{wanpeng.xie@gmail.com}
\affiliation{
$^1$ Center for High Energy Physics, Peking University, Beijing 100871, China
\\
$^2$~School of Physics, Nankai University, Tianjin 300071, China
\\
$^3$ State Key Laboratory of Theoretical Physics, Institute of Theoretical Physics,
Chinese Academy of Sciences, Beijing 100190, China}
\date{\today}

\begin{abstract}

In this work we give a comprehensive analysis on the phenomenology of a specific $\mathbb{Z}_3$ dark matter (DM) model in which neutrino mass is induced at two loops by interactions with a DM particle that can be a complex scalar or a Dirac fermion. Both the DM properties in relic density and direct detection and the LHC signatures are examined in great detail, and indirect detection for gamma-ray excess from the Galactic Center is also discussed briefly. On the DM side, both semi-annihilation and co-annihilation processes play a crucial role in alleviating the tension of parameter space between relic density and direct detection. On the collider side, new decay channels resulting from $\mathbb{Z}_3$ particles lead to distinct signals at LHC. Currently the trilepton signal is expected to give the most stringent bound for both scalar and fermion DM candidates, and the signatures of fermion DM are very similar to those of electroweakinos in simplified supersymmetric models.

\end{abstract}

\maketitle

\section{Introduction}

Neutrino mass and nonbaryonic dark matter (DM) offer two pieces of robust evidence for the existence of physics beyond standard model (SM), but their origins remain mysterious. It would be appealing if they could be understood in the same framework. At low energies neutrino mass can be accommodated by a dimension-five operator in terms of the SM Higgs and lepton fields~\cite{Weinberg:1979sa}. The operator can be realized at tree level in three different manners~\cite{Ma:1998dn} which correspond exactly to the three types of conventional seesaws. Though simple enough, these models are difficult to test experimentally since they invoke very high energy scales or very weak couplings to SM particles in order to induce tiny neutrino masses. One way to alleviate this problem is to push the neutrino mass to a radiative effect of new physics which provides additional suppression. For this purpose, an (almost) exact discrete symmetry is usually required to forbid the generation of neutrino mass at a lower order. Such a symmetry can stabilize the lightest neutral member of all particles that transform nontrivially under the symmetry, and makes it a natural DM candidate.

The above idea of DM-generated neutrino mass has been extensively exploited in the literature~\cite{Ma:2006km,Ma:2007gq,Kanemura:2010bq,Kanemura:2012rj,Restrepo:2013aga,
Hirsch:2013ola,Kajiyama:2013rla,Law:2013saa,
Fabbrichesi:2014qca,Ahriche:2014oda,Chen:2014ska,Kanemura:2014rpa,Okada:2014qsa}. The simplest discrete symmetry is a $\mathbb{Z}_2$ parity. However, if it appears as a remnant of a broken gauge group ($U(1)_X$), other $\mathbb{Z}_N$s are also possible in general~\cite{Krauss:1988zc}. The possibility with $N>2$ has been investigated in Refs~\cite{Ma:2007gq,D'Eramo:2010ep,Batell:2010bp,Belanger:2012vp,Aoki:2012ub,Belanger:2012zr,Belanger:2014bga}. Compared with the $\mathbb{Z}_2$ case, DM with $\mathbb{Z}_N$ symmetry has the following distinct features:

\begin{itemize}
  \item {New DM annihilation processes such as semi-annihilation (SE-A)~\cite{D'Eramo:2010ep} become available that allow for different numbers of DM particles to appear in the initial and final states. The processes can change significantly the evaluation of the DM relic density.}
  \item {DM particles have new interesting decay modes that result in richer phenomenology and distinguishable signatures at colliders~\cite{Agashe:2010gt,Agashe:2010tu}.}
  \item {Multi-component DM is possible. In this case, annihilation processes between different components lead to the so-called assisted freeze-out mechanism~\cite{Belanger:2011ww}, which also influences the DM relic density.}
\end{itemize}

In this paper, we focus on the first two features. We consider a specific $\mathbb{Z}_3$ DM model that induces neutrino mass at two loops. The model was originally proposed in Ref.~\cite{Ma:2007gq}, in which DM can be either a Dirac fermion or a complex scalar. Some phenomenological aspects of the model have been previously studied in Ref.~\cite{Aoki:2014cja} with emphasis on the effect of SE-A processes on relic density and direct detection. Here we aim to implement a comprehensive analysis on DM properties and collider signatures. We will show that both SE-A and co-annihilation (CO-A) processes have significant effects on the evaluation of relic density while evading stringent constraints from direct detection. Moreover, the presence of many new decay channels of $\mathbb{Z}_3$ particles induces a plenty of distinct signals at LHC for both scalar and fermion DM candidates which would be absent for $\mathbb{Z}_2$ DM.

The rest of this paper is organized as follows. In Sec. \ref{constraints}, we recall the model and discuss current experimental constraints on its parameters. Sections \ref{DM} and \ref{lhc} contain the core content of this work, in which we systematically study DM properties and LHC signatures for both scalar and fermion candidates. In Sec. \ref{DM}, we explore the vast parameter space that survives the constraints from relic density and direct detection; all important annihilation channels will be presented and discussed in detail. In the following Sec. \ref{lhc}, we first exhaust all decay patterns according to the mass spectra of new particles, and then analyze various LHC signatures and compare with the relevant LHC limits. Finally, Sec. \ref{concl} is devoted to conclusions.

\section{Model and constraints}
\label{constraints}

In the model under consideration~\cite{Ma:2007gq} a global and exact $\mZ_3$ symmetry is imposed to induce neutrino masses at the two-loop level through interactions with new particles charged under $\mZ_3$. In one minimal version of the model, one introduces two scalars $\chi_a(a=1,2)$ and one Dirac fermion $S$, both of which are neutral singlets of the SM gauge group, and one Dirac fermion doublet $\Psi=(N,E)$ of hypercharge $Y=-1$. The fermions are assumed to be vector-like to avoid chiral anomalies. The new particles transform under $\mZ_3$ in the same way as $\chi_a\to\chi_a\omega$ with $\omega=\exp{(i2\pi/3)}$, while SM particles are neutral. The Yukawa and fermion mass terms involving new fields and the SM leptons $F_{iL}=(\nu_{iL},\ell_{iL}),~\ell_{iR}$ and Higgs boson $\Phi=(G^+,\phi^0)$ are:
\begin{eqnarray}
\mathcal{L}&\supset& -y^{\prime ij} \bar{F}_{iL}\Phi \ell_{jR}
 - m_S \bar{S}_L S_R - m_\Psi \bar{\Psi}_L\Psi_R
- \frac{1}{2}x_L^{\prime a} \chi_a \bar{S}_L^C S_L
- \frac{1}{2}x_R^{\prime a} \chi_a \bar{S}_R^C S_R
\nonumber
\\
&& - z_L' \bar{S}_L \tilde{\Phi}^\dag \Psi_R
 - z_R' \bar{S}_R \tilde{\Phi}^\dag \Psi_L
 - h^{\prime ai} \chi_a^\dag \bar{F}_{iL} \Psi_R
 + \mbox{H.c.},
\label{Yukawa}
\end{eqnarray}
where $\tilde{\Phi}=i\sigma_2\Phi^*$. And the scalar potential is
\begin{eqnarray}
V & = & - m^2 \Phi^\dag\Phi + (M^2)^{ab}\chi^\dag_a\chi_b+ \frac{1}{2}\lambda_1(\Phi^\dag\Phi)^2 +  \frac{1}{2}\lambda^{ab;cd}_{2} (\chi^\dag_a\chi_b)(\chi^\dag_c\chi_d) \\\nonumber
 & &+ \lambda_{3}^{ab}(\Phi^\dag\Phi)(\chi^\dag_a\chi_b) + \frac{1}{6}(\mu^{abc}\chi_a\chi_b\chi_c+\mbox{H.c.}),
\end{eqnarray}
where $M^2$ and $\lambda_3$ are Hermitian, $\mu^{abc}$ is complex and symmetric in the three indices, and $\lambda_2^{ab;cd}=\lambda_2^{cd;ab}=(\lambda_2^{ba;dc})^*$ is complex as well. Some of the phases in the above couplings can be removed by field redefinitions but there are still many physical ones. To make the number of independent parameters under control for our later numerical analysis, we will simply assume that $\lambda^{ab;cd}_{2}=\lambda_2$, $\lambda_{3}^{ab}=\lambda_3$, $\mu^{abc}=\mu$, and that they are all real.

There are some theoretical considerations that can be used to set a bound on the parameters in the scalar potential, such as perturbativity, unitarity \cite{Lerner:2009xg}, and $\mZ_3$ not to be spontaneously broken \cite{Belanger:2012zr}. These constraints are easily respected in our numerical analysis. Since $\mZ_3$ is exact, new particles do not mix with SM particles but can mix among themselves. We assume that $\chi_{1,2}$ are diagonalized by an angle $\alpha$ to $\chi_{L,H}$ of masses $M_{\chi_L}\leq M_{\chi_H}$. The electrically neutral fermions $S$ and $N$ also mix due to the Yukawa couplings $z'_{L,R}$ by an angle $\beta$ to $N_{1,2}$ of masses $M_{N_{1,2}}$. Our convention is that $N_1$ ($N_2$) is dominantly a singlet $S$ (doublet $N$) for small $\beta$ but either mass order is possible. In terms of the mass-eigenstate fields the couplings will involve the mixing angles. For the Yukawa couplings, we simply replace the primed couplings by unprimed ones, e.g., $x_{L,R}^a$ and $h^{ai}$. For the scalar self-couplings, the angle $\alpha$ enters explicitly; e.g., the $\chi_a\chi_b\chi_c$ coupling (now $a,~b,~c=L,~H$) is proportional to $\mu g_{abc}$, in which an index $L$ ($H$) is associated with a factor of $\alpha_-=\cos\alpha-\sin\alpha$ ($\alpha_+=\cos\alpha+\sin\alpha$), for instance,
\begin{eqnarray}
g_{HHH}=\alpha_+^3,~g_{LHH}=\alpha_+^2\alpha_-,~\textrm{etc.}
\end{eqnarray}
Therefore we can take $M_{\chi_{L,H}}$, $m_{N_{1,2}}$, $\alpha$, $\beta$, $h^{ai}$, $x^a_{L,R}$, $\lambda_2$, $\lambda_3$, and $\mu$ as our input parameters.

\begin{figure}[!htbp]
\begin{center}
\includegraphics[width=0.45\linewidth]{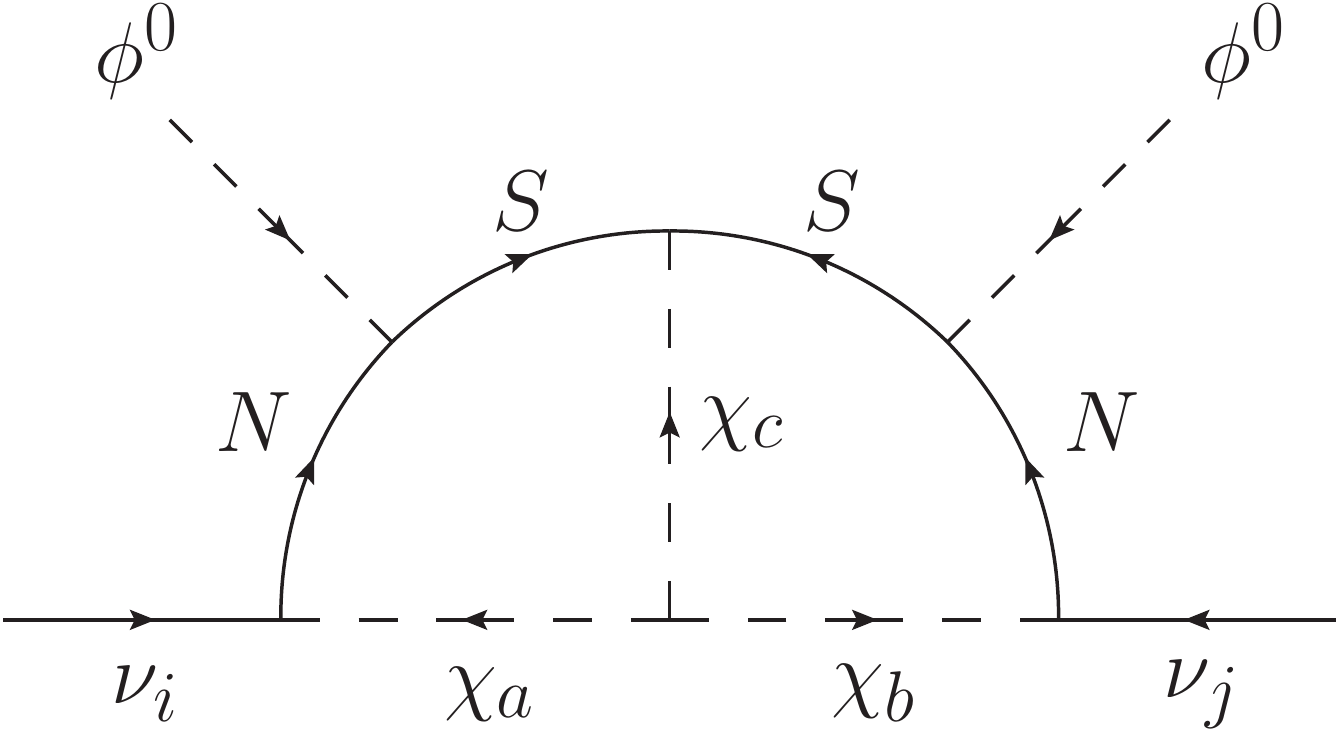}
\end{center}
\caption{Feynman diagram for neutrino mass.
\label{Fig:mv}}
\end{figure}

The above interactions induce neutrino masses at the two-loop level \cite{Ma:2007gq} as shown in Fig. \ref{Fig:mv}:
\begin{eqnarray}\label{Eqn:mv}
(m_{\nu})^{ij}= \frac{\mu\sin^2 2\beta}{4(4\pi)^4}\sum_{a,b,c}\sum_{k,l} h^{ai} h^{bj}g_{abc}(-1)^{k+l}\left(x_L^c I_{Lkl}^{abc}+x_R^c I_{Rkl}^{abc}\right),
\label{loop}
\end{eqnarray}
where $a,b,c=L,H$ refer to scalars $\chi_{L,H}$ and $k,l=1,2$ to fermions $N_{1,2}$. The loop functions are
\begin{eqnarray}
I_{Lkl}^{abc}&=&\frac{M_l}{M_k}\int_0^1dxdydz\frac{\delta(x+y+z-1)}{z(1-z)}
\left[\frac{\xi_k^a\ln\xi_k^a}{(1-\xi_k^a)(\xi_k^a-\xi_{kl}^{bc})}
+     \frac{\xi_{kl}^{bc}\ln\xi_{kl}^{bc}}{(1-\xi_{kl}^{bc})(\xi_{kl}^{bc}-\xi_k^a)}\right],
\nonumber
\\
I_{Rkl}^{abc}&=&\int_0^1dxdydz\frac{\delta(x+y+z-1)}{(1-z)}
\left[\frac{(\xi_k^{a})^2\ln\xi_k^a}{(1-\xi_k^a)(\xi_k^a-\xi_{kl}^{bc})}
+     \frac{(\xi_{kl}^{bc})^2   \ln\xi_{kl}^{bc}}{(1-\xi_{kl}^{bc})(\xi_{kl}^{bc}-\xi_l^a)}\right],
\label{loop1}
\end{eqnarray}
with
\begin{equation}
\xi_k^a=\frac{M_a^2}{M_k^2},~\xi_{kl}^{bc}=\frac{xM_l^2+yM_b^2+zM_c^2}{z(1-z)M_k^2}.
\label{loop2}
\end{equation}
Our loop functions agree with Ref. \cite{Okada:2014qsa} which shares the same topology of Feynman diagrams in a different model, but the relative sign of the two terms differs from that in Ref. \cite{Aoki:2014cja} which computes neutrino mass from a Feynman diagram of same topology in another scenario of the $\mZ_3$ model. The induced $3\times 3$ neutrino mass matrix has a degenerate structure implying a massless neutrino in either normal or inverted hierarchy. With a two-loop suppression factor of $(4\pi)^{-4}\sim 10^{-5}$, it is easy to accommodate a mass of order $0.1~\eV$ for the other two neutrinos by assuming reasonable values of new parameters. For heavy masses of same order, the loop functions are of order $0.1$. As will be shown below, the constraints from lepton flavor violating (LFV) transitions can be trivially fulfilled with $h^{ai}\sim 0.01$. This then requires $\mu x_{L,R}^a\sin^2(2\beta)\sim 0.1~\GeV$.

\begin{figure}[!htbp]
\begin{center}
\includegraphics[width=0.45\linewidth]{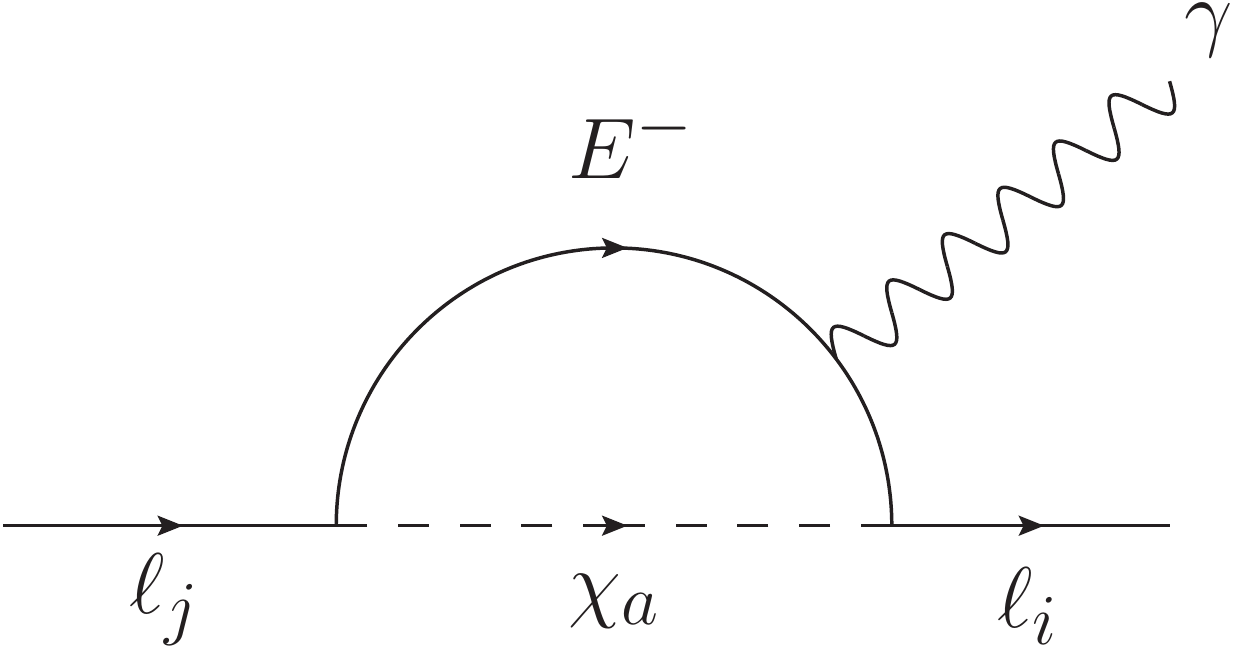}
\end{center}
\caption{Feynman diagram for LFV process $\ell_j\to \ell_i\gamma$.
\label{lfv}}
\end{figure}

As is well known, precise measurements of LFV transitions set strong constraints on relevant interactions. The diagram in Fig. \ref{lfv} for the lepton radiative decay $\ell_j\to\ell_i\gamma$ yields the branching ratio \cite{Liao:2009fm,Ding:2014nga}:
\begin{equation}
\mbox{BR}(\ell_j\to \ell_i\gamma)=\mbox{BR}(\ell_j\to \ell_i \bar{\nu}_i\nu_j)\frac{3\alpha}{16\pi G_F^2M_E^4}
\left|\sum_{a=L,H}h^{ai*}h^{aj}F\left(\frac{M_{\chi_a}^2}{M_E^2}\right)\right|^2,
\end{equation}
where the loop function $F(x)$ is given by
\begin{equation}
F(x)=-\frac{1}{12(1-x)^4}[1-6x+3x^2+2x^3-6x^2\ln x].
\end{equation}
Currently, the most stringent limit comes from BR$(\mu\to e\gamma)<5.7\times10^{-13}(90\%~\mbox{C.L.})$ \cite{Adam:2013mnn}, while the limits on $\tau$ decays are less stringent, BR$(\tau\to e\gamma)<3.3\times10^{-8}(90\%~\mbox{C.L.})$ \cite{Aubert:2009ag} and BR$(\tau\to e\gamma)<4.4\times10^{-8}(90\%~\mbox{C.L.})$ \cite{Aubert:2009ag}. For $\mZ_3$ particles of similar masses, $M_{\chi_{L,H}}\sim M_{E}$, we have $F(x)\sim1/24$; then the above bounds translate into the constraints on the Yukawa couplings $h^{ai}$:
\begin{eqnarray}
|h^{ae*} h^{a\mu}|\lesssim 5.1\times 10^{-5} \left(\frac{M_{E}}{100~\GeV}\right)^2,
\nonumber
\\
|h^{ae*} h^{a\tau}|\lesssim 3.0\times 10^{-2} \left(\frac{M_{E}}{100~\GeV}\right)^2,
\nonumber
\\
|h^{a\mu*} h^{a\tau}|\lesssim 3.4\times 10^{-2} \left(\frac{M_{E}}{100~\GeV}\right)^2.
\end{eqnarray}
For instance, when $M_{E}\sim 200~\GeV$, the above become $|h^{ae*} h^{a\mu}|\lesssim 0.0002$, $|h^{ae*} h^{a\tau}|\lesssim 0.12$, and $|h^{a\mu*} h^{a\tau}|\lesssim 0.14$. Thus, without requiring a special flavor structure we can assume safely a universal bound $|h^{ai}|\lesssim 0.01$.
\footnote{We recall that in the case of fermion DM with a $\mathbb{Z}_2$ symmetry it is hard to provide correct relic density with such small Yukawa couplings~\cite{Vicente:2014wga}.}

The mass splitting between the charged fermion $E$ and neutral ones $N_{1,2}$ will contribute to the custodial symmetry breaking measured by the parameter $\Delta T$. Using the formulas in, e.g., Refs.~\cite{Cynolter:2008ea,Lavoura:1992np} and the fitting result $|\Delta T|<0.2$ \cite{Baak:2014ora}, we can set a bound on the mass splitting. In the most stringent case for a large mixing $\sin\beta\sim 0.66$, one gets $|M_{N_2}-M_{N_1}|<250~\GeV$. In the opposite case of a small mixing $\sin\beta<0.1$, $|M_{N_2}-M_{E}|$ is restricted to be less than a few GeV \cite{Bhattacharya:2015qpa}.
In our numerical analysis we will always work with a small $\beta$ and assume $N_2,~E$ are degenerate. Furthermore, for a light DM particle $\chi_L$ or $N_1$ there are collider constraints on invisible Higgs decays, which will be examined in the next section.

\section{Dark matter phenomenology}
\label{DM}

In this section, we investigate DM phenomenology of the $\mZ_3$ model. For this purpose, we generate the {\tt CalcHEP}~\cite{Belyaev:2012qa} file by using the {\tt FeynRules}~\cite{feynrules} package, which is used by {\tt micrOMEGAs4.1}~\cite{Belanger:2014vza} to calculate the DM relic density and DM-nucleon scattering cross section. We implement random scans over a vast parameter space (with a total of $3\times10^5$ samples), for the ranges or values of the input parameters shown in Table~\ref{tab:ranges}. The constraints from relic density and direct detection are then imposed on each sample. For relic density, we use the combined Planck+WP+highL+BAO $2\sigma$ range, $0.1153<\Omega_\DM h^2<0.1221$~\cite{Ade:2013zuv}. As for direct detection, we adopt the currently most restrictive spin-independent limit provided by the LUX experiment~\cite{Akerib:2013tjd}~\footnote{Since the exclusion limit varies with the DM particle mass $M_\DM$, we interpolate the LUX data to obtain the corresponding exclusion limit for each randomly generated $M_\DM$.}. We make some comments before presenting our numerical results:
\begin{itemize}
  \item {There are four new neutral particles in this model, $\chi_{L,H}$ and $N_{1,2}$, resulting in a rich annihilation pattern. In addition to the standard annihilation (ST-A) processes, the SE-A and CO-A processes play a crucial role in the DM relic density. The latter two processes are sensitive to the mass relations of new particles and should be thoroughly examined. We will discuss this issue in great detail in Sec.~\ref{aps}.}
  \item {For DM-nucleon scattering, it is sufficient to include tree-level contributions since one-loop terms are subleading.
      In the case of scalar DM, $\chi_L$ interacts with quarks through the Higgs-portal $\lambda_3$ term, so that direct detection can set a stringent constraint on $\chi_L\chi_L^\dagger\to b\bar{b}$ annihilation channel. This is a common feature of Higgs-portal models as we will discuss in Sec.~\ref{aac}. For fermion DM, $N_1$ can scatter with quarks via the $t$-channel $Z$ exchange due to the singlet-doublet mixing. Direct detection then imposes a very stringent bound on the mixing angle $\beta$. The additional $t$-channel Higgs exchange also contributes via the $z_{L,R}$ Yukawa terms but is subleading to the $Z$ exchange. The angle $\beta$ is therefore varied in a much narrower interval than $\alpha$. The constraints from the Higgs and $Z$ invisible decays and electroweak precision measurements also prefer a small $\beta$.}
  \item {Since the quartic coupling $\lambda_2$ is only related to the DM self-interaction and has no further phenomenological effect, we assume a fixed value for it in the scan.}
  \item {According to our discussion in Sec.~\ref{constraints}, we assume a universal $|h^{ai}|\simeq0.01$ to avoid dangerous LFV processes, though a relatively large $|h^{a\tau}|$ can be accommodated by a specific flavor structure.}
\end{itemize}

\begin{table*}[hbt]
\begin{tabular}{|c|c|c|c|c|c|c|c|c|c|c|c|c|}
\hline
$M_{\chi_L}$ & $M_{N_1}$ & $M_{\chi_H}-M_{\chi_L}$     & $M_{N_2}-M_{N_1}$ & $\alpha$        & $\beta$     & $\lambda_2$   & $\lambda_3$  & $\mu$ &  $x_{L,R}^a$  & $|h^{ai}|$  \\
\hline
$[1,1000]$    & $[1,1000]$            & $[1,500]$ & $[1,500]$ & $[0,\pi]$ & $[0.001,0.2]$ & $0.1$ & $[0.001,1]$         & $10$       &  $[0.1,1]$          & $0.01$ \\
\hline
\end{tabular}
\caption{The ranges or values of the input parameters used in our scan. All masses and $\mu$ are in the units of GeV. The SM Higgs has the mass $M_h=125~\GeV$.}
\label{tab:ranges}
\end{table*}

\subsection{Analysis of parameter space}\label{aps}

\begin{figure}[!htbp]
\begin{center}
\includegraphics[width=0.45\linewidth]{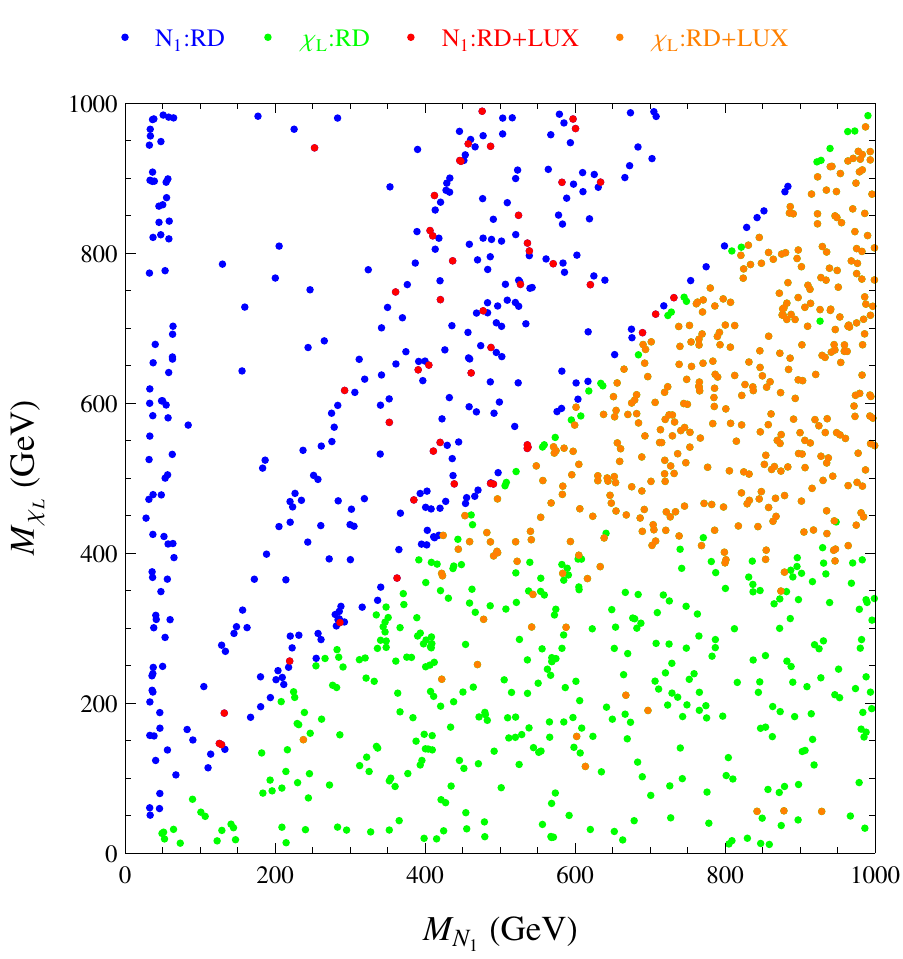}
~~
\includegraphics[width=0.45\linewidth]{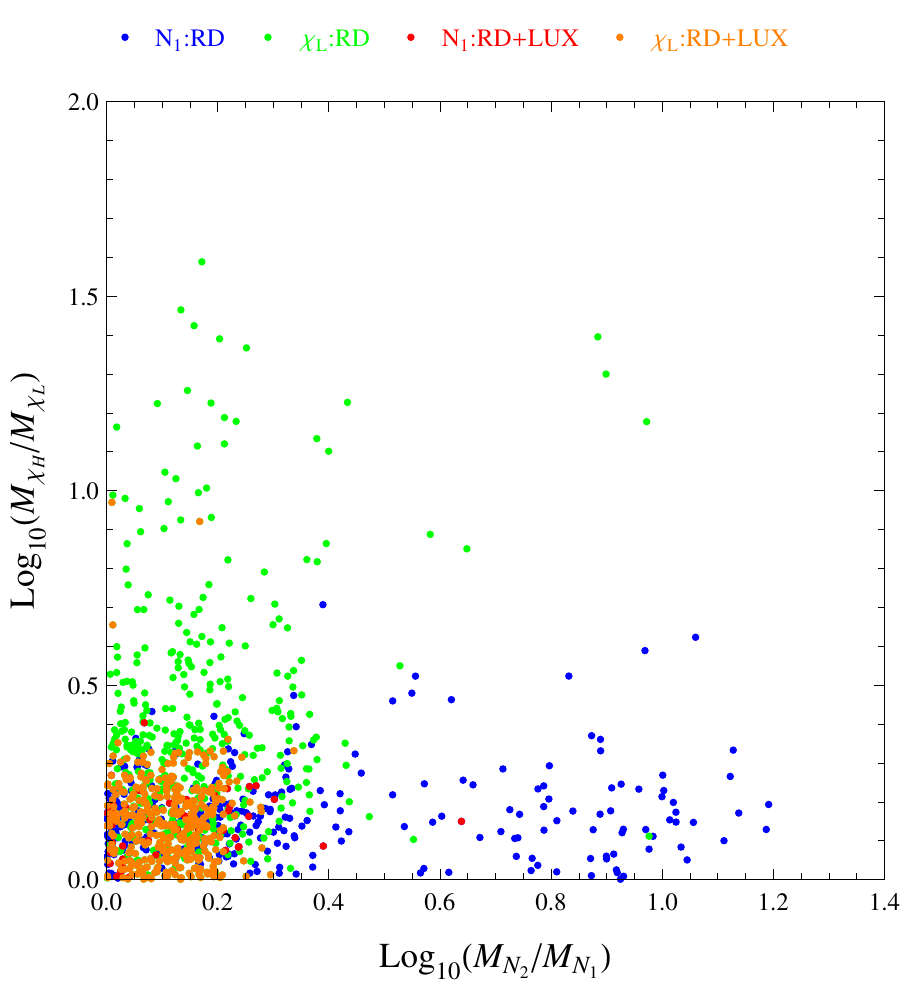}
~~
\end{center}
\caption{Distributions of survived samples in the $(M_{N_1},~M_{\chi_L})$ (left panel) and $[\log(M_{N_2}/M_{N_1}),~\log(M_{\chi_H}/M_{\chi_L})]$ (right) plane. The blue (green) points only pass the relic density (RD) constraint in the case of $N_1$ ($\chi_L$) DM, while the red (orange) points satisfy both RD and LUX constraints.
\label{DM1}}
\end{figure}

We present our numerical scans in this subsection. Figure~\ref{DM1} displays the distributions of survived samples in the $(M_{N_1},~M_{\chi_L})$ and $[\log(M_{N_2}/M_{N_1})
,~\log(M_{\chi_H}/M_{\chi_L})]$ plane.
From the figure, we learn some important features:
\begin{itemize}
  \item {It is clear from the left panel that both scalar $\chi_L$ and fermion $N_1$ DM
    samples can survive in the whole mass regions that we explored, but the scalar one has a much more number. This can be explained as follows. Since both CO-A and SE-A processes depend on specific mass relations, the survived samples resulting from them do not distinguish much between $\chi_L$ and $N_1$. The difference originates instead from ST-A processes. While $\chi_L$ annihilates into gauge boson pairs and produces the correct relic density in a vast mass region, $N_1$ mostly annihilates into light and $b$ quark pairs through the Higgs exchange, which can only give the correct relic density in a relatively low mass region.}

  \item {The right panel shows that survived samples tend to cluster in regions of small mass splitting for RD+LUX points, where CO-A and SE-A processes are generally dominant. Since ST-A channels are not sensitive to mass splitting and tend to cause a more scattered distribution in the plane, the clustering indicates that direct detection imposes a more severe constraint on ST-A processes. We will illustrate this feature explicitly in section~\ref{aac}.}
\end{itemize}

\begin{figure}[!htbp]
\begin{center}
\includegraphics[width=0.5\linewidth]{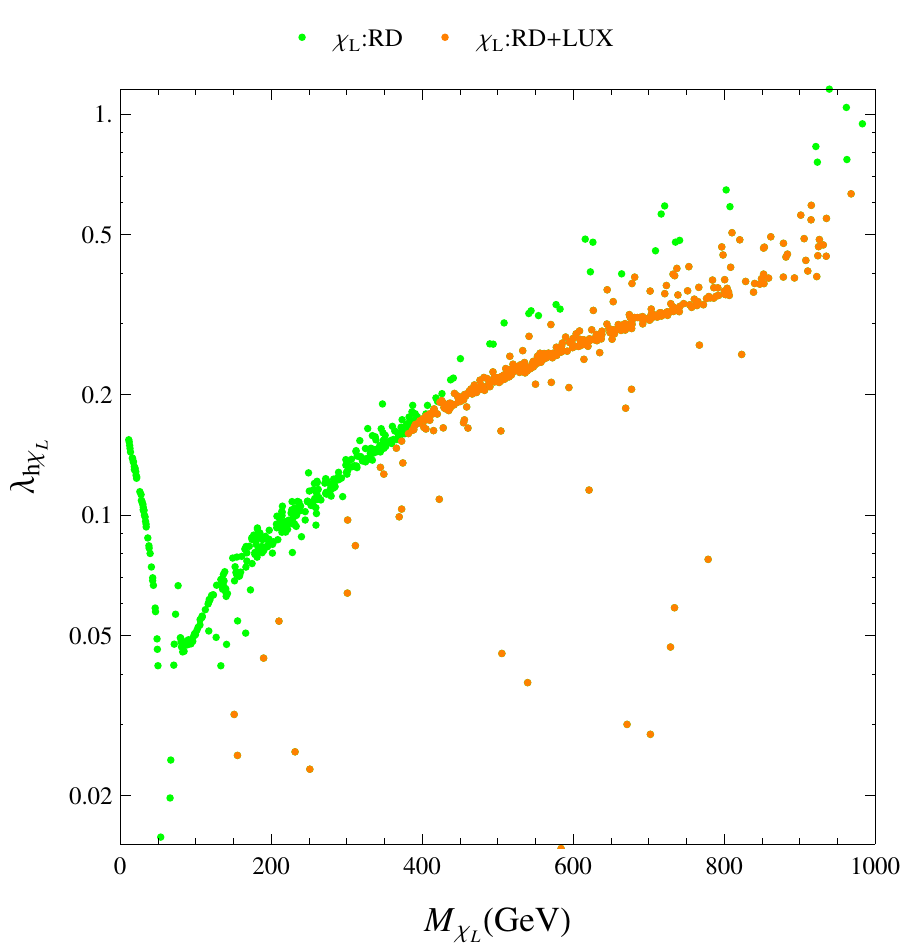}
\end{center}
\caption{Distributions of survived samples for $\chi_L$ DM in the $[M_{\chi_L},~\lambda_{h\chi_L}]$ plane. The symbols of points are the same as in Fig.~\ref{DM1}.
\label{DM2}}
\end{figure}

\begin{figure}[!htbp]
\begin{center}
\includegraphics[width=0.45\linewidth]{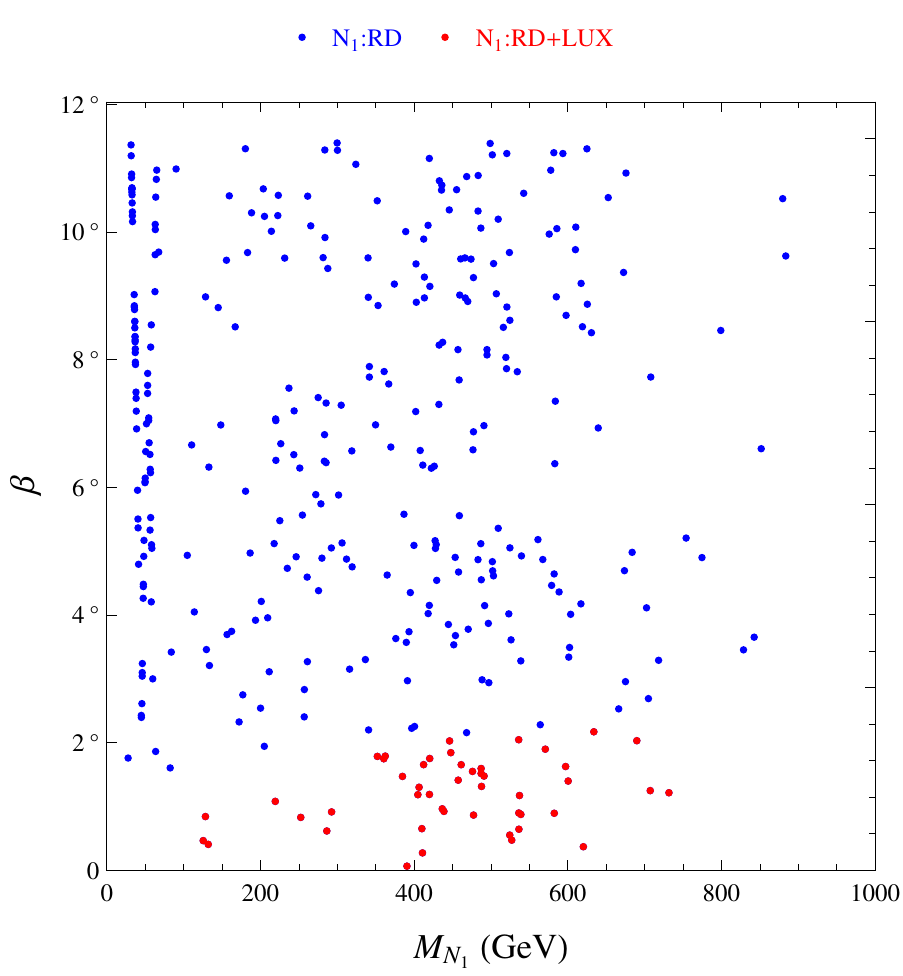}
~~
\includegraphics[width=0.45\linewidth]{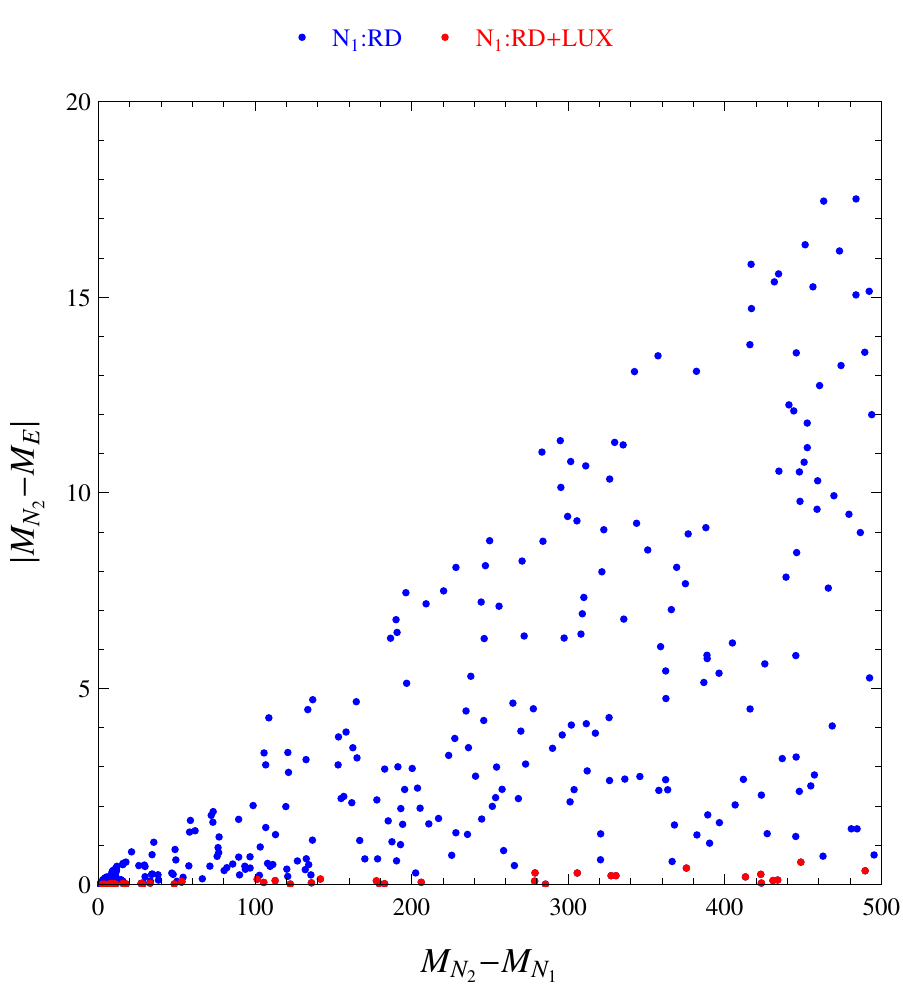}
\end{center}
\caption{Distributions of survived samples for $N_1$ DM in the $[M_{N_1},~\beta]$ (left panel) and $[M_{N_2}-M_{N_1},~|M_{N_2}-M_{E}|]$ (right) plane. The symbols of points are the same as in Fig.~\ref{DM1}.
\label{DM3}}
\end{figure}

In order to investigate the parameter space more comprehensively, the distributions of survived samples in the $[M_{\chi_L},\lambda_{h\chi_L}]$ for $\chi_L$ DM are shown in Fig.~\ref{DM2}. Here $\lambda_{h\chi_L}=\lambda_3\alpha_{-}^2$ is the $h\chi_L\chi_L^\dagger$ coupling. Similarly, the distributions for $N_1$ DM samples are displayed in the $[M_{N_1},~\beta]$ and $[M_{N_2}-M_{N_1},~|M_{N_2}-M_{E}|]$ plane
in Fig.~\ref{DM3}; here both $M_{N_2}-M_{N_1}$ and $\beta$ enter the charged fermion
mass $M_E$. We summarize the properties seen in the figures:
\begin{itemize}
  \item {As shown in Fig.~\ref{DM2} for $\chi_L$ DM, survived samples are distributed in a narrow band in the $[M_{\chi_L},\lambda_{h\chi_L}]$ plane, and most RD+LUX samples prefer a heavy DM.}
  \item {Direct detection indeed imposes a stringent limit on the mixing angle $\beta$ in the case of $N_1$ DM. For samples passing the LUX bound, $\beta$ cannot exceed $2^{\circ}$. In addition, the charged fermion mass $M_E$ is dominated by $M_{N_2}$, with a mass splitting determined by $M_{N_2}-M_{N_1}$ and $\sin\beta$. The maximal splitting reaches $17$ GeV when $M_{N_2}-M_{N_1}\simeq 500$ GeV. However, upon imposing the LUX constraint, $M_E$ always stays nearly degenerate with $M_{N_2}$.}
 \end{itemize}

\begin{figure}[!htbp]
\begin{center}
\includegraphics[width=0.5\linewidth]{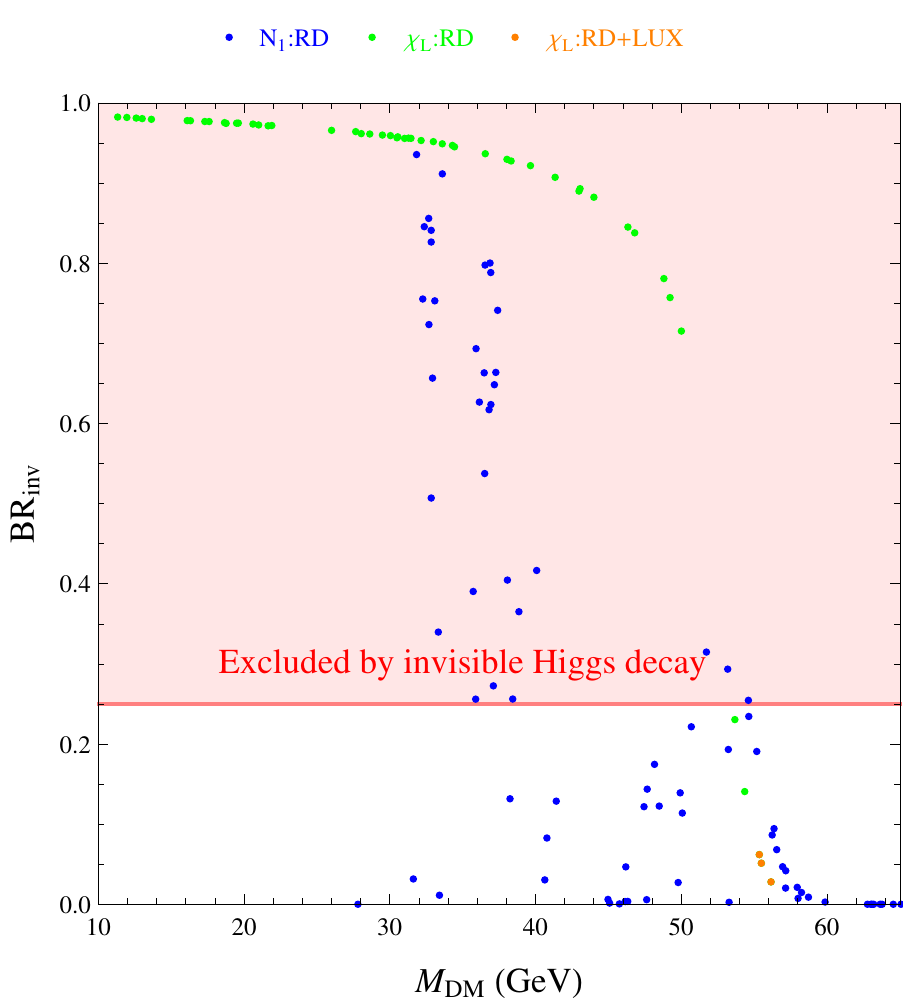}
~~
\end{center}
\caption{Distribution of $\rm{BR}_{inv}$ as a function of $M_{\rm DM}$ for RD and RD+LUX survived samples. The shaded area is excluded by the $95\%$ CL upper bound from Ref~\cite{Aad:2015pla}.
\label{inv1}}
\end{figure}

The direct searches for invisible Higgs decays have been carried out by ATLAS and CMS in the weak boson fusion (WBF)~\cite{Aad:2015txa,Chatrchyan:2014tja} and $Zh$ associated production channels \cite{Aad:2014iia,Chatrchyan:2014tja}, with the $95\%$ CL upper bounds on BR$(h\to\mbox{invisible})$ of $28\%$(ATLAS), $65\%$(CMS) in the WBF channel and $75\%$(ATLAS), $83\%$(CMS) in the $Zh$ channel, respectively. Alternatively, invisible Higgs decays also get constrained by fitting to visible Higgs decays \cite{Bechtle:2014ewa,Corbett:2015ksa,Aad:2015pla}. The upper bound thus found is stronger, BR$(h\to\mbox{invisible})<25\%$ ($95\%$ CL) \cite{Aad:2015pla}, which will be employed in our following discussion. The decay width to scalar or fermion DM reads
\begin{eqnarray}
\nonumber
\Gamma(h\to \chi_L \chi_L^\dag)&=& \frac{\lambda_{h\chi_L}^2 v^2}{16\pi M_h}
\sqrt{1-4\frac{M_{\chi_L}^2}{M_h^2}},
\\
\Gamma(h\to N_1 \bar{N}_1) &=& \frac{M_h}{32\pi v^2}(M_{N_2}-M_{N_1})^2\sin^4 2\beta
\left(1-4 \frac{M_{N_1}^2}{M_h^2}\right)^{\frac{3}{2}},
\label{gammainv}
\end{eqnarray}
with $v\approx 246~\GeV$. We take $\Gamma_{\rm{vis}}=4.07$ MeV for the visible decay width at $M_h=125~\GeV$ and Eq.~\ref{gammainv} for the invisible one $\Gamma_{\rm{inv}}$. The scatter plot of the invisible decay branching fraction $\rm{BR}_{\rm{inv}}=\Gamma_{\rm{inv}}/(\Gamma_{\rm{vis}}+\Gamma_{\rm{inv}})$ is presented in Fig.~\ref{inv1} as a function of $M_{\rm DM}$ for RD and RD+LUX survived samples, where the shaded area indicates the region excluded by the upper bound from Ref.~\cite{Aad:2015pla}. We found that for $\chi_L$ DM, samples with $M_{\chi_L}<55$ GeV are totally excluded while for $N_1$ DM the corresponding bound can be relaxed to about $28~\GeV$.

\begin{figure}[!htbp]
\begin{center}
\includegraphics[width=0.45\linewidth]{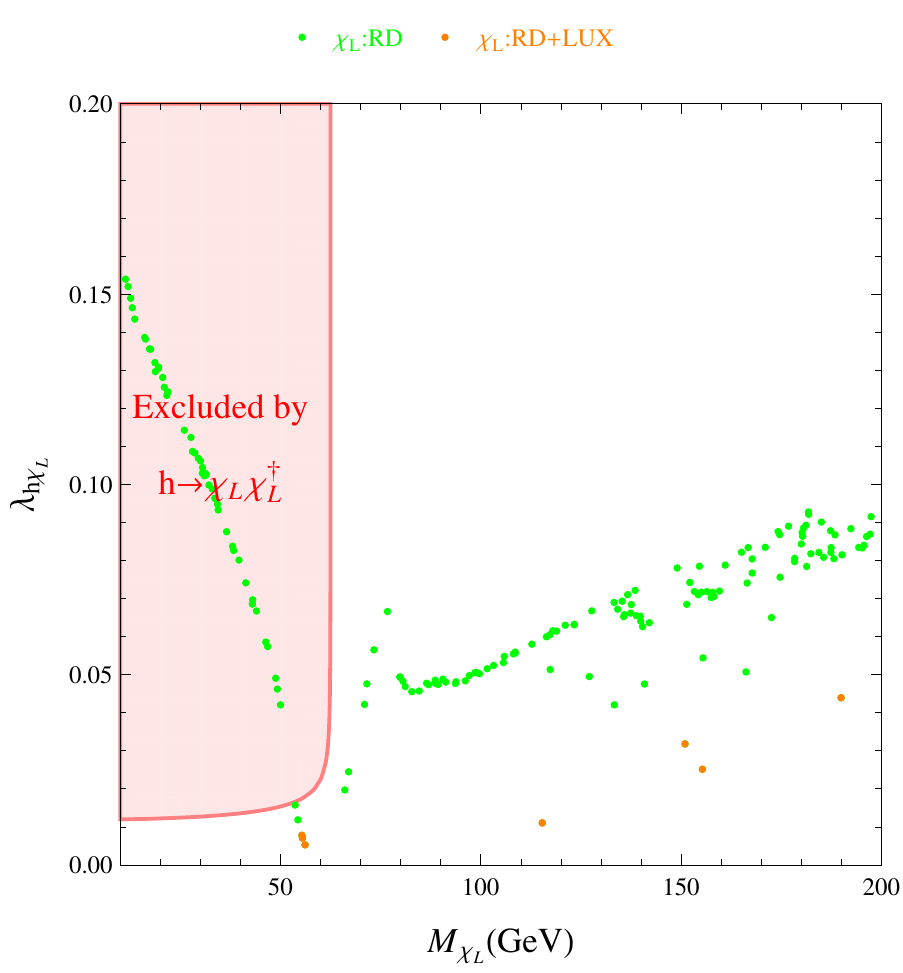}
~~
\includegraphics[width=0.45\linewidth]{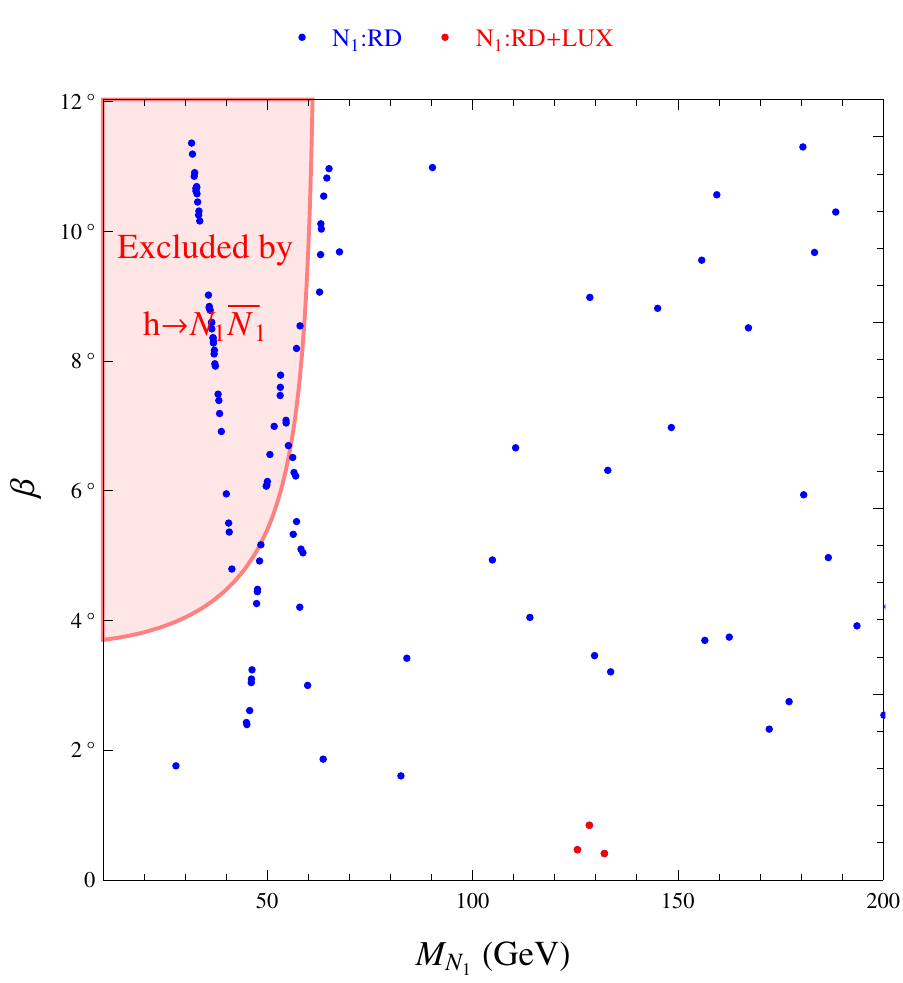}
~~
\end{center}
\caption{Left (right) panel: exclusion region from invisible decay $h\to \chi_L \chi_L^\dag$ ($N_1 \bar{N}_1$) in the $[M_{\chi_L},~\lambda_{h\chi}]$ ($[M_{N_1},~\beta]$) plane for $\chi_L$ ($N_1$) DM. For $N_1$ DM, $M_{N_2}-M_{N_1}=500~\GeV$ is fixed,yielding the most stringent limit.
\label{inv2}}
\end{figure}

One can also convert the upper bound on invisible Higgs decays into excluded regions in the $[M_{\chi_L},\lambda_{h\chi_L}]$ ($[M_{N_1},~\beta]$) plane for $\chi_L$ ($N_1$) DM. As shown in Fig.~\ref{inv2}, the correlations among parameters manifest themselves more explicitly. In this manner, we obtain the most stringent bound $\lambda_{h\chi_L}\lesssim 0.01$ with $M_{\chi_L}<55~\GeV$ for $\chi_L$ DM, or $\beta\lesssim 4^{\circ}$ with $M_{N_2}-M_{N_1}=500~\GeV$ for $N_1$ DM in the most stringent case. We notice that these constraints are less stringent than from direct detection in the same mass regions, such that all of RD+LUX samples easily survive for either $\chi_L$ or $N_1$ DM. Finally, $N_1$ DM also contributes to the invisible $Z$ decay if kinematically allowed,
\begin{equation}
\Gamma(Z\to N_1 \bar{N}_1)= \frac{M_Z^3\sin^4\beta}{12\pi v^2}
\left(1+2\frac{M_{N_1}^2}{M_Z^2}\right)\sqrt{1-4\frac{M_{N_1}^2}{M_Z^2}}.
\end{equation}
The LEP $95\%$ CL upper bound of $3~\MeV$~\cite{ALEPH:2005ab} translates to $\beta\lesssim 17^{\circ}$, which is weaker than from invisible Higgs decays. For light $N_2$, the decays
$h\to N_1 \bar{N}_2/N_2 \bar{N}_1,~N_2 \bar{N}_2$ and
$Z\to N_1 \bar{N}_2/N_2 \bar{N}_1,~N_2 \bar{N}_2$ may also be allowed.
These decay modes could provide more severe constraints~\cite{Bhattacharya:2015qpa}, but are still weaker than the LUX constraint. Therefore we will not consider invisible decays as an individual constraint in our later analysis.

\subsection{Analysis of annihilation channels}
\label{aac}

\begin{figure}[!htbp]
\begin{center}
\includegraphics[width=0.25\linewidth]{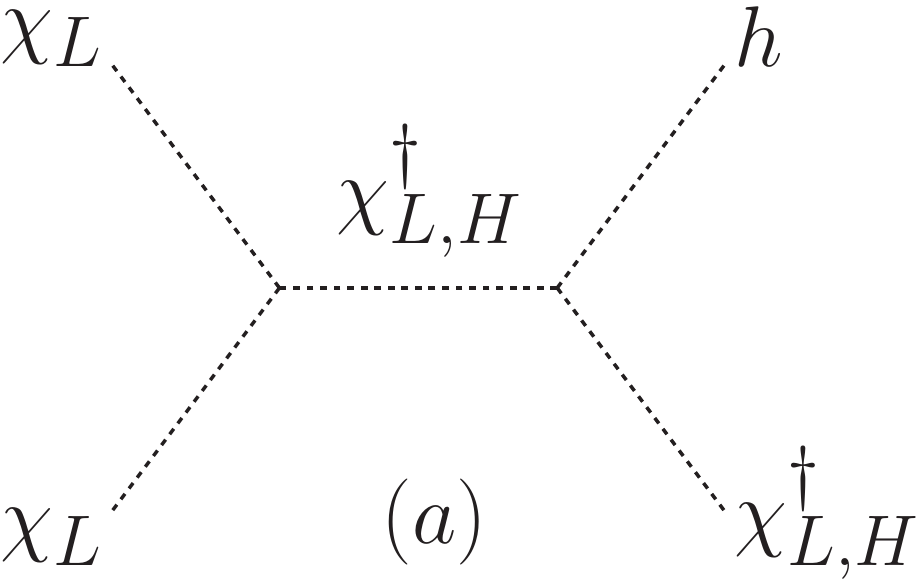}
~~
\includegraphics[width=0.24\linewidth]{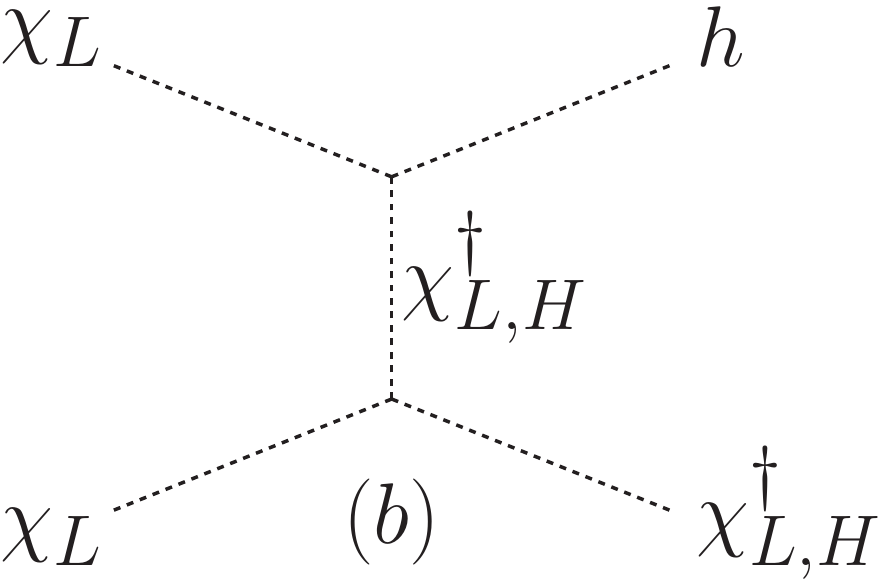}
~~
\includegraphics[width=0.28\linewidth]{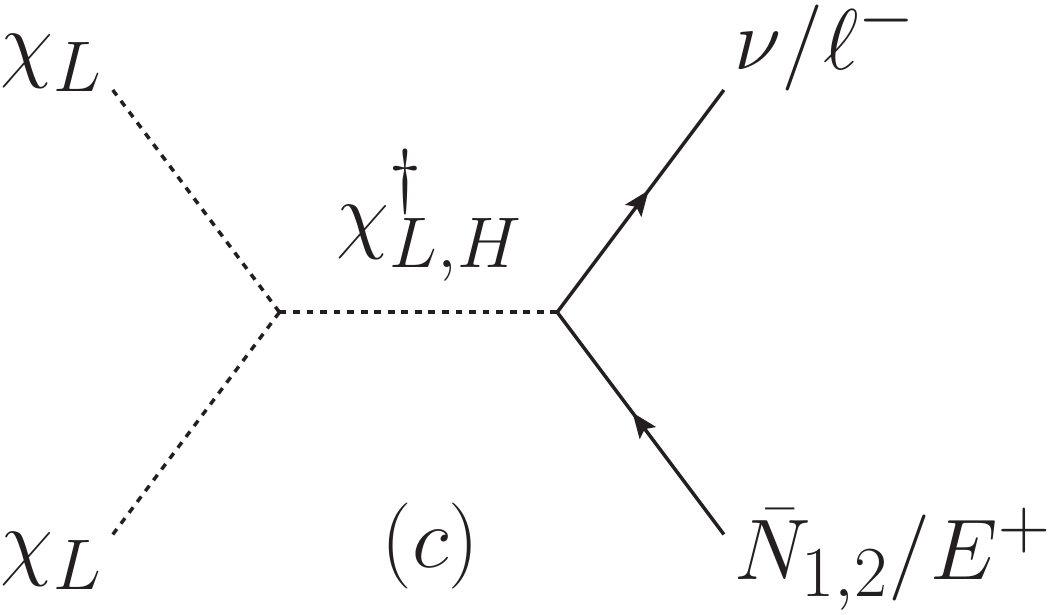}
\end{center}
\begin{center}
\includegraphics[width=0.25\linewidth]{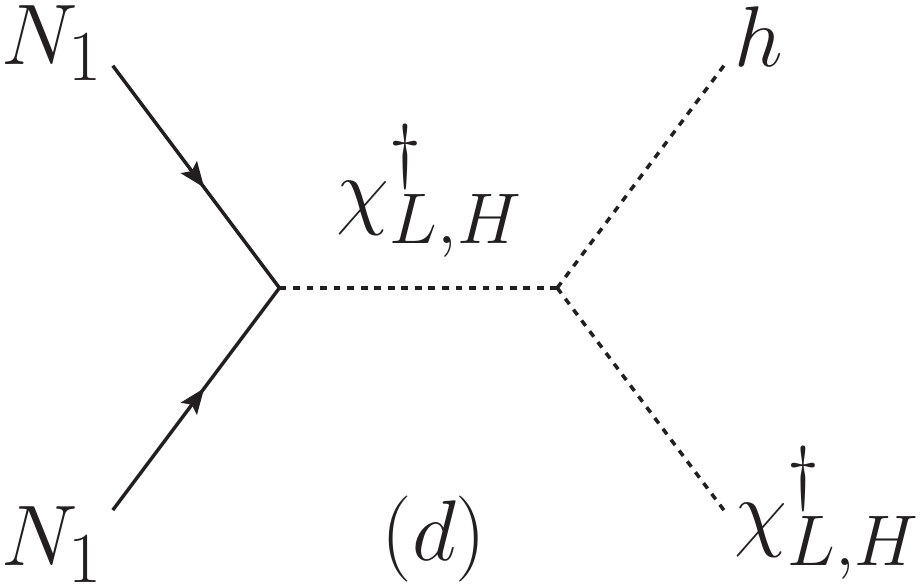}
~~
\includegraphics[width=0.24\linewidth]{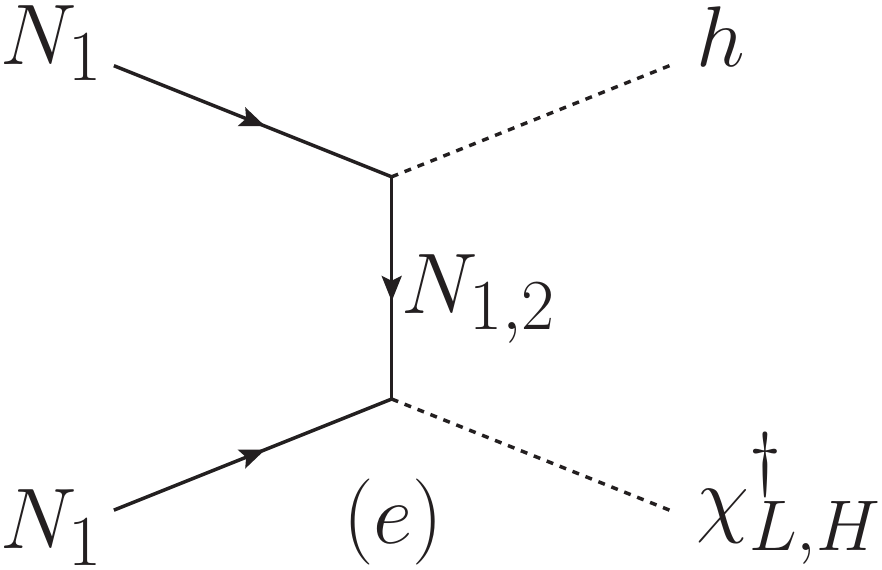}
~~
\includegraphics[width=0.28\linewidth]{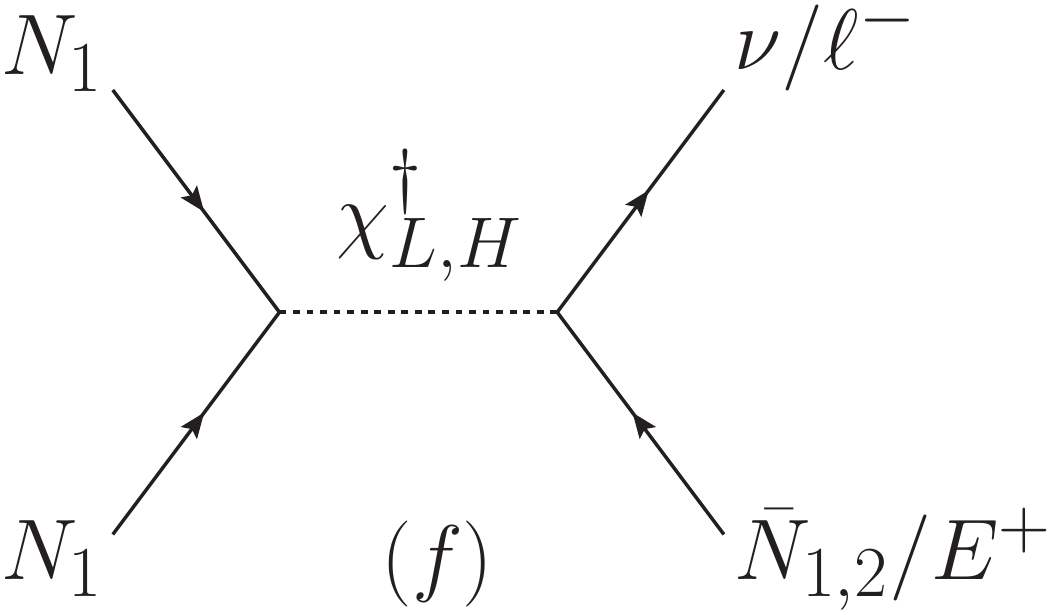}
\end{center}
\caption{SE-A processes for $\chi_L$ DM (upper panel) and $N_1$ DM (lower panel) respectively. \label{semi}}
\end{figure}

The aim of this subsection is to demonstrate the effects of various annihilation channels on relic density and direct detection, especially the crucial roles played by CO-A and SE-A processes. For this purpose, we first list all SE-A processes for both $\chi_L$ and $N_1$ DM. As seen in Fig.~\ref{semi}, a $\chi_L$ pair can annihilate into $\chi^{\dag}_{L,H} h$, $\bar{N}_{1,2}\nu$ and $E^+\ell^-$ final states via the $s$- or $t$-channel exchange of $\chi^{\dag}_{L,H}$. Similarly, an $N_1$ pair annihilates into $\chi_{L,H}^\dagger h$, $\bar N_{1,2}\nu$ and $E^+\ell^-$ final states via the exchange of an $s$-channel $\chi^{\dag}_{L,H}$ or a $t$-channel $N_{1,2}$. Therefore, the $s$-channel annihilation may dominate when $M_{\chi_H}\simeq 2 M_{\chi_L}$ for $\chi_L$ DM or $M_{\chi_{L,H}}\simeq 2 M_{N_1}$ for $N_1$ DM. As we will show, the $t$-channel annihilation can also dominate in some regions. Moreover, CO-A processes are important in this model, which occur and even dominate in the case of $M_{\chi_H}\simeq M_{\chi_L}$ for $\chi_L$ DM or of $M_{\chi_{L,H},~N_{2}}\simeq M_{N_{1}}$ for $N_1$ DM. Finally, ST-A processes still have significant contributions in certain parameter regions.

\begin{figure}[!htbp]
\begin{center}
\includegraphics[width=0.48\linewidth]{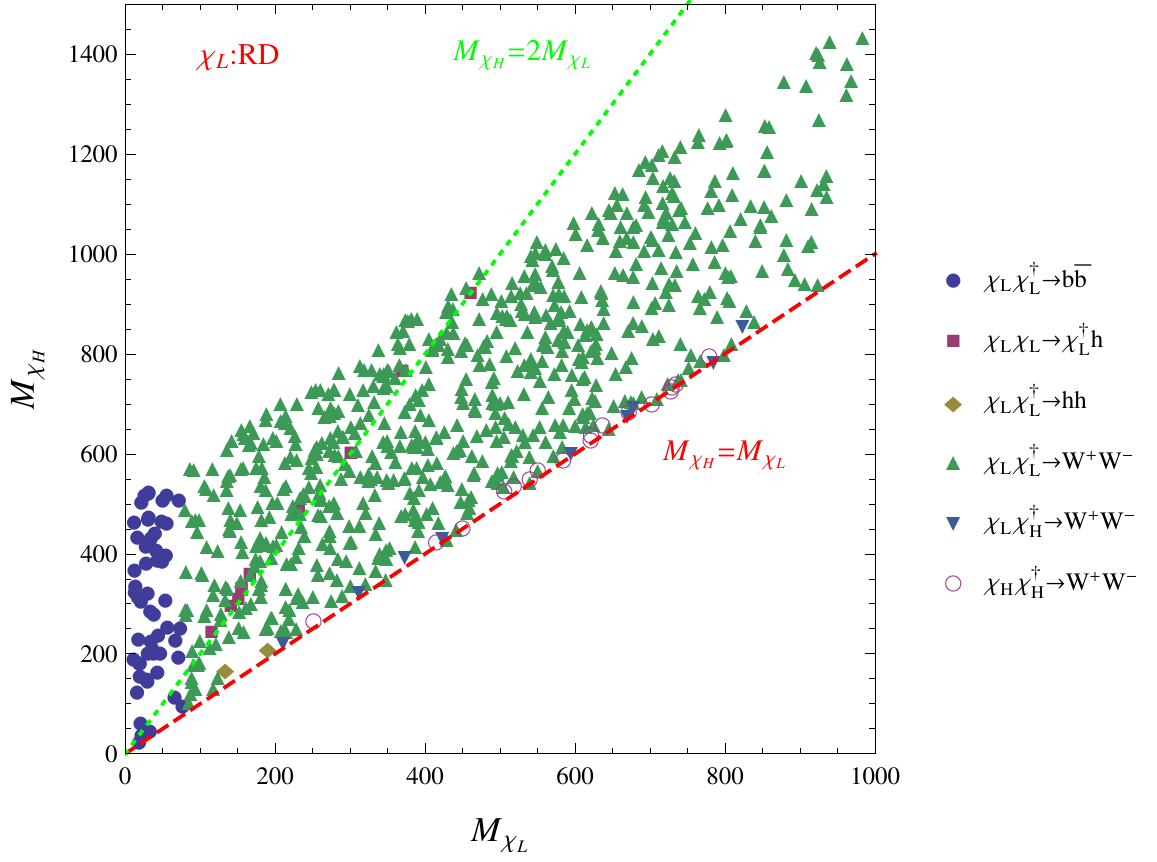}
~~
\includegraphics[width=0.48\linewidth]{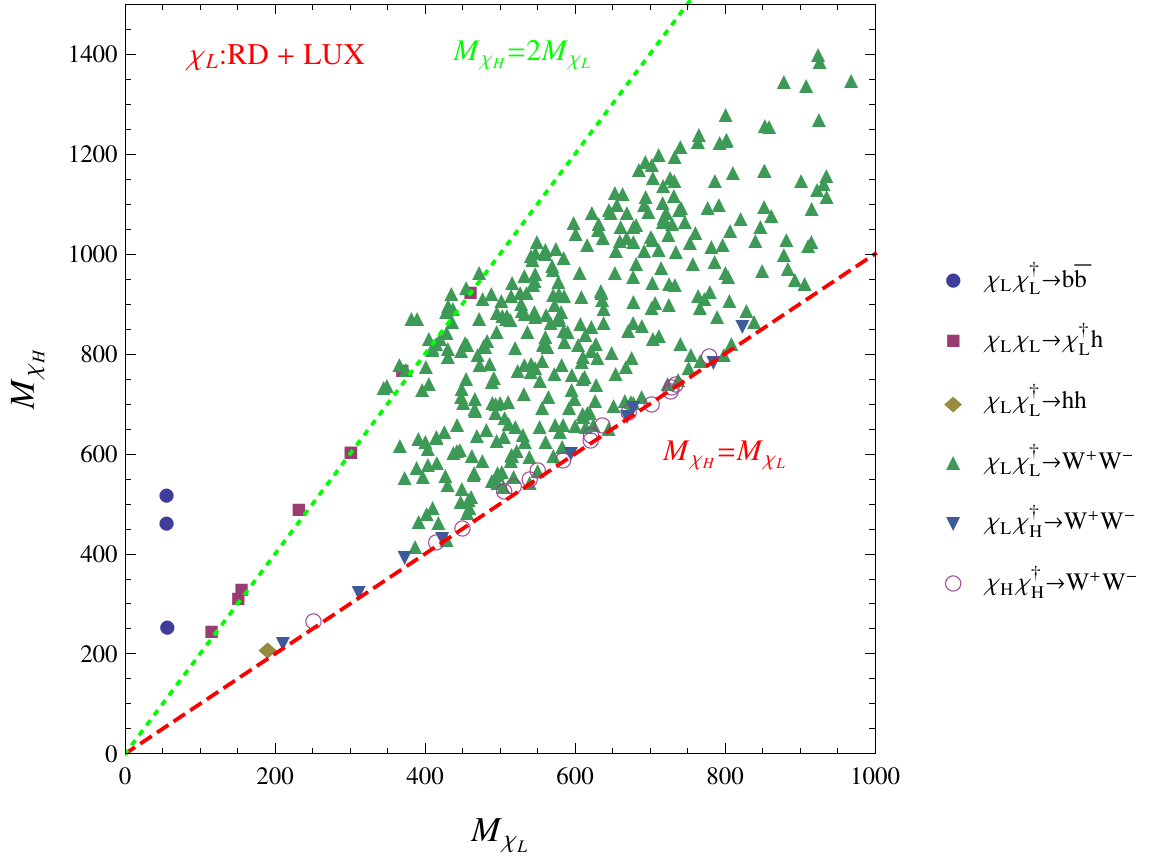}
~~
\end{center}
\caption{Distribution of dominant annihilation channels for $\chi_L$ DM in the $[M_{\chi_L},M_{\chi_H}]$ plane. Left (right) panel corresponds to RD (RD+LUX) survived samples.
\label{channel1}}
\end{figure}

\begin{figure}[!htbp]
\begin{center}
\includegraphics[width=0.48\linewidth]{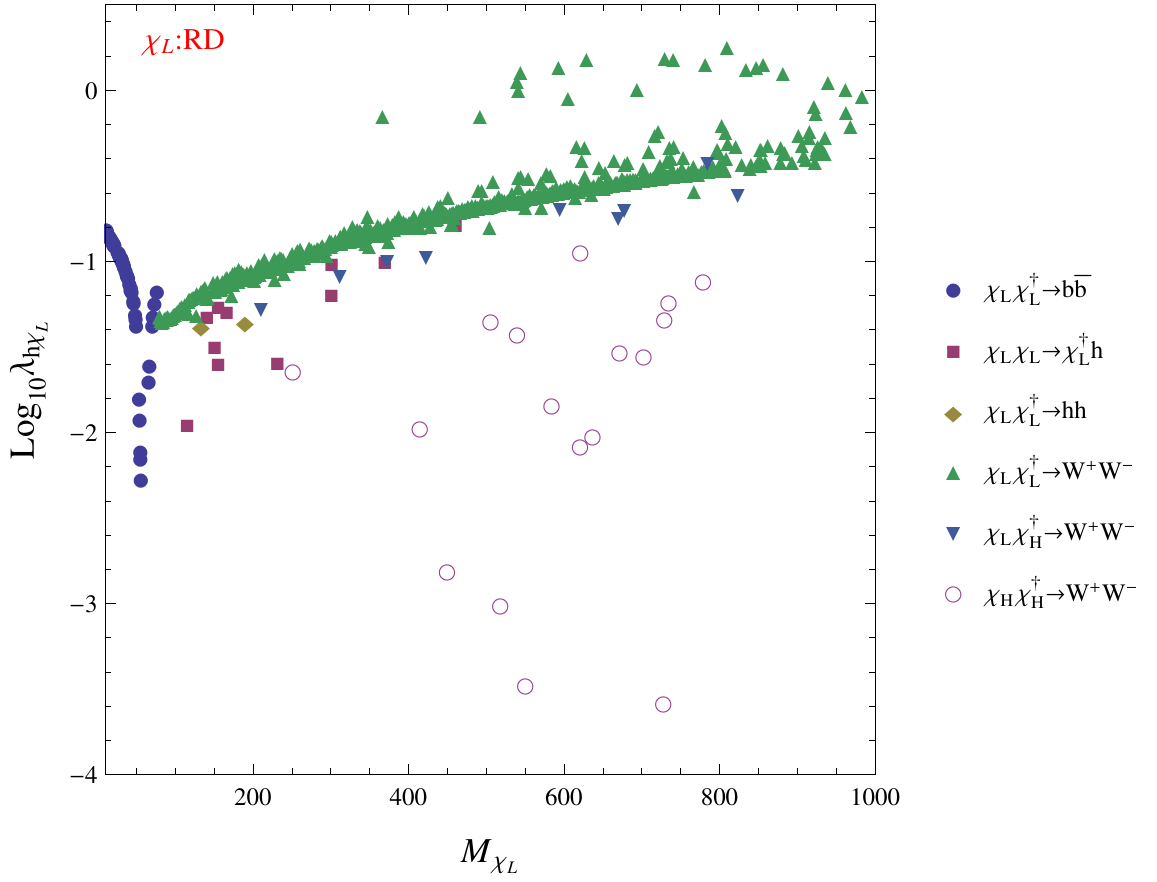}
~~
\includegraphics[width=0.48\linewidth]{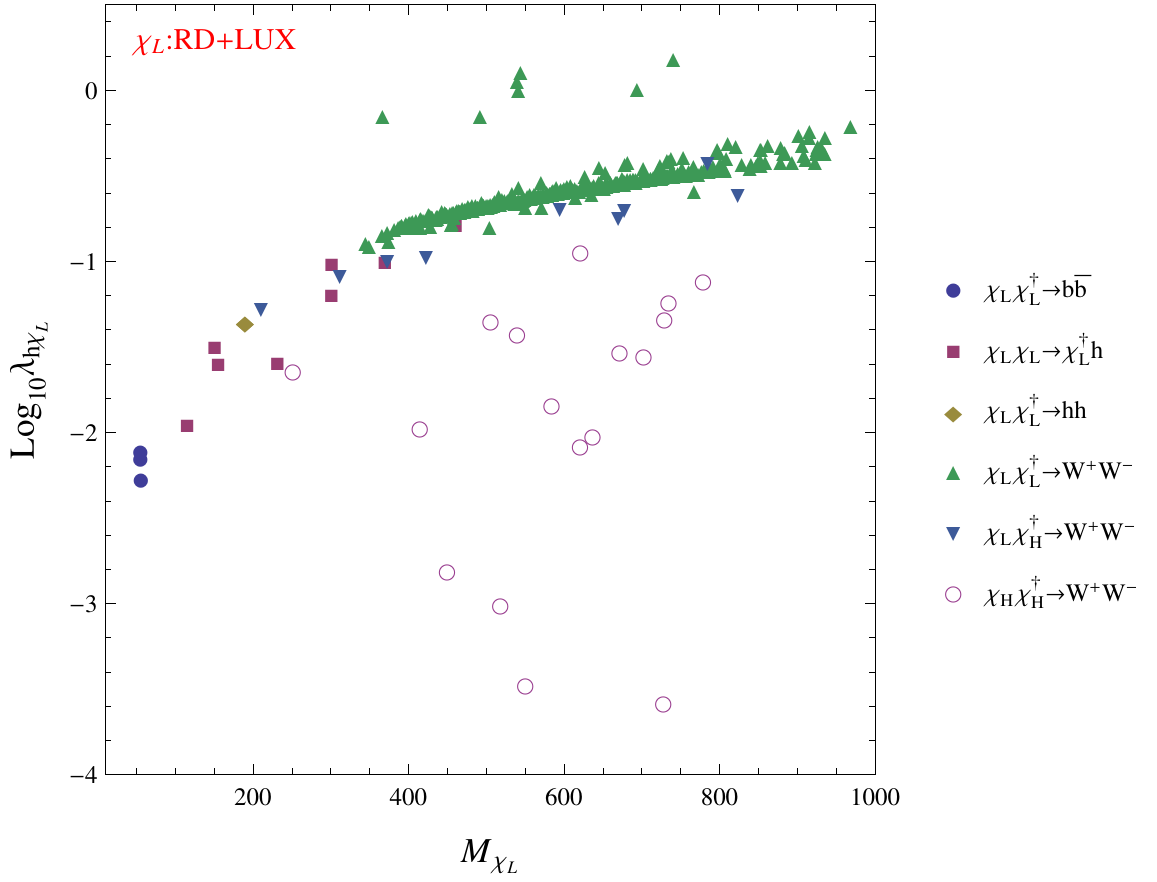}
~~
\end{center}
\caption{Same as Fig.~\ref{channel1} but in the $[M_{\chi_L},~\log_{10}\lambda_{h\chi_L}]$ plane.
\label{channel2}}
\end{figure}

\begin{figure}[!htbp]
\begin{center}
\includegraphics[width=0.5\linewidth]{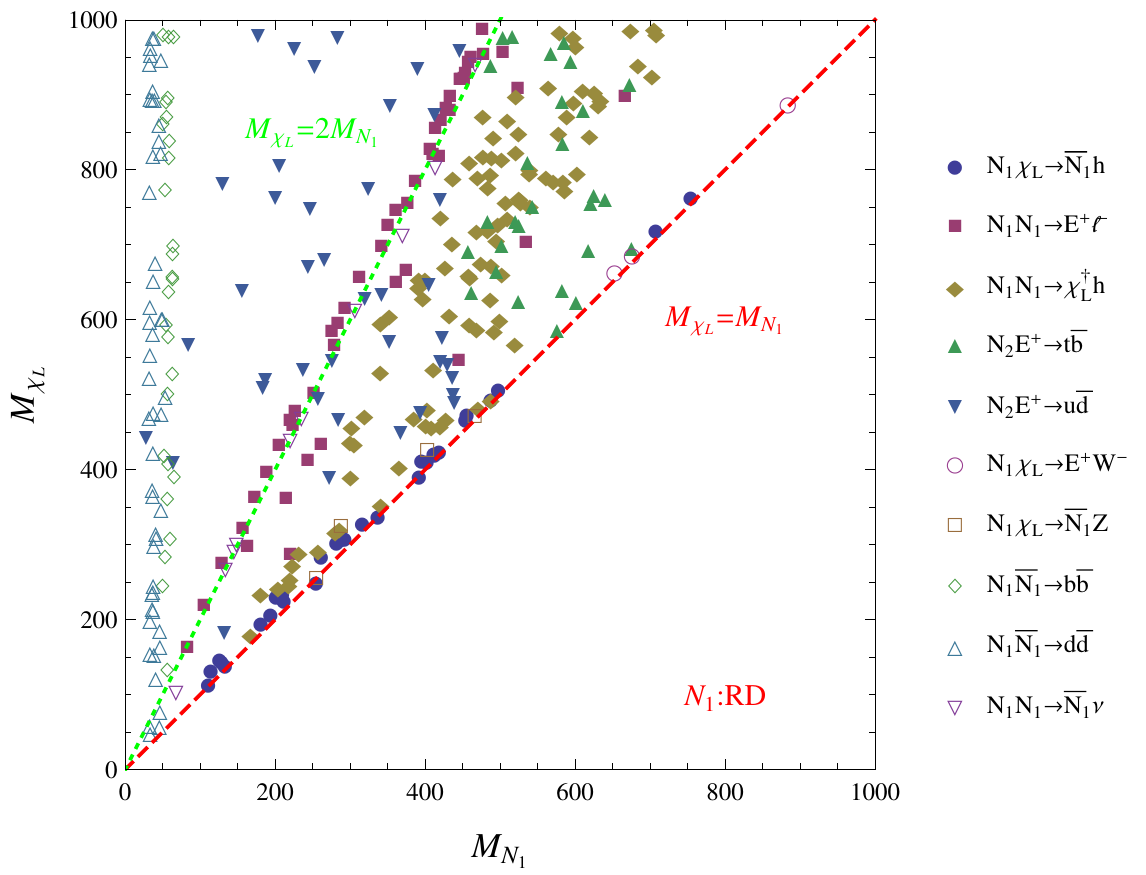}
~~
\includegraphics[width=0.48\linewidth]{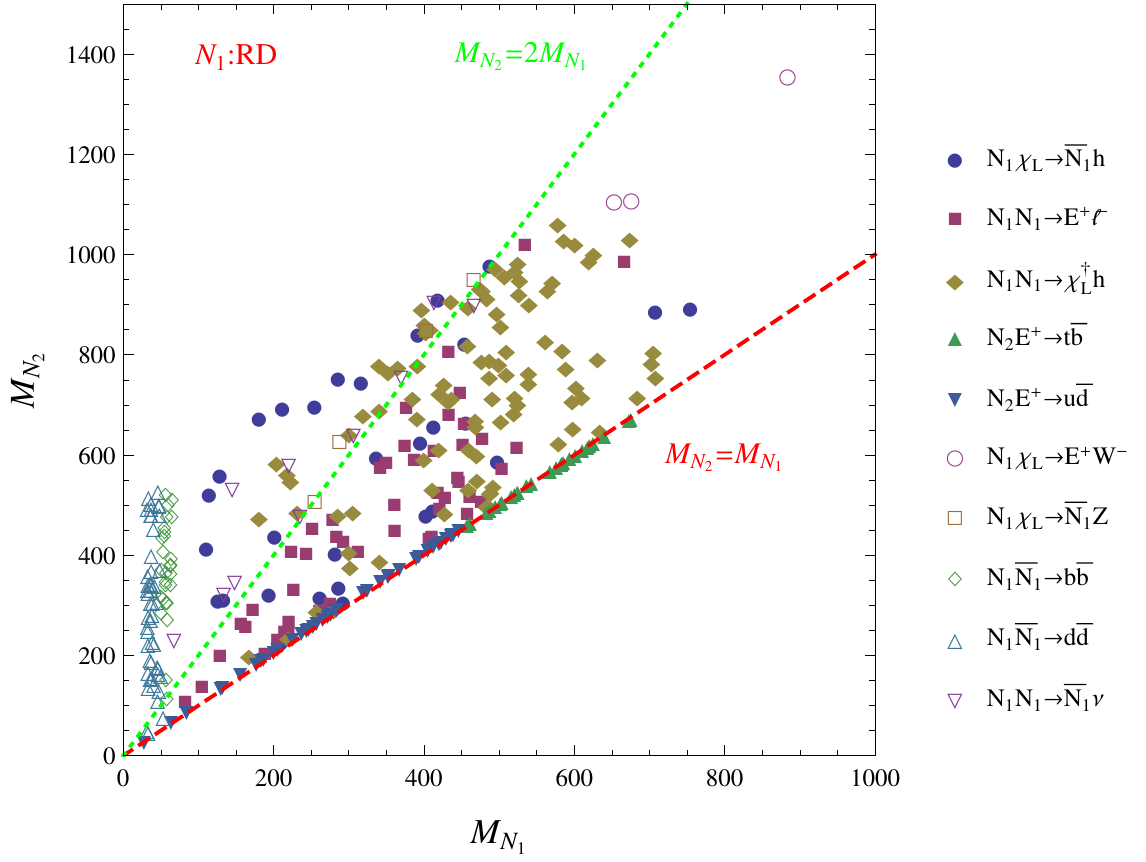}
~~
\includegraphics[width=0.48\linewidth]{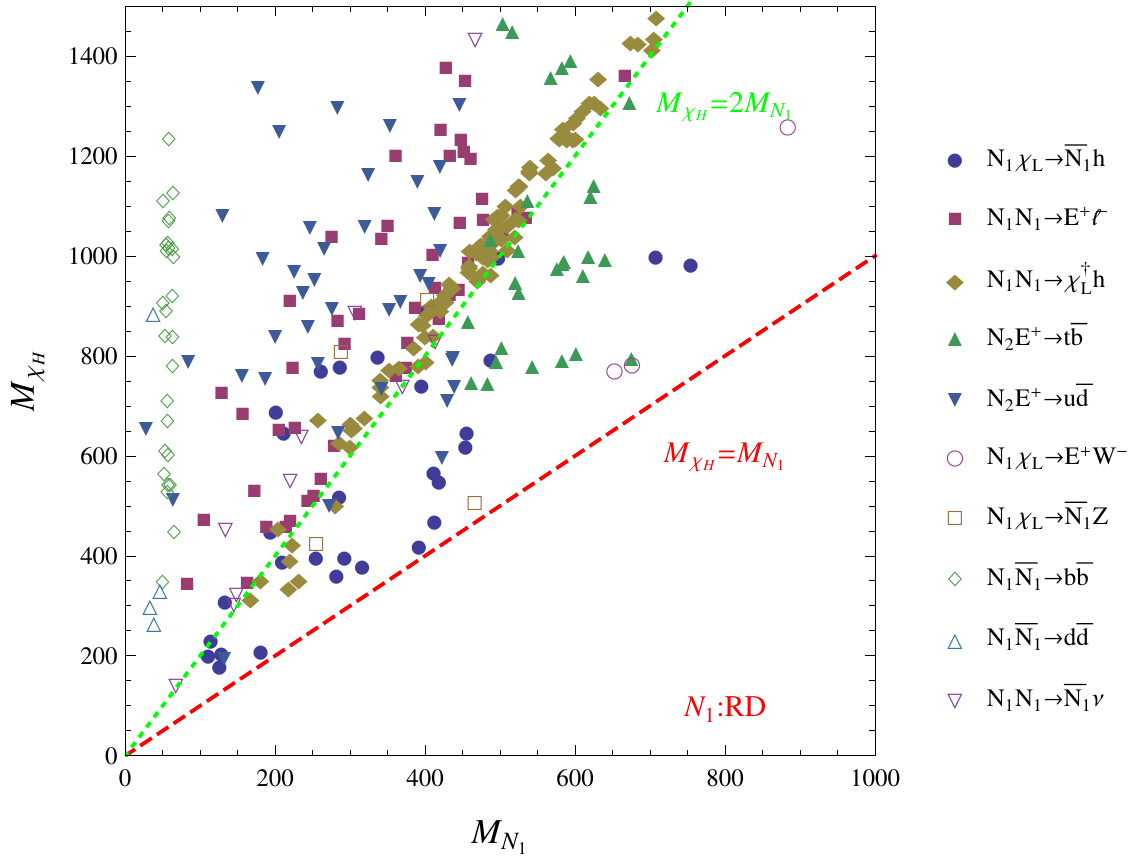}
\end{center}
\caption{Distribution of dominant annihilation channels for $N_1$ DM with RD survived samples in the plane of $[M_{N_1},M_{\chi_L}]$ (top panel), $[M_{N_1},M_{N_2}]$ (bottom-left), and $[M_{N_1},M_{\chi_H}]$ (bottom-right) respectively.
\label{channel3}}
\end{figure}

\begin{figure}[!htbp]
\begin{center}
~~
\includegraphics[width=0.5\linewidth]{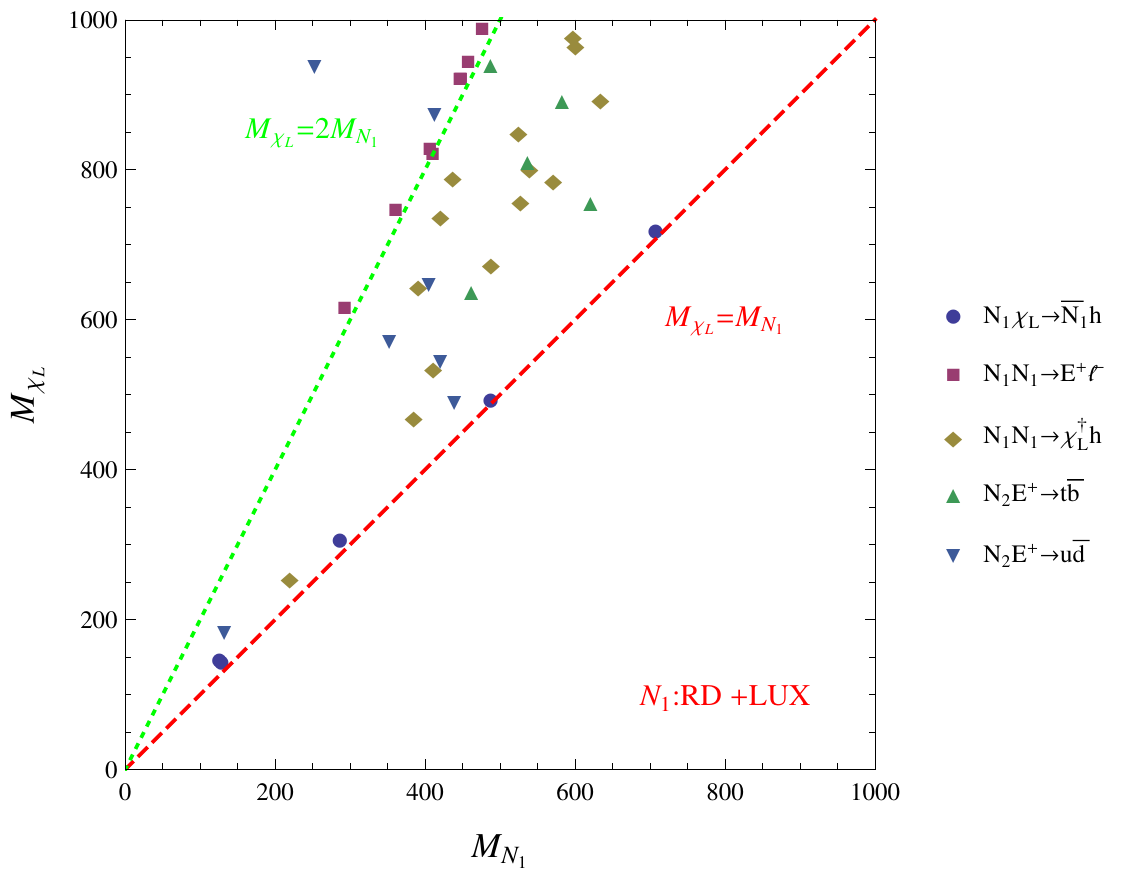}
~~
\includegraphics[width=0.48\linewidth]{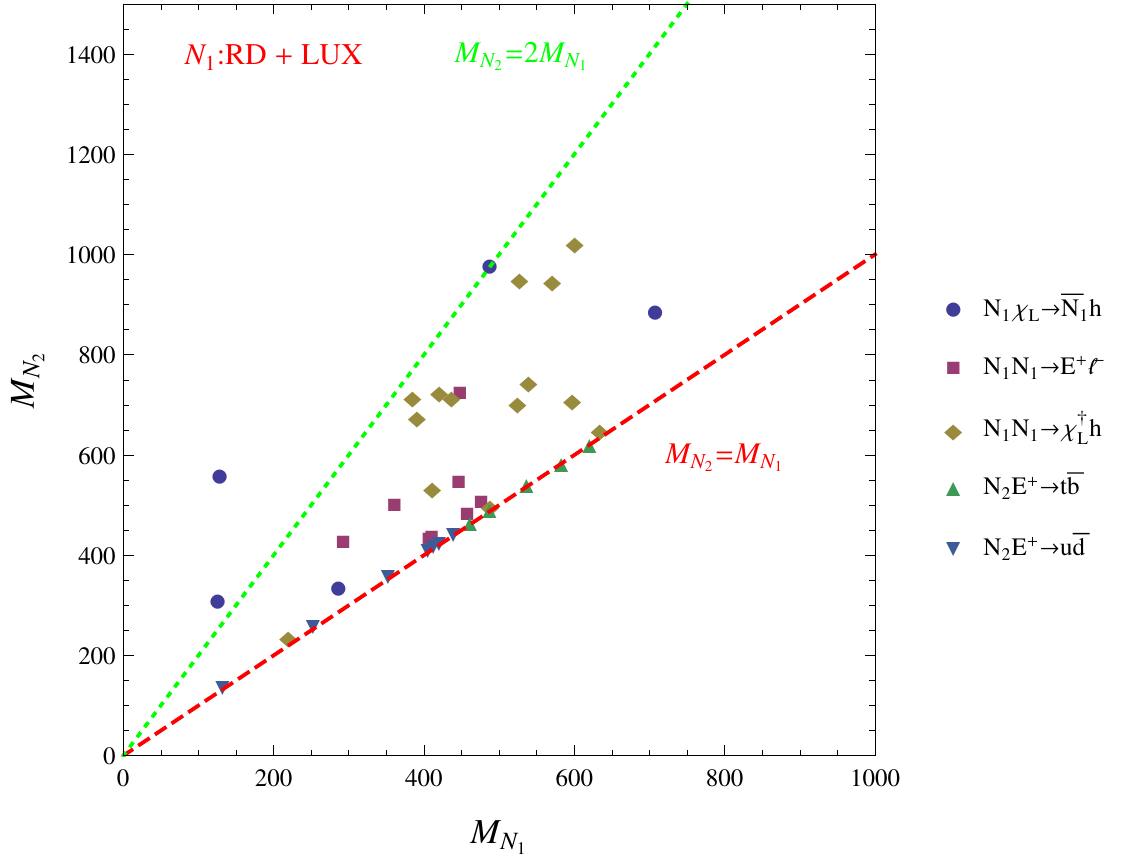}
~~
\includegraphics[width=0.48\linewidth]{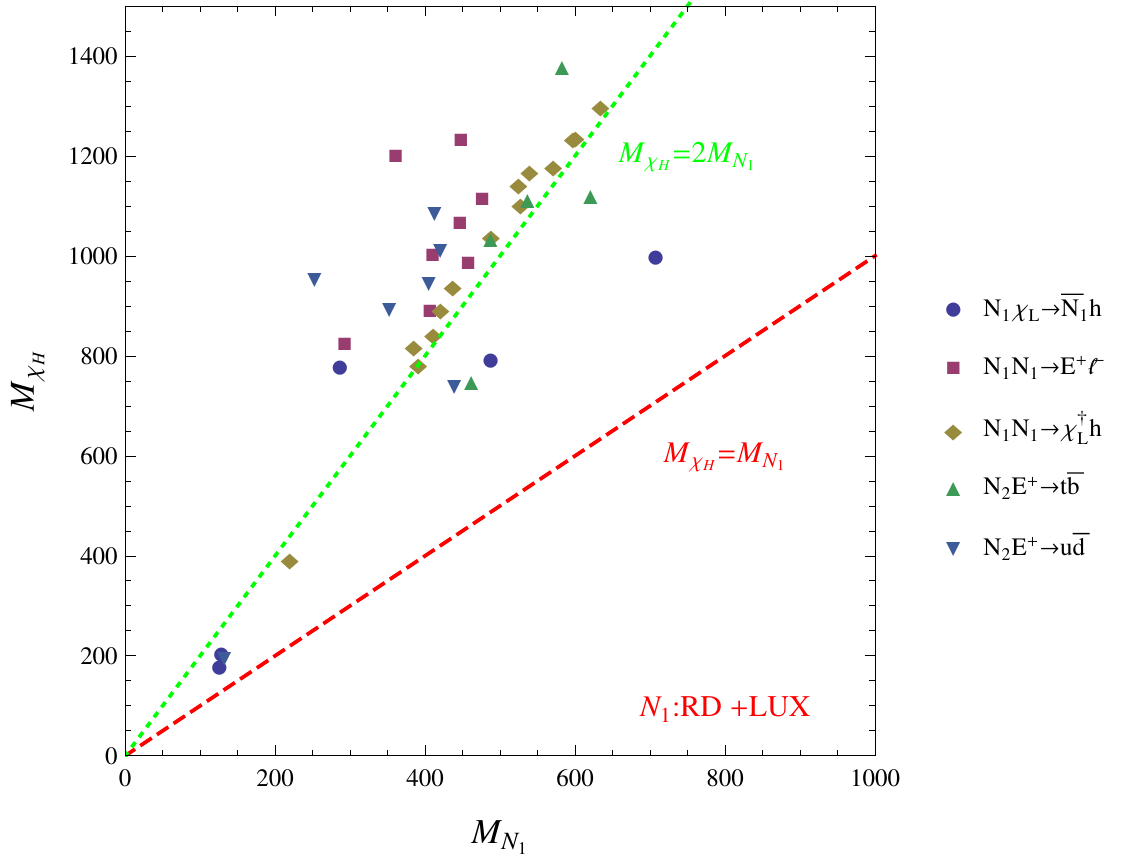}
\end{center}
\caption{Same as Fig.~\ref{channel3} but for RD+LUX survived samples.
\label{channel4}}
\end{figure}

With such an involved annihilation pattern as described above, a clear way of investigation is to bookkeep the most dominant annihilation channels for each survived sample and examine their distributions in the parameter space. These distributions are displayed in Figs.~\ref{channel1} and \ref{channel2} for $\chi_L$ DM and in Figs.~\ref{channel3} and~\ref{channel4} for $N_1$ DM. For comparison, the lines of $M_{\chi_H}=M_{\chi_L},~2 M_{\chi_L}$ and $M_{\chi_{H,L}}=M_{N_1},~2 M_{N_1}$ are also shown respectively. (The line of $M_{N_2}=2 M_{N_1}$ for the latter is only plotted for completeness.) The fractions of various channels in survived samples are listed in Table~\ref{fraction}. We can gain some useful information from the figures and table.

For $\chi_L$ DM, we have the following observations:
\begin{itemize}
  \item {As seen in Fig.~\ref{channel1}, $\chi_L$ DM has three ST-A channels. For light DM ($M_{\chi_L}<M_h/2$), it dominantly annihilates into $b\bar{b}$, while for heavy DM ($M_{\chi_L}>M_W$), the dominant annihilation processes are into gauge boson and Higgs pairs. Since DM annihilating through the Higgs portal type always tends to produce more gauge boson than Higgs pairs, the majority of samples is from the $W^{+}W^{-}$ channel with rare samples coming from the $hh$ channel. Furthermore, SE-A (CO-A) processes occur only when $M_{\chi_L}>M_h$ ($M_{\chi_L}>M_W$) for kinematical reasons. As expected, $\chi_L\chi_L\to \chi_L^\dagger h$ or $\chi_{L,H}\chi_H^\dagger\to W^{+}W^{-}$ dominates when $M_{\chi_H}\simeq2 M_{\chi_L}$ or $M_{\chi_H}\simeq M_{\chi_L}$, but all of them take a small fraction.}

  \item {For light DM, since annihilation cross section for the $b\bar{b}$ channel is suppressed by the Yukawa coupling of $b$, one first requires a relatively large $\lambda_{h\chi_L}$ to saturate relic density. As $M_{\chi_L}$ approaches $M_h/2$, resonance enhancement and phase space suppression compete. Since the former dominates, the overall effect is to require a decline in $\lambda_{h\chi_L}$. After $M_{\chi_L}$ climbs over the $h$ resonance, the opposite takes place, resulting in the valley structure in the left panel of Fig.~\ref{channel2}. This is indeed a common feature of Higgs-portal models. On the other hand, for heavy DM, the annihilation cross sections for the $W^{+}W^{-}$ and $hh$ channels are respectively proportional to the gauge coupling and Higgs self-coupling, so that relic density selects a narrow band in the $[M_{\chi_L},~\lambda_{h\chi_L}]$ plane.}

  \item {Upon imposing the direct detection constraint, most samples with the $b\bar{b}$ channel are excluded since $\lambda_{h\chi_L}$ as required by relic density is too large to evade the LUX bound for such light DM. The only exception is a DM mass near the resonance area, where a few samples survive due to a much smaller $\lambda_{h\chi_L}$. In contrast, most of  samples with SE-A and CO-A processes are safe in this case. This feature is mainly because, when relic density is determined by these two processes, a smaller $\lambda_{h\chi_L}$ is still allowed for the same order of DM mass, therefore alleviating the tension from direct detection. For instance, the mass interval $M_{\chi_L}\in[80,~350]~\GeV$ is excluded by the LUX limit for the ST-A channel $\chi_L\chi_L^\dagger\to W^{+}W^{-}$ alone, but is allowed when the SE-A and CO-A channels $\chi_L\chi_L\to \chi_L^\dagger h$ and $\chi_L\chi_H^\dagger\to W^{+}W^{-}$ are taken into account. When $M_{\chi_L}>350~\GeV$, all above channels satisfy the LUX bound, but SE-A and CO-A channels still keep smaller couplings.}
\end{itemize}

For $N_1$ DM, we observe the following features:
\begin{itemize}
  \item {Compared with $\chi_L$ DM, $N_1$ DM has a more complicated annihilation pattern since more particles are involved in annihilation processes. As shown in Fig.~\ref{channel3}, there are two ST-A channels in the RD survived samples, $N_1\bar{N_1}\to d\bar{d},~b\bar{b}$, both dominating for $M_{N_1}<M_h/2$. For SE-A processes, $N_1N_1\to \chi_L^\dagger h$ dominates when $M_{\chi_H}\simeq 2M_{N_1}$, or $N_1N_1\to E^{+}\ell^{-},~\bar{N_1}\nu$ when $M_{\chi_L}\simeq2 M_{N_1}$. Finally, for CO-A processes, $N_1\chi_L\to \bar{N_1}h$ and $N_1\chi_L\to E^{+}W^{-},~\bar{N_1}Z$ channels dominate when $M_{\chi_L}\simeq M_{N_1}$, and $N_2E^{+}\to u\bar{d},~t\bar{b}$ do when $M_{N_2}\simeq M_E\simeq M_{N_1}$.}
  \item {Including the LUX limit, there are only five SE-A/CO-A annihilation channels that survive the combined RD+LUX constraints: $N_1\chi_L\to \bar{N_1}h,~N_1N_1\to E^{+}\ell^{-},~\chi_L^\dagger h$ and $N_2E^{+}\to u\bar{d},~t\bar{b}$, as shown in Fig.~\ref{channel4}. This is due to the similar reason as for $\chi_L$ DM, i.e., they benefit from smaller couplings compared with ST-A channels, which breaks the tight correlation between relic density and direct detection.}
\end{itemize}

\begin{table*}[hbt]
\begin{tabular}{|c|c|c|c|c|c|c|}
\hline
Channels ($\chi_L$)& $\chi_L\chi_L^\dagger\to b\bar{b}$ & $\chi_L\chi_L\to \chi_L^\dagger h$ & $\chi_L\chi_L^\dagger\to hh$    & $\chi_L\chi_L^\dagger\to W^{+}W^{-}$ & $\chi_L\chi_H^\dagger\to W^{+}W^{-}$  & $\chi_H\chi_H^\dagger\to W^{+}W^{-}$      \\
\hline
RD (\%) & $7.39$    & $1.43$            & $0.26$ & $87.55$ & $1.17$ & $2.2$ \\
\hline
RD+LUX (\%) & $0.78$    & $2.09$            & $0.26$ & $90.08$ & $2.35$ & $4.44$  \\
\hline
\hline
Channels ($N_1$) & $N_1\chi_L\to \bar{N_1}h$ & $N_1N_1\to E^{+}\ell^{-}$ & $N_1N_1\to \chi_L^\dagger h$   & $N_2E^{+}\to t\bar{b}$ & $N_2E^{+}\to u\bar{d}$  & $N_1\chi_L\to E^{+}W^{-}$       \\
\hline
RD (\%) & $8.9$    & $14.54$            & $28.49$ & $8.31$ & $11.87$ & $0.89$ \\
\hline
RD+LUX (\%) & $12.82$    & $20.51$            & $35.9$ & $12.82$ & $17.95$ & $\times$ \\
\hline
Channels ($N_1$)& $N_1\chi_L\to \bar{N_1}Z$   & $N_1\bar{N_1}\to b\bar{b}$  & $N_1\bar{N_1}\to d\bar{d}$ &  $N_1N_1\to \bar{N_1}\nu$ & & \\
\hline
RD (\%) & $1.19$   & $8.01$  & $14.84$ & $2.97$ &  & \\
\hline
RD+LUX (\%) & $\times$    & $\times$   & $\times$ & $\times$ &  & \\
\hline
\end{tabular}
\caption{Fractions of dominant annihilation channels in RD and RD+LUX survived samples. The slot with a symbol $\times$ indicates its channel has been excluded by direct detection.}
\label{fraction}
\end{table*}

\subsection{Comment on gamma-ray excess from the Galactic Center}
\label{SecGEC}

While a complete analysis on indirect detection constraints is beyond the scope of this paper,
we discuss briefly one of the most interesting anomalies in DM searches, namely a possible gamma-ray excess from the Galactic Center (GCE). The excess has been reported by a series of theoretical analyses using public data from the Fermi-LAT since 2009~\cite{Daylan:2014rsa,Calore:2014xka,Goodenough:2009gk,Hooper:2010mq,
Boyarsky:2010dr,Hooper:2011ti,Abazajian:2012pn}. Very recently, the Fermi collaboration has also released their own analysis~\cite{TheFermi-LAT:2015kwa}. This has attracted great attention in both astrophysics and particle physics communities. When using a ST-A process to interpret the excess, the spectrum is best fit by $b\bar{b}$ final states for a DM mass of $30-50~\GeV$ with $\langle\sigma v\rangle_{b\bar{b}} \in [1.4,~2]\times10^{-26}~\rm{cm}^3~s^{-1}$~\cite{Daylan:2014rsa}, and the morphology of DM density distribution matches the canonical Navarro-Frenk-White (NFW) halo profile. The $\tau^+\tau^-$, $q\bar{q}$ and $c\bar{c}$ ($gg$, $W^+W^-$, $ZZ$, $hh$ and $t\bar{t}$~) final states with a lighter (heavier) DM mass and a slightly different annihilation cross section are also acceptable~\cite{Calore:2014nla,Agrawal:2014oha,Cline:2015qha,Elor:2015tva}. Furthermore, it does not conflict with current limits from dwarf spheroidal, antiproton and CMB observations when taking into account uncertainties in the DM halo profile and propagation model~\cite{Calore:2014nla,Cirelli:2014lwa,Ade:2015xua,Slatyer:2015jla}. As usual, the excess can also be incorporated by astrophysical phenomena, including millisecond pulsars or unresolved gamma-ray point sources~\cite{Gordon:2013vta,Abazajian:2014fta,
Yuan:2014rca,Bartels:2015aea,Lee:2015fea}. However, astrophysical interpretations encounter some challenges on matching the spectrum and morphology of the excess. In any case, GCE has triggered extensive model building studies in both general and specific frameworks~\cite{Berlin:2014tja,Alves:2014yha,Agrawal:2014una,Abdullah:2014lla,
Martin:2014sxa,Berlin:2014pya,Basak:2014sza,Cline:2014dwa,Wang:2014elb,Cheung:2014lqa,
Ko:2014loa,Bell:2014xta,Okada:2014usa,Borah:2014ska,
Cahill-Rowley:2014ora,Guo:2014gra,Cao:2014efa,Freytsis:2014sua,Buckley:2014fba,
Hooper:2014fda,Dolan:2014ska,Cerdeno:2015ega,Alves:2015pea,Kaplinghat:2015gha,
Chen:2015nea,Modak:2015uda,Gherghetta:2015ysa,Rajaraman:2015xka,Mondal:2015rba,
Butter:2015fqa,Buckley:2015cia,Freese:2015ysa,Williams:2015bfa,Duerr:2015bea,Cai:2015zza,
Cai:2015tam}. These models can be divided into two scenarios from a model-independent viewpoint: one-step direct annihilation and multi-step cascade annihilation~\cite{Abdullah:2014lla,Martin:2014sxa,Elor:2015tva}. In the first scenario, DM annihilates directly into SM final states, so that its mass and cross sections are tightly bounded with the resulting photon spectrum. More critically, this scenario usually suffers from stringent constraints from direct detection and collider searches on DM or exchanged particles.
On the other hand, in the second scenario, DM annihilates into lighter mediators which subsequently decay to SM particles. Since cascade decays modify observed signals of DM annihilation, shift SM final states (and thus the resulting photons) to lower energies and broaden their spectra, the corresponding parameter space will be considerably extended and could evade bounds from direct detection.

\begin{figure}[!htbp]
\begin{center}
~~
\includegraphics[width=0.5\linewidth]{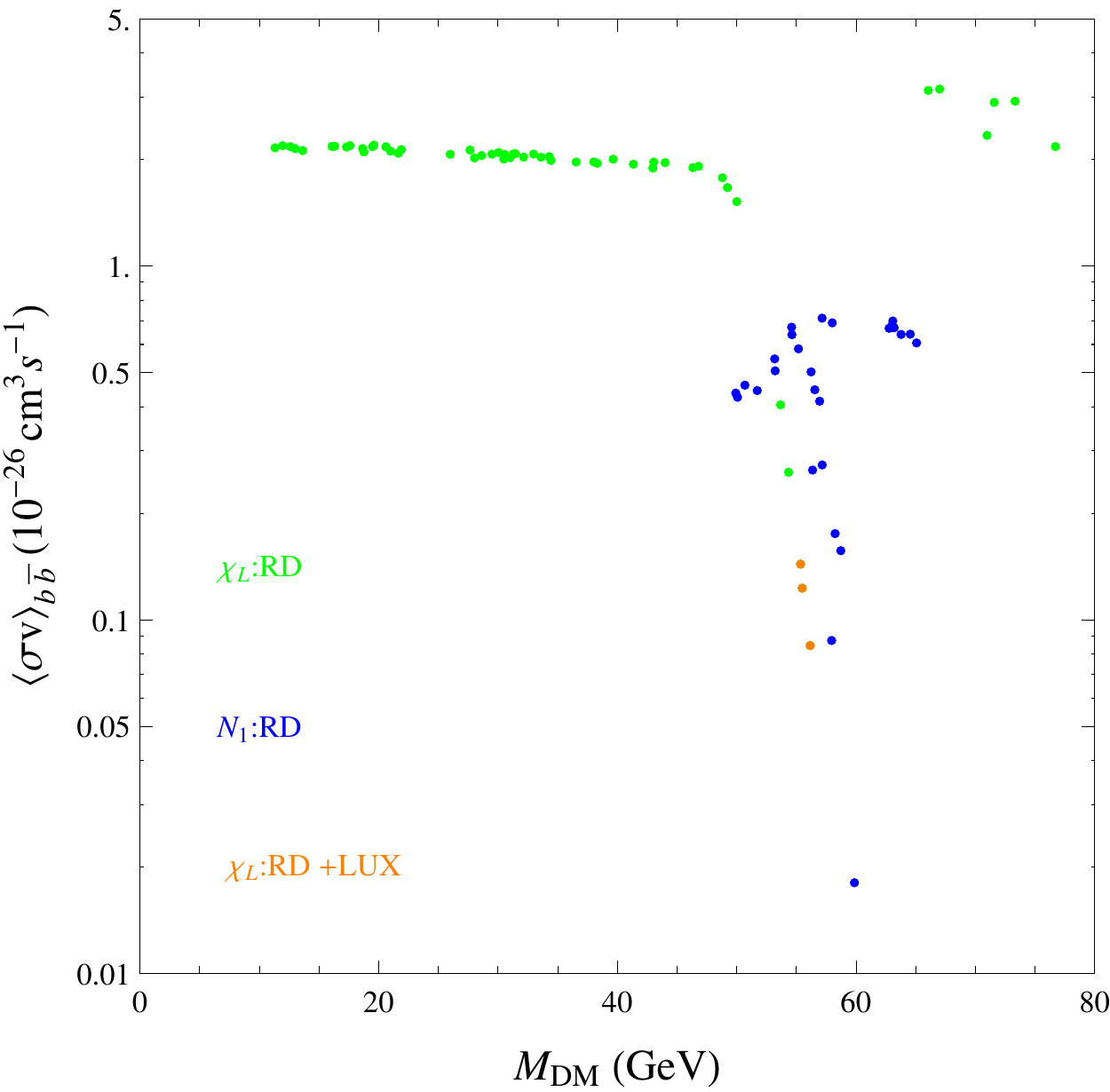}
\end{center}
\caption{Distribution of $\langle\sigma v\rangle_{b\bar{b}}$ (in unit of $10^{-26}~\rm{cm}^3~s^{-1}$) as a function of $M_{\rm DM}$ for survived samples. The region roughly consistent with the GCE is $M_{\rm DM}\sim 30-50$ GeV with $\langle\sigma v\rangle_{b\bar{b}} \in [1.4,~2]\times10^{-26}~\rm{cm}^3~s^{-1}$~\cite{Daylan:2014rsa}. Notice that for $N_1$ DM, all of samples with $b\bar{b}$ final states are excluded by direct detection constraint.
 \label{gamma}}
\end{figure}

In the $\mathbb{Z}_3$ model under study, ST-A channels also face the same difficulty mentioned above. In Fig.~\ref{gamma} we plot the distribution of $\langle\sigma v\rangle_{b\bar{b}}$ for all survived samples, except for RD+LUX samples in the case of $N_1$ DM, which are entirely excluded by the LUX constraint. We see that the parameter region consistent with GCE is excluded by direct detection. In order to avoid this conflict, some recent papers proposed a class of DM models with a local $\mathbb{Z}_3$ symmetry~\cite{Ko:2014nha,Ko:2014loa,Choi:2015bya}, which often arises as a remnant of a spontaneously broken hidden gauge symmetry. The GCE may then be explained using semi-annihilation channels associated with new Higgs/gauge bosons. More interestingly, as pointed out in Ref.~\cite{Cai:2015zza,Cai:2015tam}, singlet models with a global $\mathbb{Z}_3$ symmetry can also fit the GCE signal when taking into account SE-A contributions properly. In such models, DM candidates can be either a scalar or a two-component scalar and fermion. It has been shown that the GCE signal can be accommodated in either case when the DM mass is close to the SM Higgs so that the produced single Higgs through SE-A processes is nearly at rest. This mechanism also works for the $\mathbb{Z}_3$ model under consideration, and the relevant SE-A channels correspond to $\chi_L\chi_L\to \chi^\dagger_Lh$ and $N_1\chi_L\to \bar N_1h$. However, since the model content here is richer, the parameter space required by GCE could be very different. A comprehensive and highly efficient analysis of this issue would employ the MCMC method, which we hope to address in the future.

\section{LHC phenomenology}
\label{lhc}

\begin{figure}[!htbp]
\begin{center}
\includegraphics[width=0.45\linewidth]{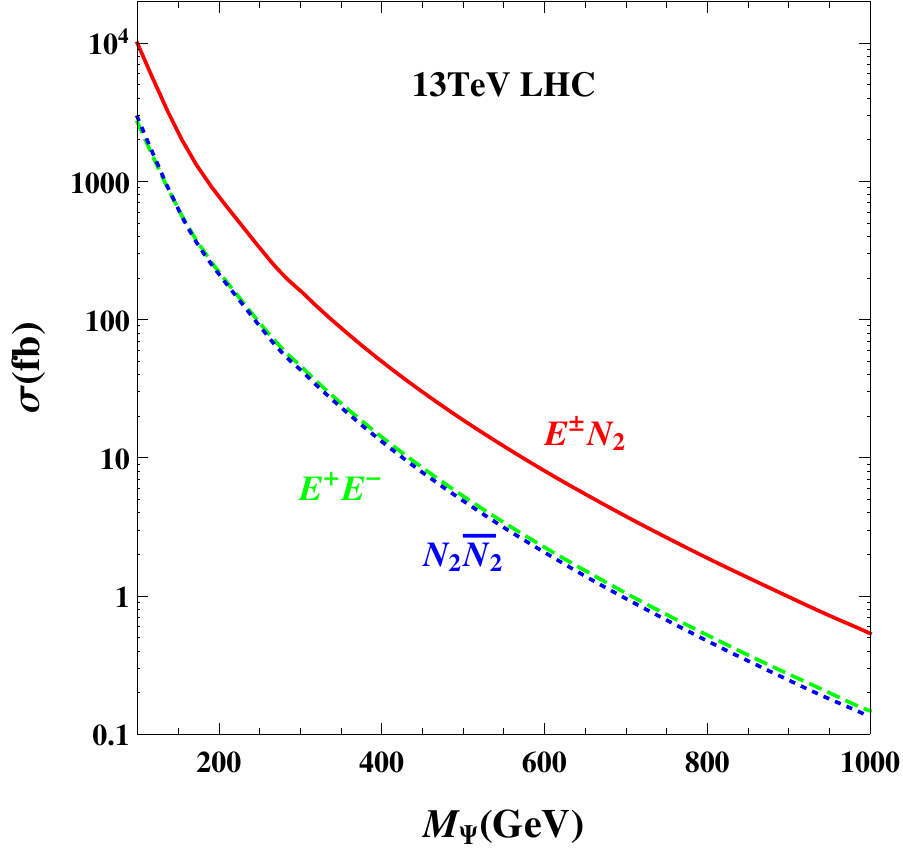}
\includegraphics[width=0.45\linewidth]{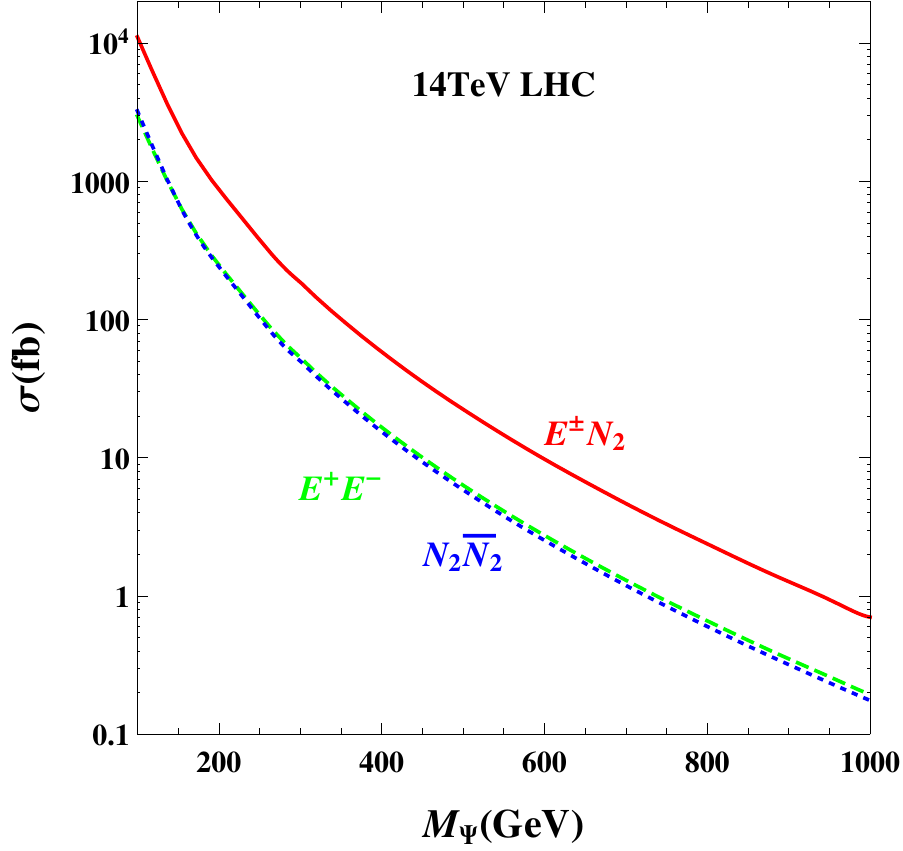}
\end{center}
\caption{Pair and associated production of doublet fermions at $13~\TeV$(Left) and $14~\TeV$(Right) LHC. Here we assume $\sin\beta=0$ and thus $M_{E}=M_{N_2}=M_{\Psi}$.
\label{cs}}
\end{figure}

In carrying out the LHC study of the $\mathbb{Z}_3$ model, we use {\tt MadGraph5\_aMC@NLO}~\cite{MG5} to calculate the production cross sections of $\mathbb{Z}_3$ particles with {\tt CTEQ6L1} ~\cite{Nadolsky:2008zw} parton distribution functions (PDFs). The leading contributions under consideration are the pair and associated production of the doublet fermions via the $s$-channel Drell-Yan process:
\begin{equation}
pp\to E^\pm N_2, E^+E^-, N_2 \bar{N}_2.
\end{equation}
The total cross sections of these processes are plotted in Fig. \ref{cs} as a function of the mass $M_\Psi$, where an overall $K$-factor of 1.2 is applied to both $13~\TeV$ and $14~\TeV$ cases \cite{Ruiz:2015zca}. For simplicity we assume $\beta=0$ and thus degenerate doublet fermions ($M_{E}=M_{N_2}=M_{\Psi}$) in the calculation of cross sections. The singlet fermion $N_1\simeq S$ and scalars $\chi_{L,H}$ can be produced through the decays of the doublet fermions which will be computed for a small $\beta$. The cross sections at LHC $13~\TeV$ of the doublet range from $10~\pb$ to $0.1~\fb$ in the mass interval $100-1000~\GeV$, and become slightly bigger at $14~\TeV$. This is also typical of the production of electroweakinos (charginos and neutralinos) in the minimal supersymmetric standard model (MSSM) \cite{Han:2013kza}. 

\subsection{Decay properties}
\label{SecDP}

\begin{figure}[!htbp]
\begin{center}
\boxed{\includegraphics[width=0.25\linewidth]{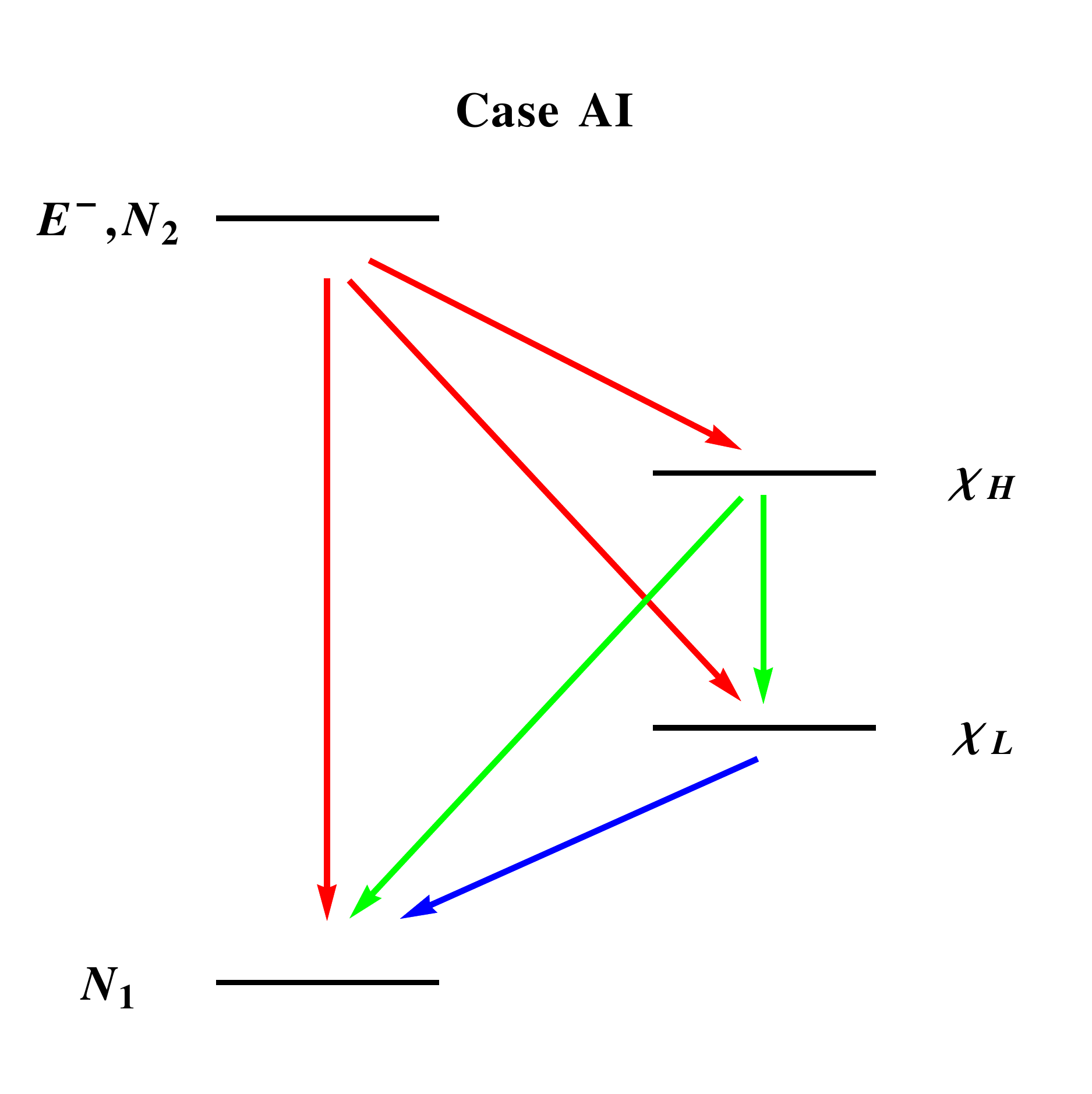}}
\boxed{\includegraphics[width=0.25\linewidth]{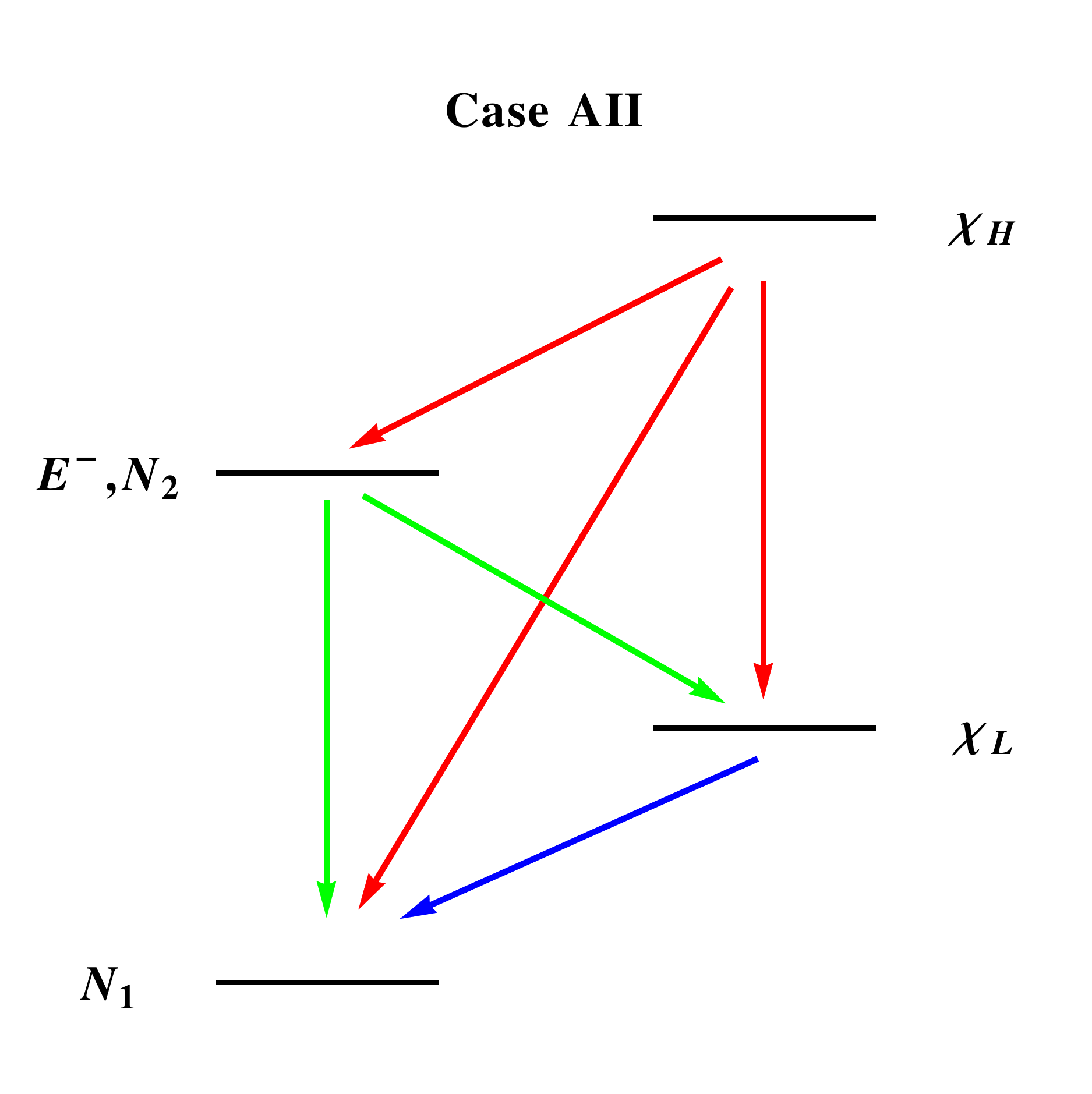}}
\boxed{\includegraphics[width=0.25\linewidth]{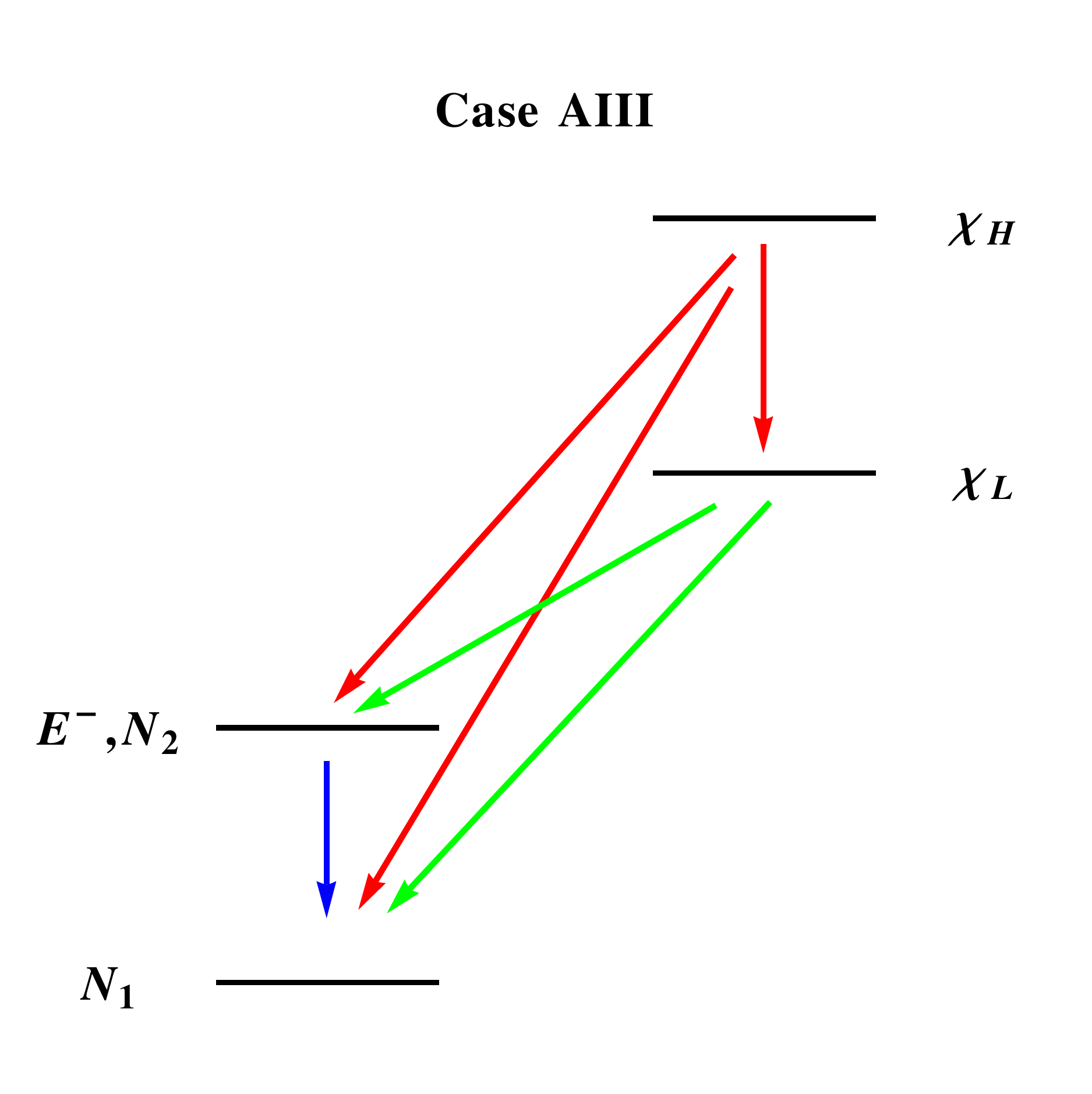}}
\end{center}
\begin{center}
\boxed{\includegraphics[width=0.25\linewidth]{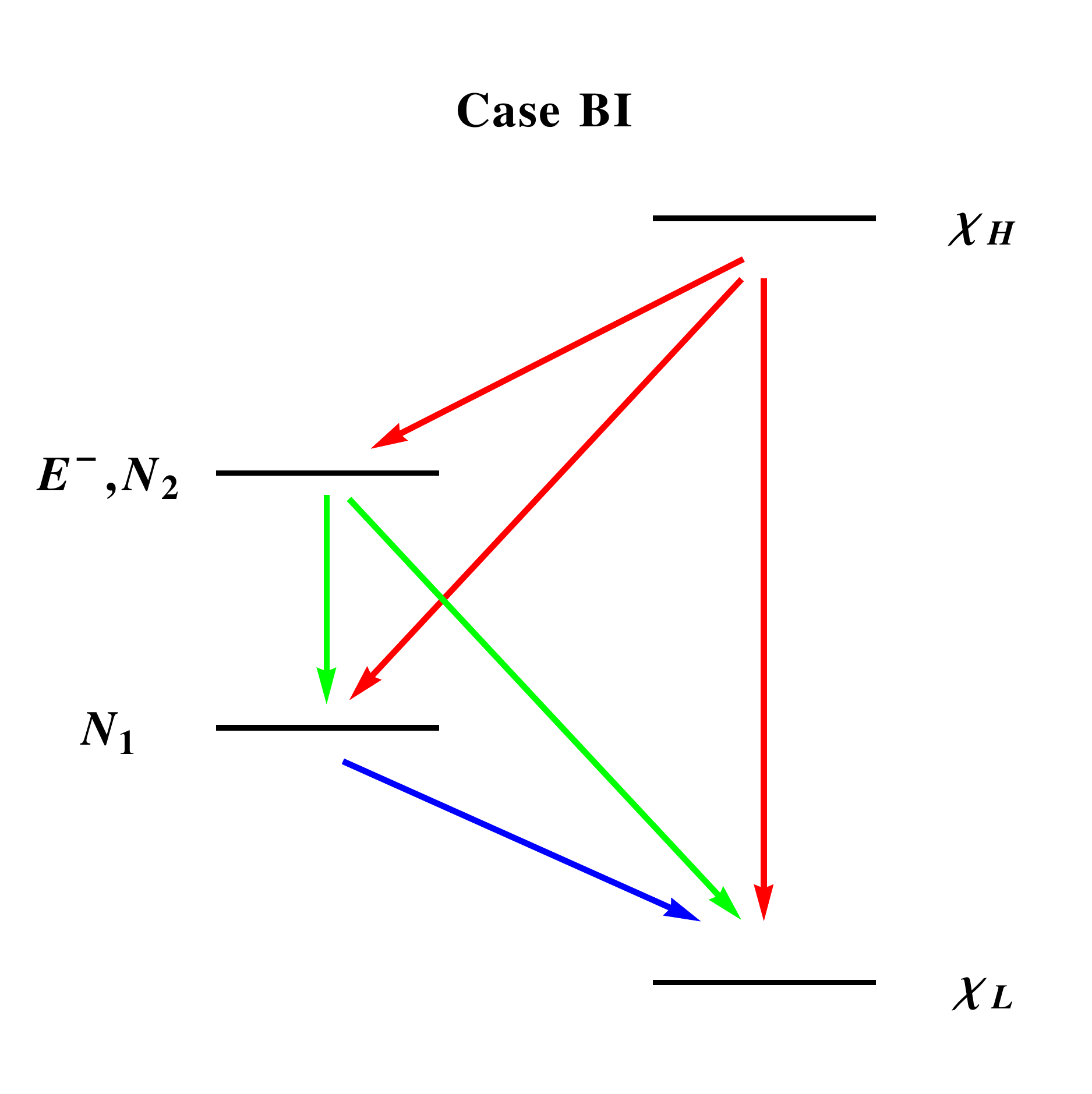}}
\boxed{\includegraphics[width=0.25\linewidth]{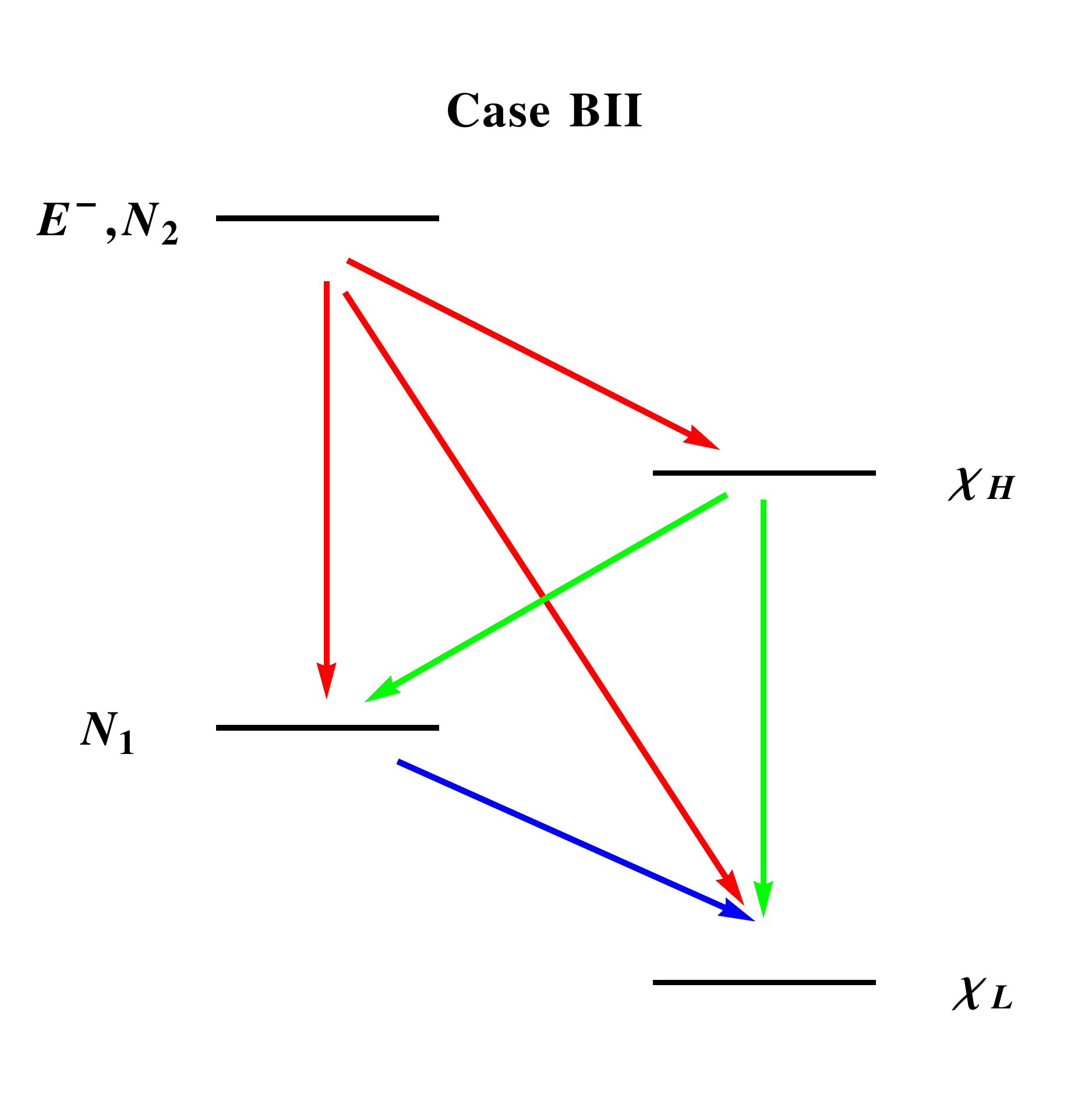}}
\boxed{\includegraphics[width=0.25\linewidth]{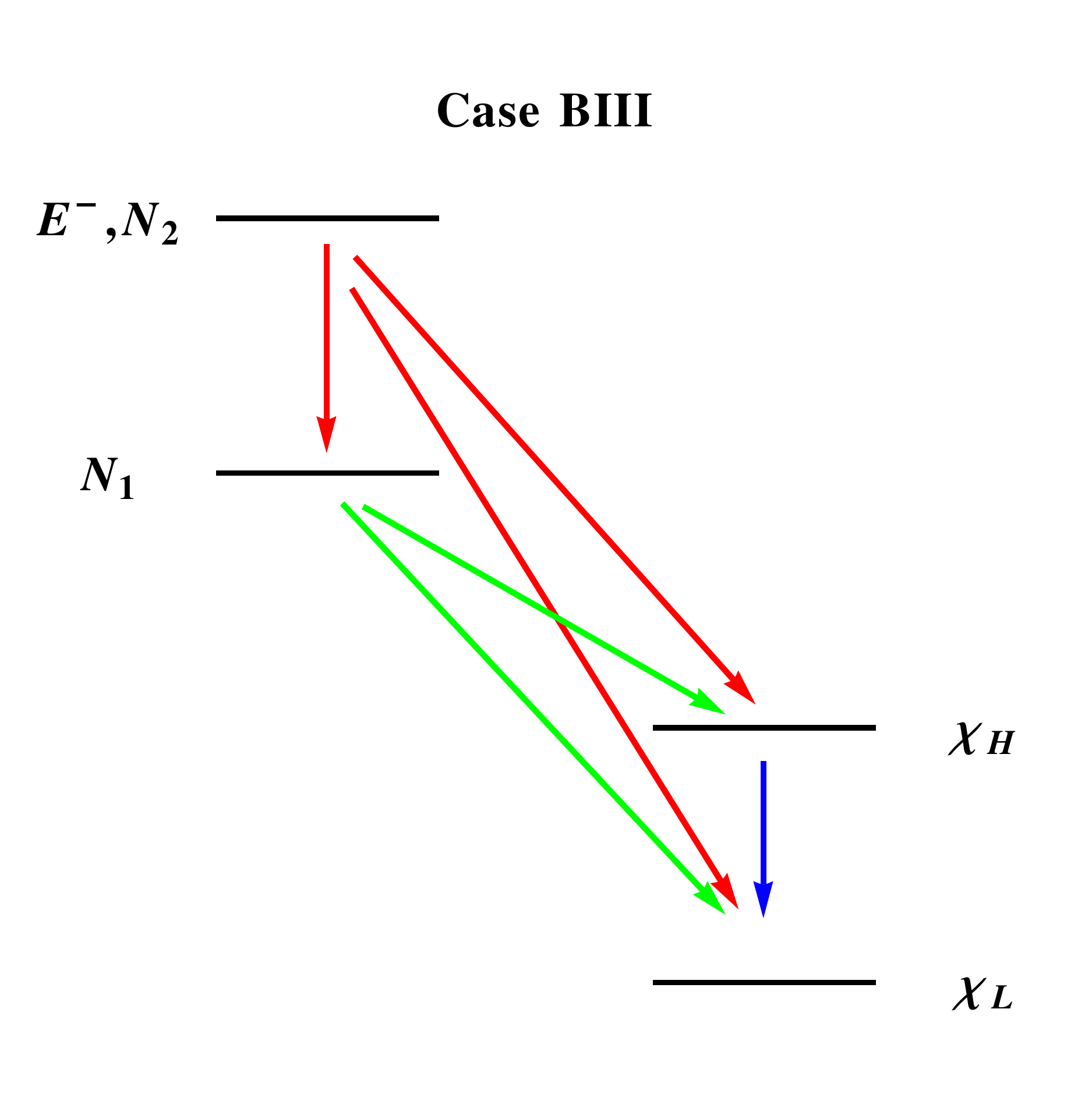}}
\end{center}
\begin{center}
\boxed{\includegraphics[width=0.25\linewidth]{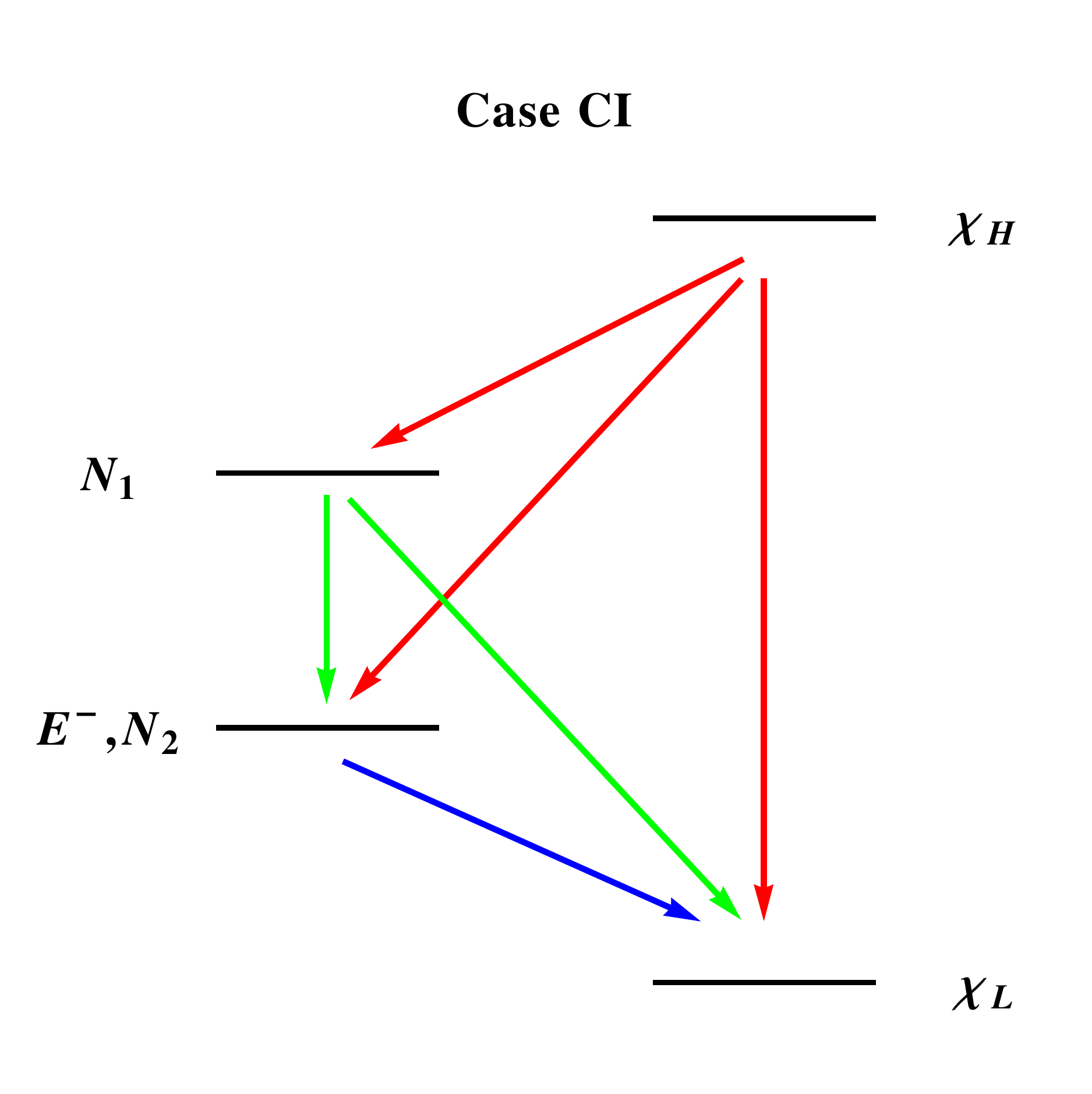}}
\boxed{\includegraphics[width=0.25\linewidth]{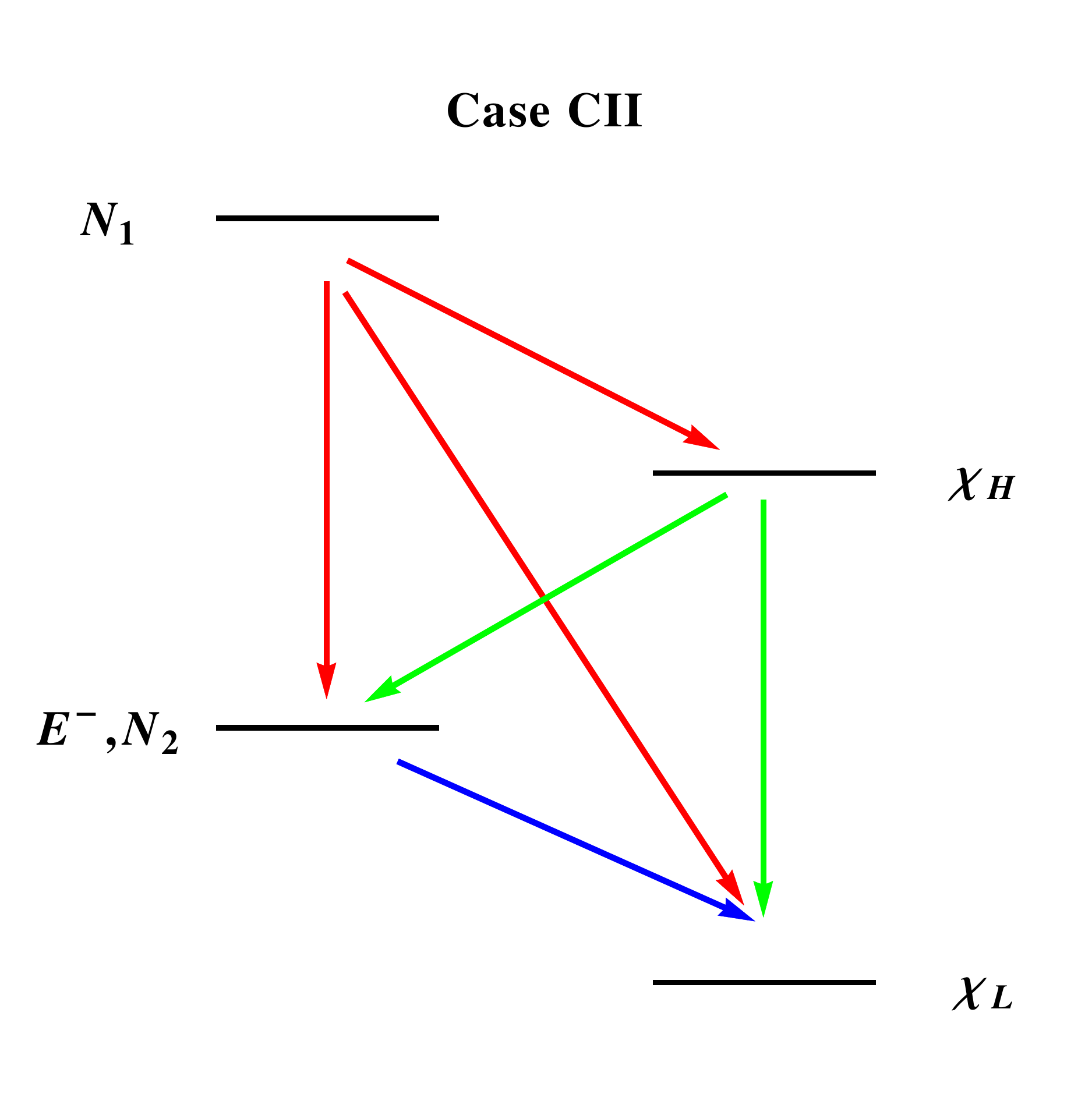}}
\boxed{\includegraphics[width=0.25\linewidth]{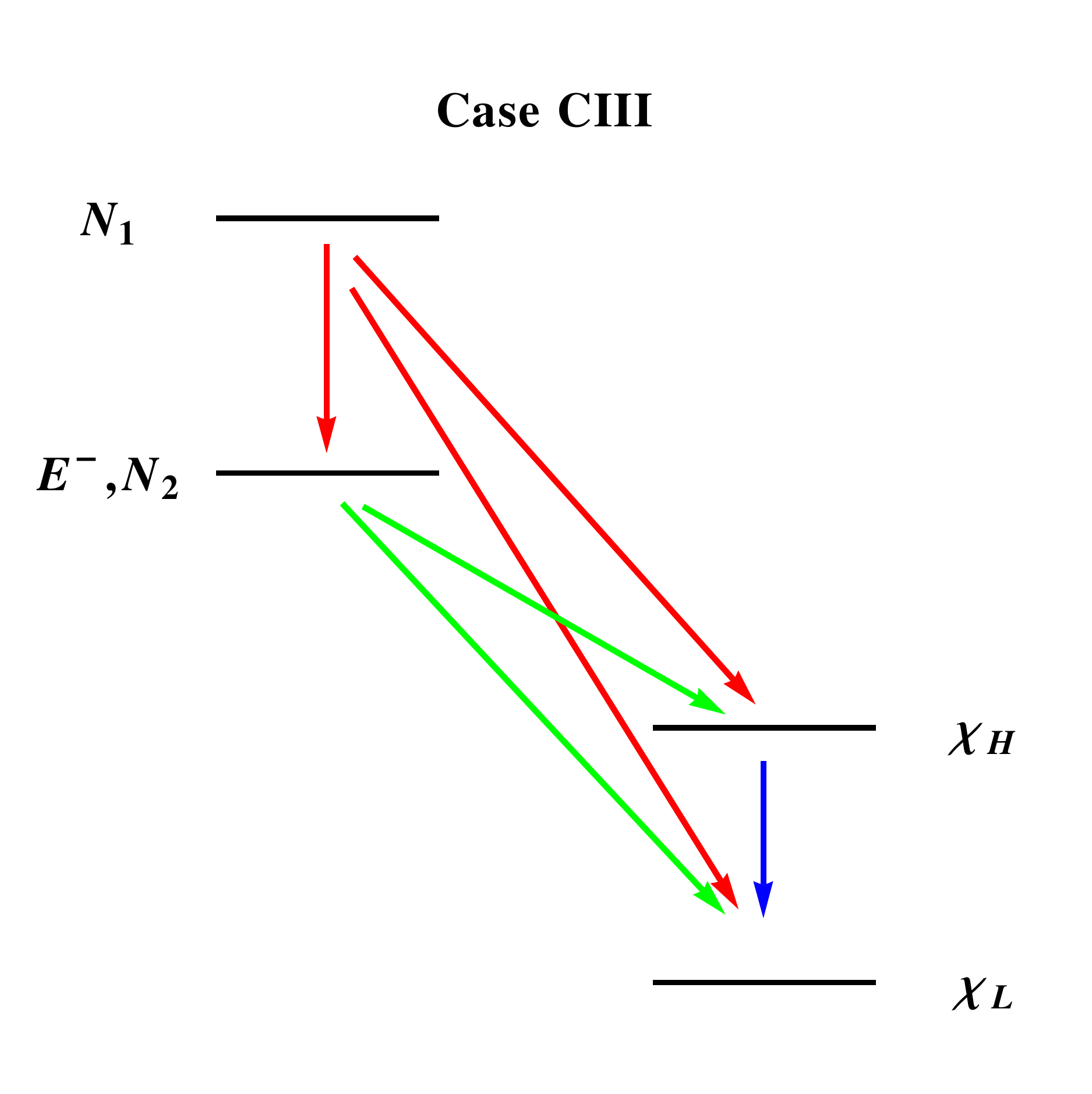}}
\end{center}
\caption{Decay patterns of $\mathbb{Z}_3$ particles for all the nine cases AI-CIII assuming degeneracy of $N_2$ and $E$.
\label{DP}}
\end{figure}

\begin{figure}[!htbp]
\begin{center}
\includegraphics[width=0.3\linewidth]{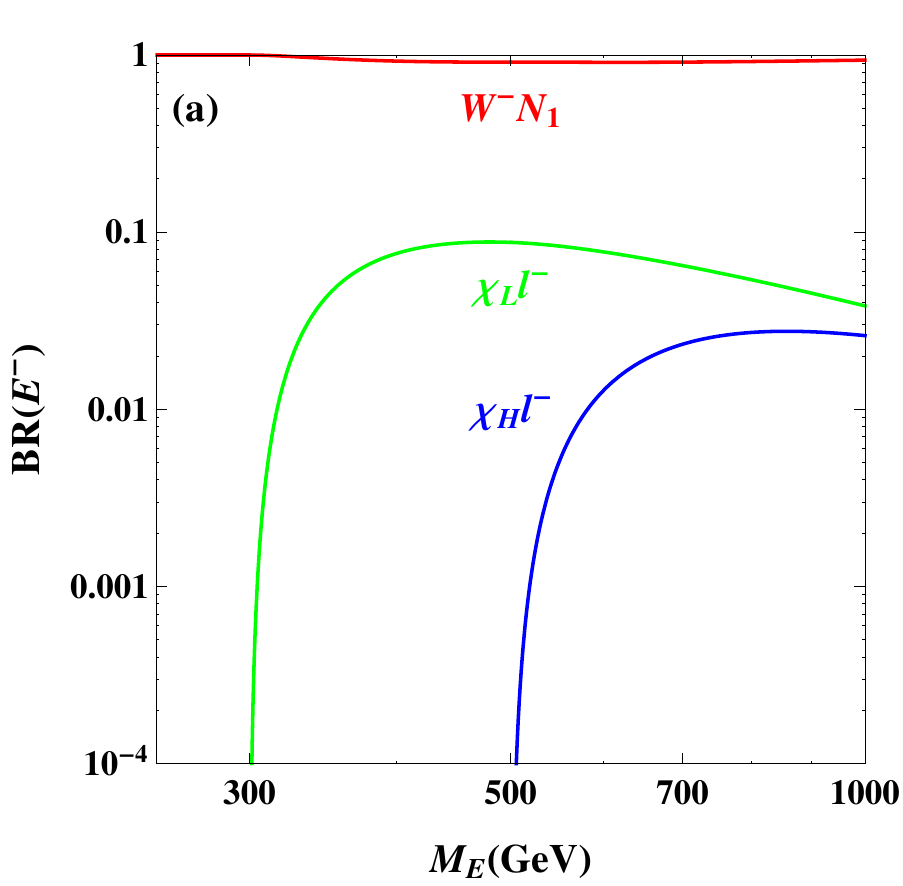}
\includegraphics[width=0.3\linewidth]{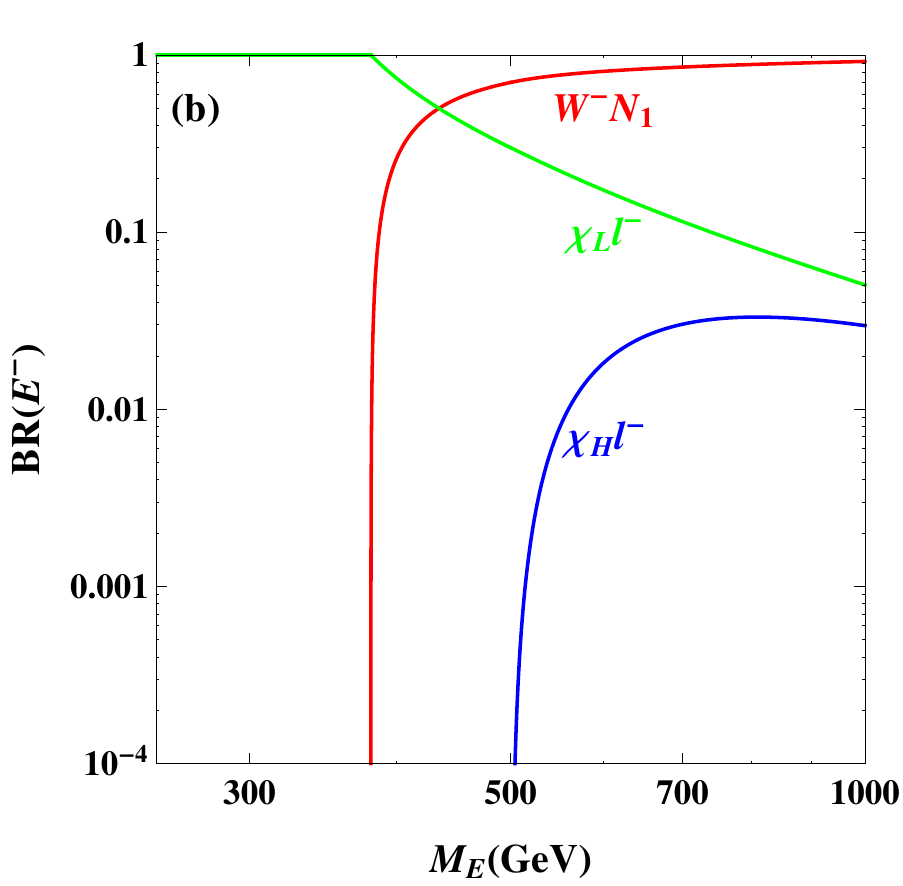}
\includegraphics[width=0.3\linewidth]{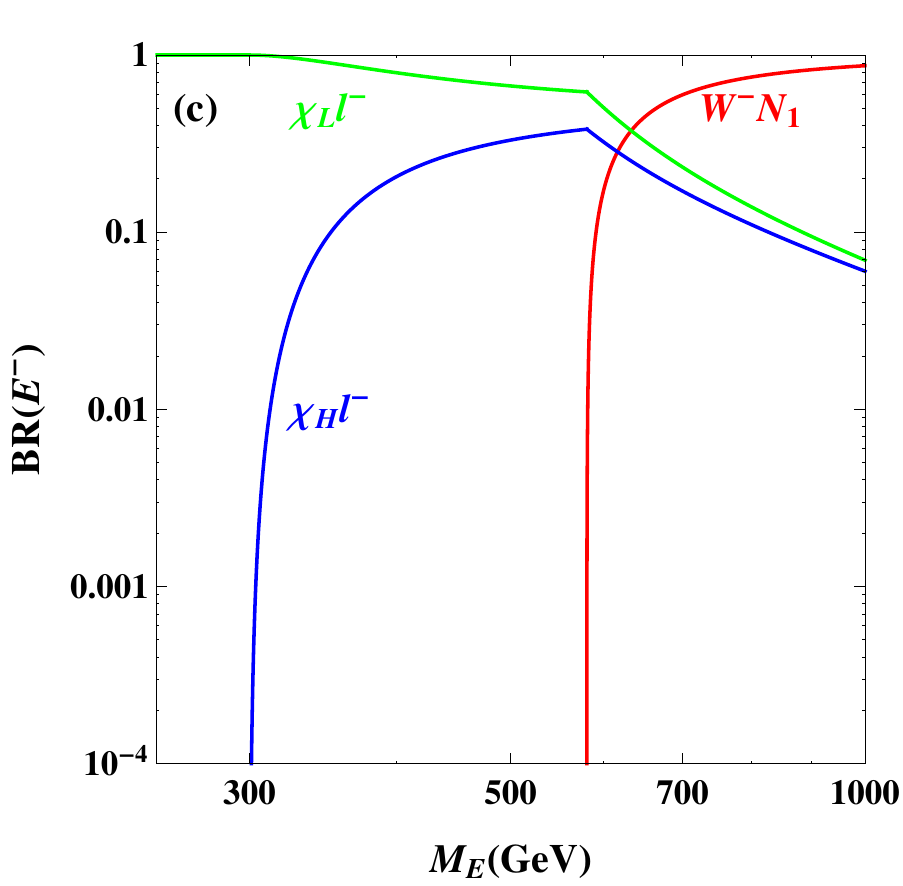}
\end{center}
\caption{Branching ratios of $E^-$ as a function of $M_{E}$. The masses of $(N_1,~\chi_L,~\chi_H)$ are, in units of GeV: (a) $(150,~300,~500)$; (b) $(300,~150,~500)$; (c) $(500,~150,~300)$.
\label{BREc}}
\end{figure}

\begin{figure}[!htbp]
\begin{center}
\includegraphics[width=0.3\linewidth]{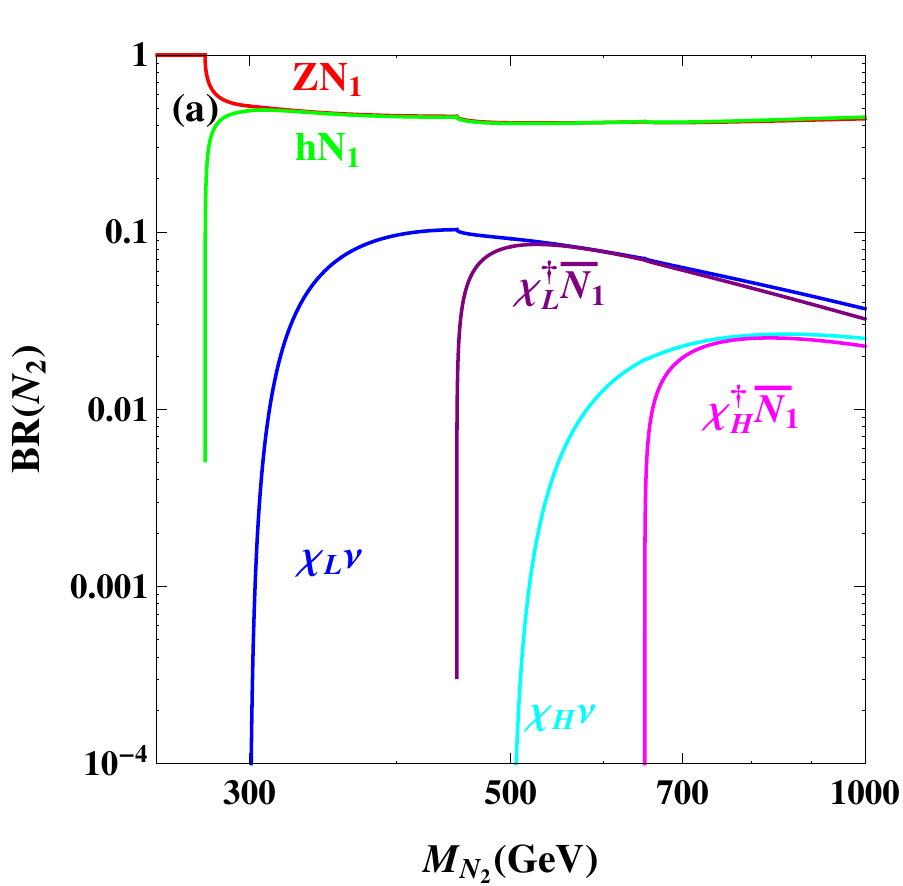}
\includegraphics[width=0.3\linewidth]{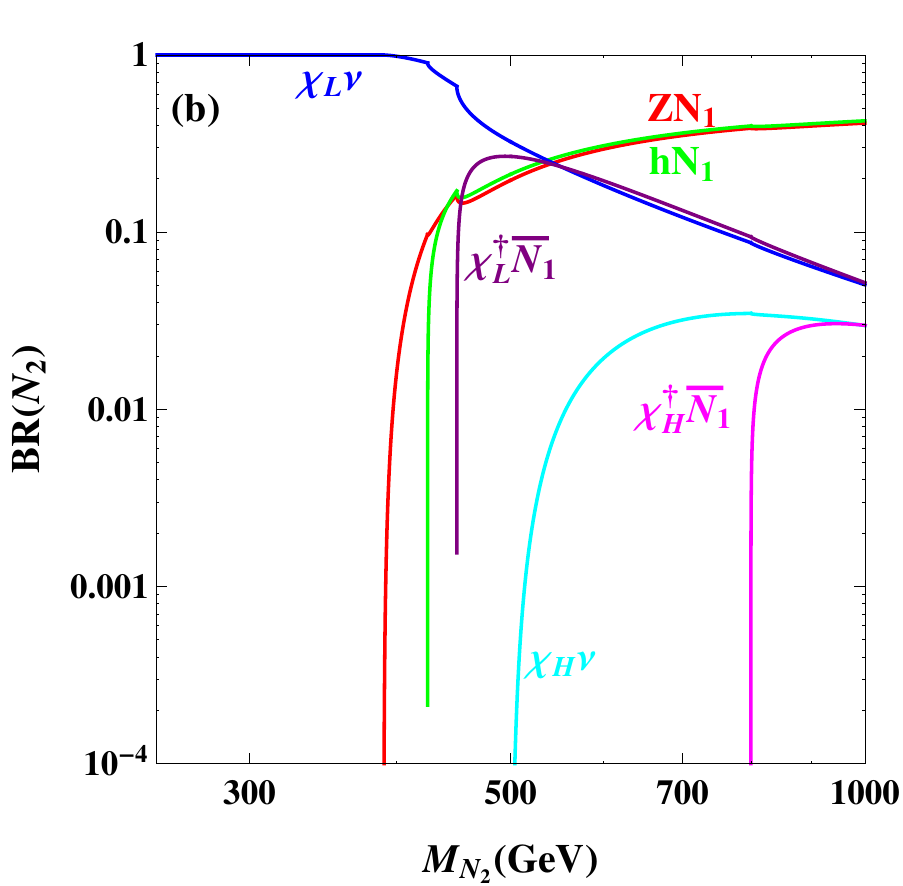}
\includegraphics[width=0.3\linewidth]{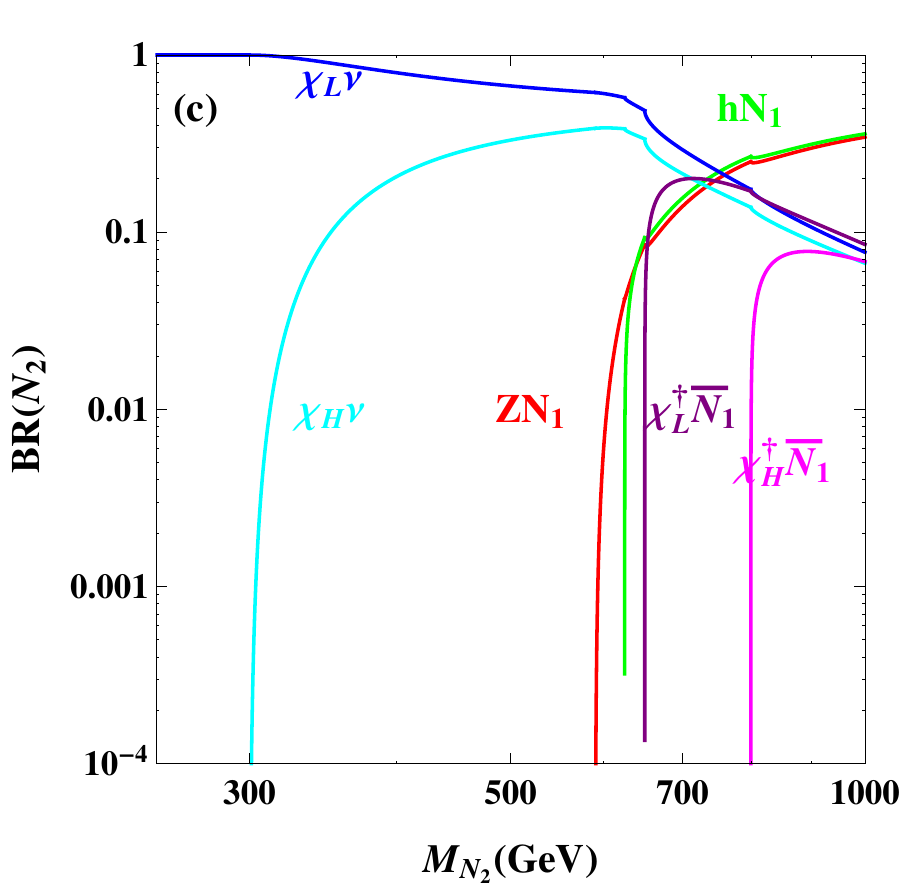}
\end{center}
\caption{Branching ratios of $N_2$ as a function of $M_{N_2}$ for the same sets of $(N_1,~\chi_L,~\chi_H)$ as in Fig.~\ref{BREc}.
\label{BRN2}}
\end{figure}

\begin{figure}[!htbp]
\begin{center}
\includegraphics[width=0.3\linewidth]{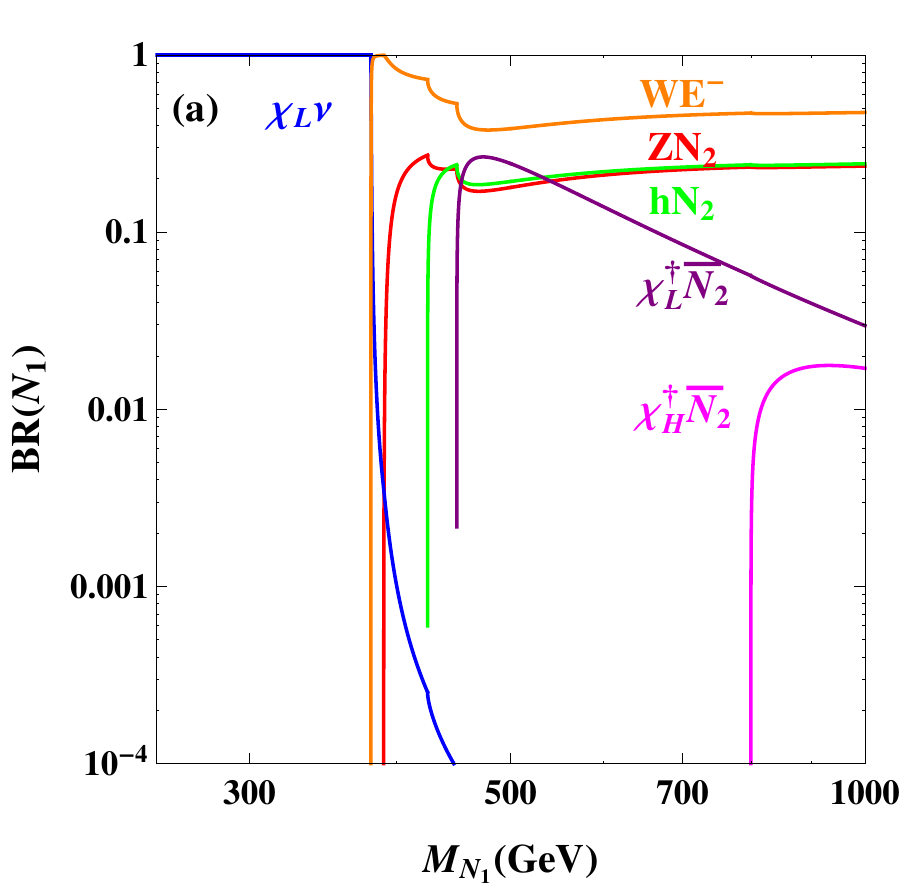}
\includegraphics[width=0.3\linewidth]{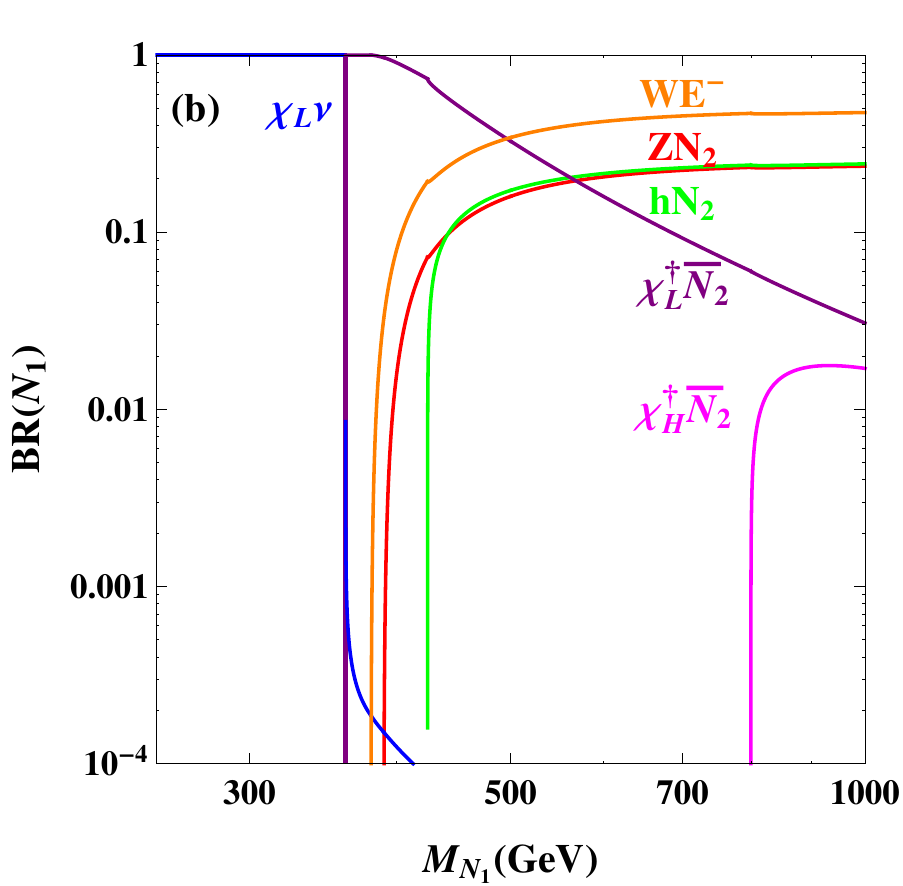}
\includegraphics[width=0.3\linewidth]{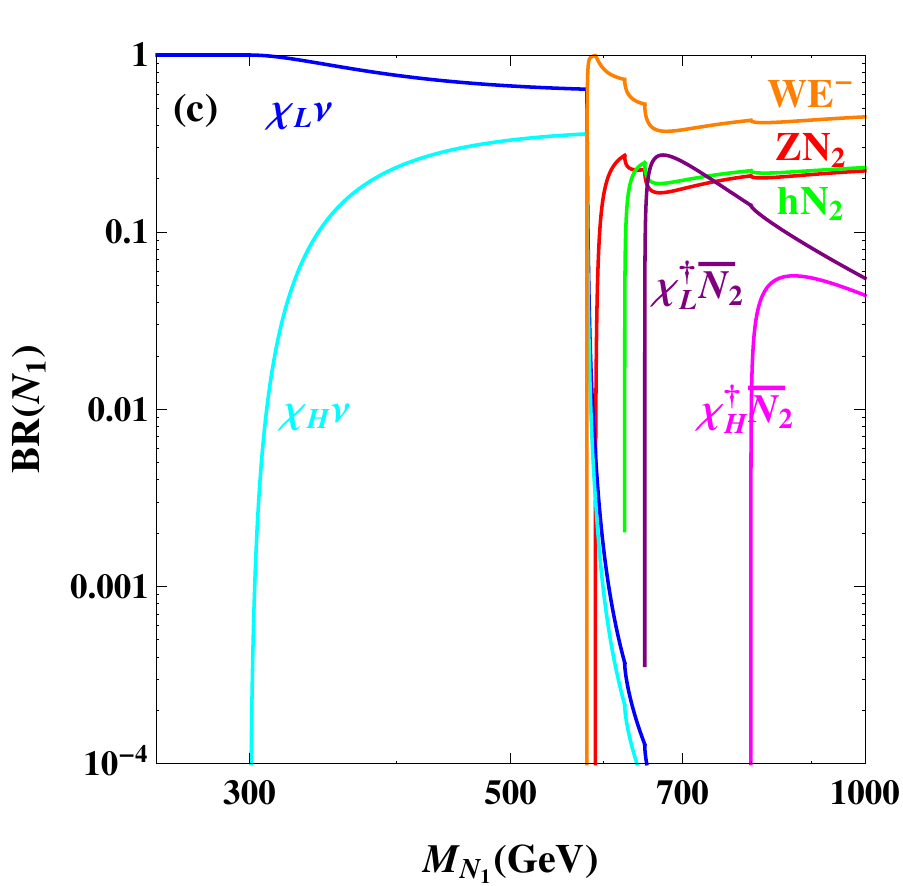}
\end{center}
\caption{Branching ratios of $N_1$ as a function of $M_{N_1}$. The masses of $(N_2,~\chi_L,~\chi_H)$ are, in units of GeV: (a) $(300,~150,~500)$; (b) $(300,~62,~500)$; (c) $(500,~150,~300)$.
\label{BRN1}}
\end{figure}

\begin{figure}[!htbp]
\begin{center}
\includegraphics[width=0.3\linewidth]{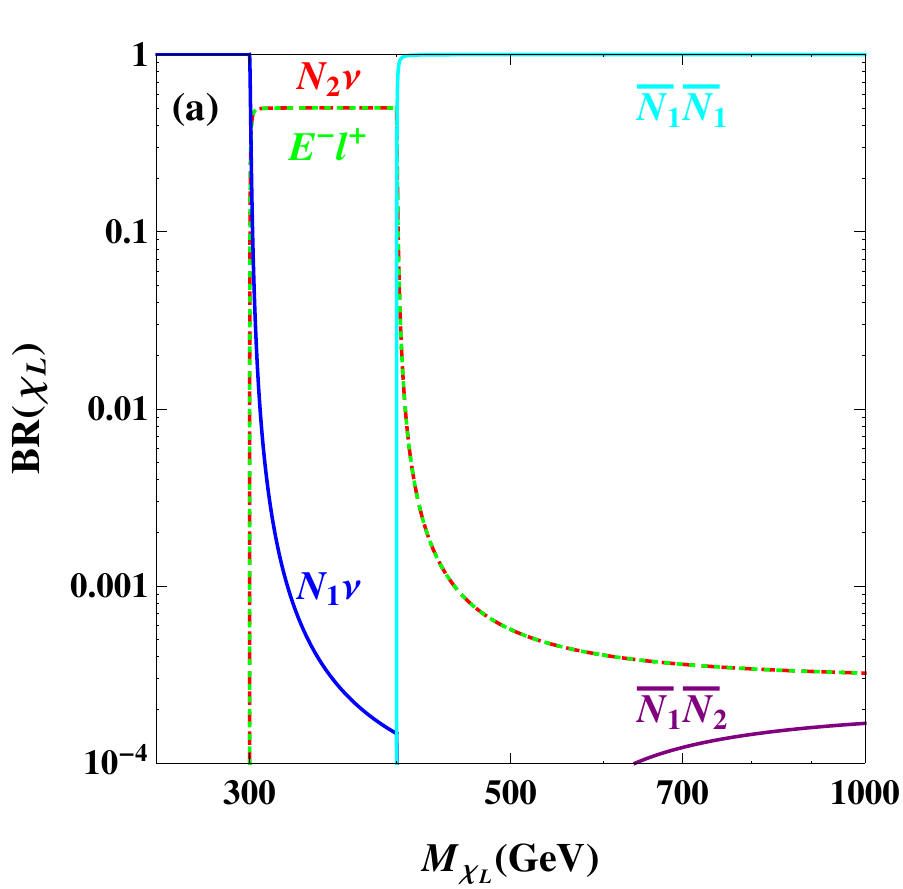}
\includegraphics[width=0.3\linewidth]{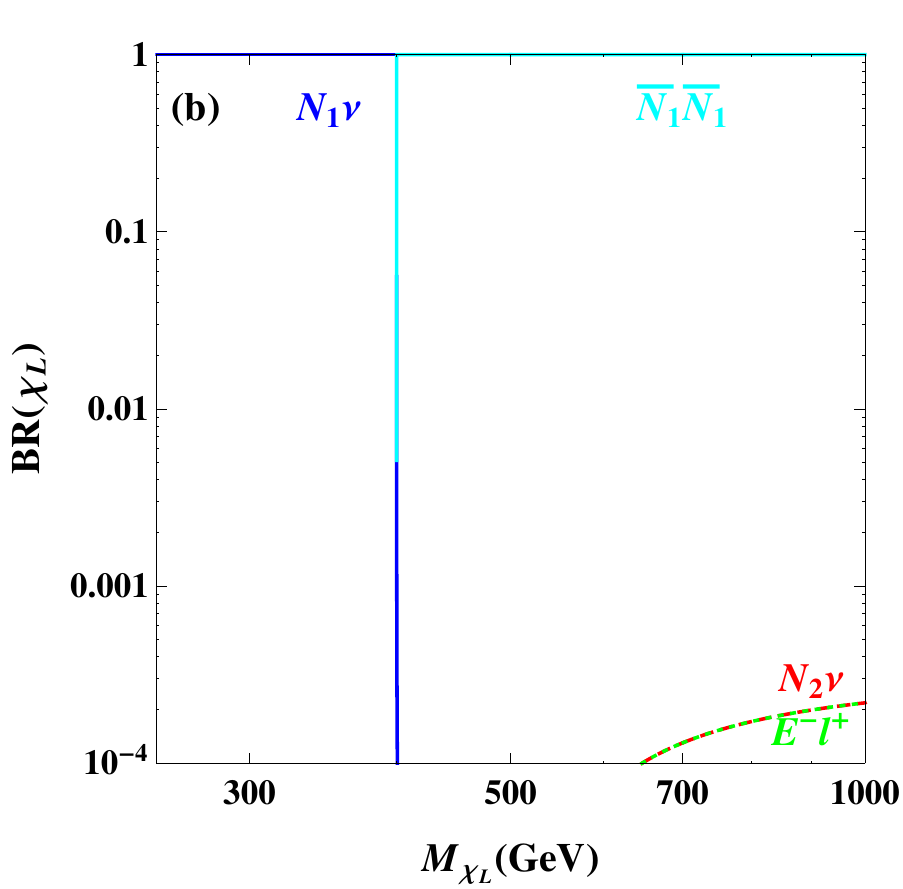}
\end{center}
\caption{Branching ratios of $\chi_L$ as a function of $M_{\chi_L}$. The masses of $(N_1,~N_2,~\chi_H)$ are, in units of GeV: (a) $(200,~300,~1000)$; (b) $(200,~500,~1000)$.
\label{BRXL}}
\end{figure}

\begin{figure}[!htbp]
\begin{center}
\includegraphics[width=0.3\linewidth]{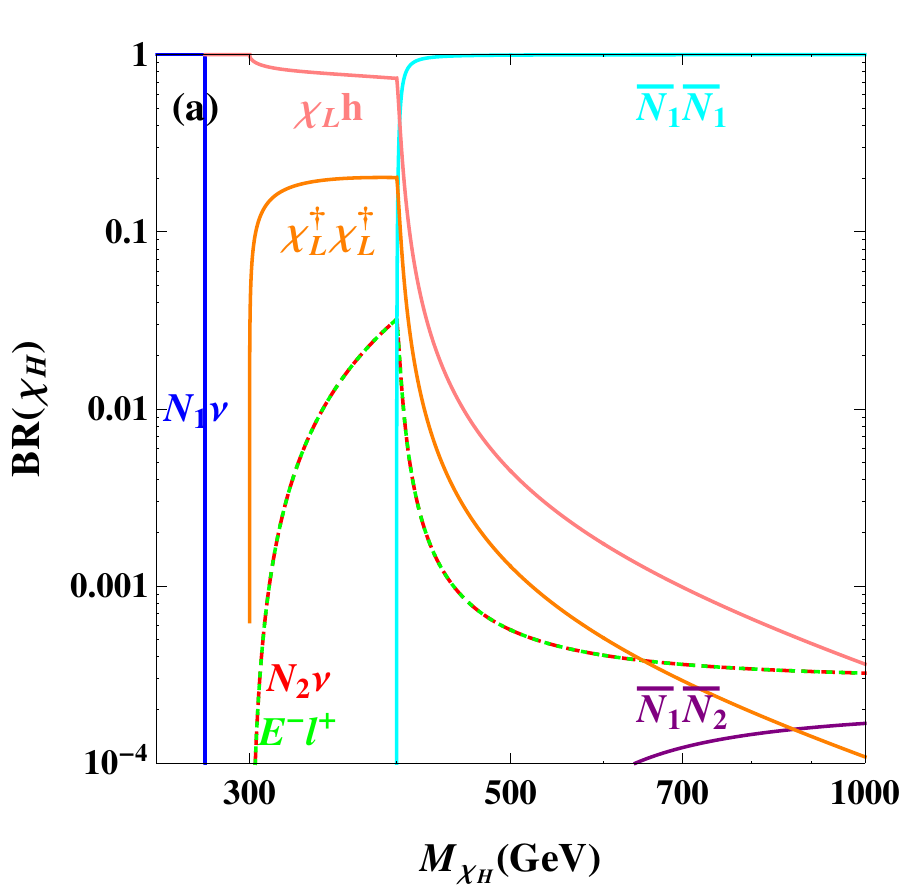}
\includegraphics[width=0.3\linewidth]{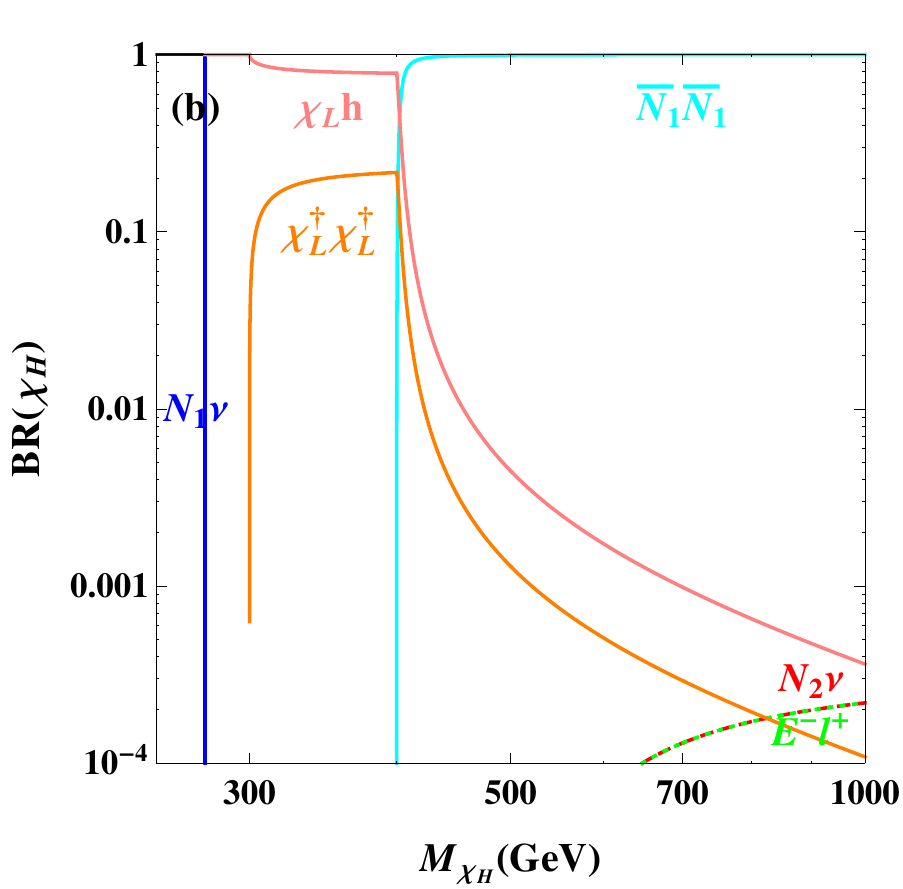}
\includegraphics[width=0.3\linewidth]{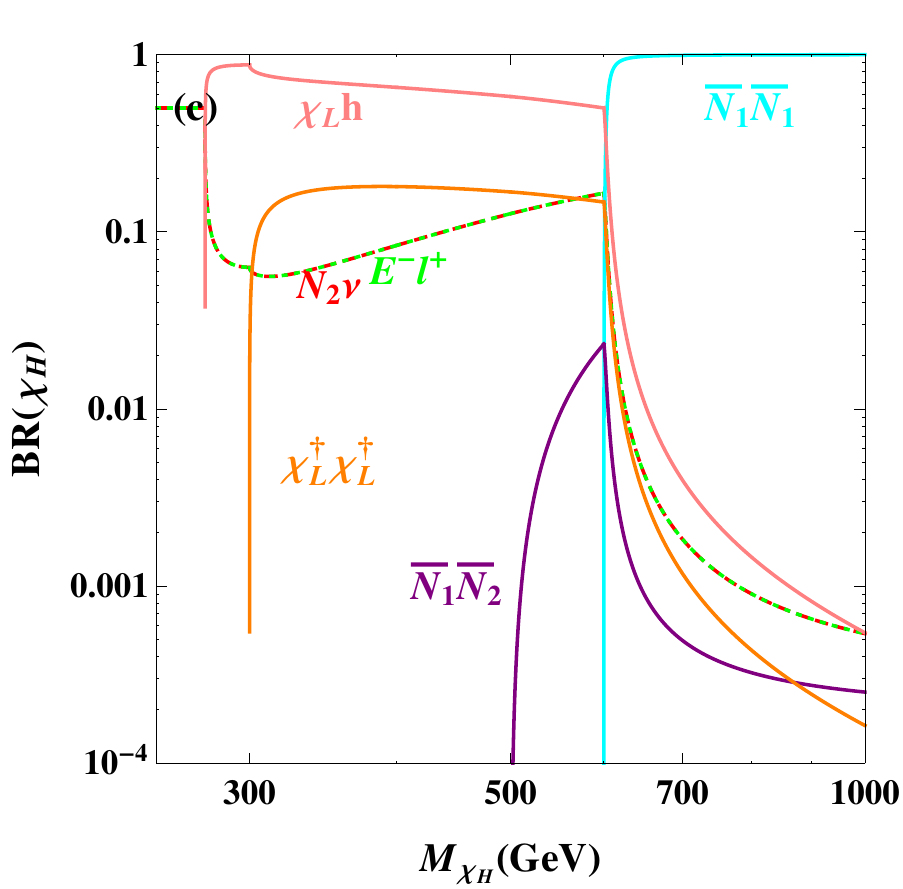}
\end{center}
\begin{center}
\includegraphics[width=0.3\linewidth]{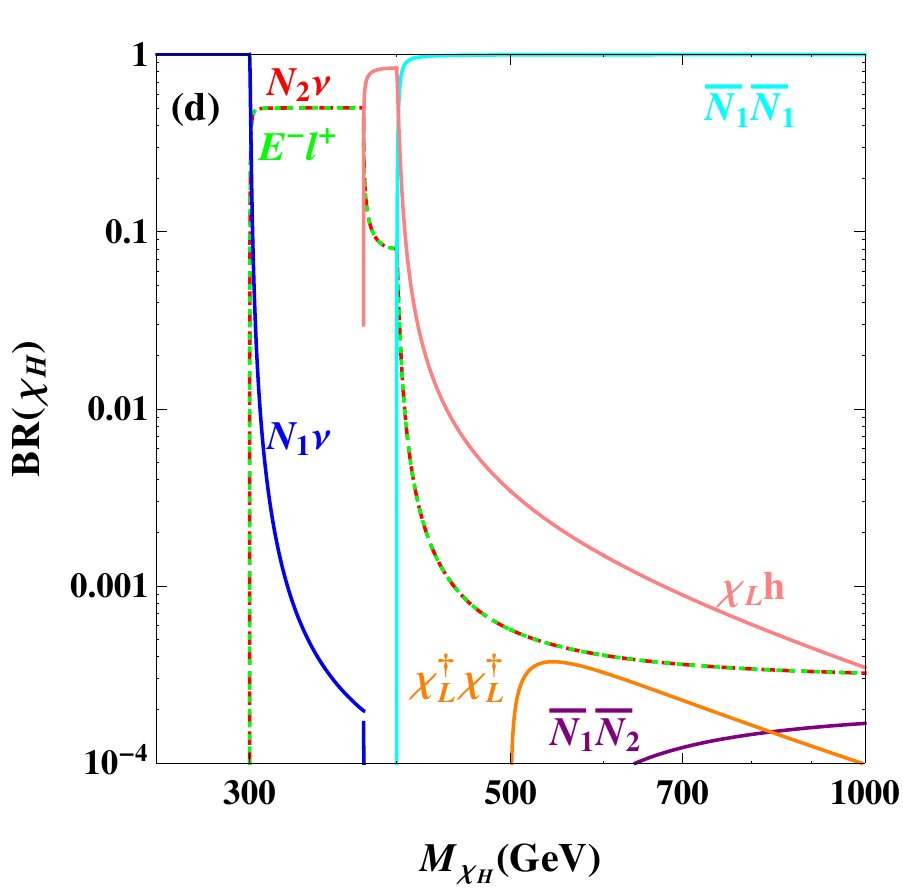}
\includegraphics[width=0.3\linewidth]{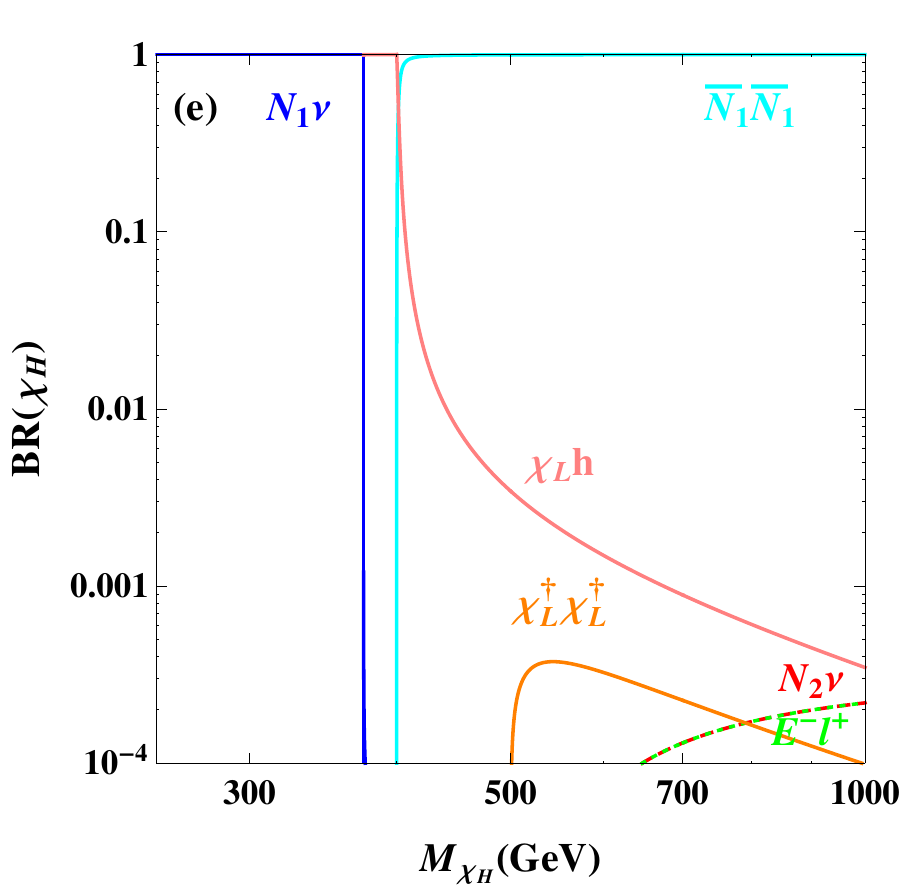}
\includegraphics[width=0.3\linewidth]{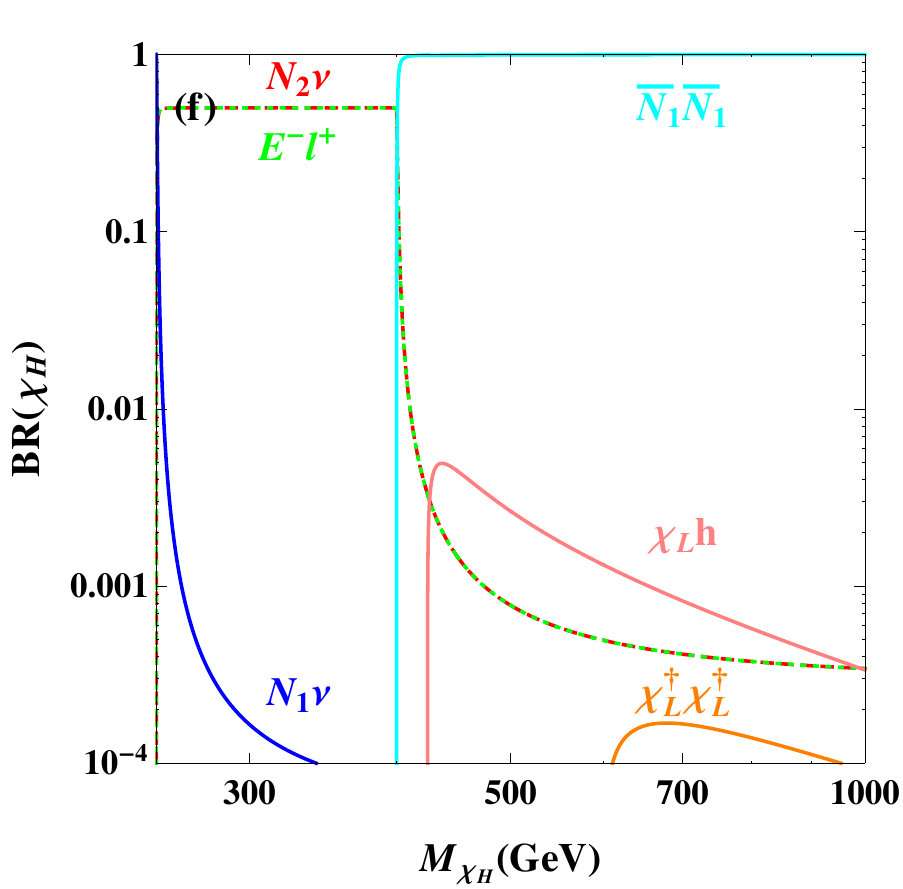}
\end{center}
\caption{Branching ratios of $\chi_H$ as a function of $M_{\chi_H}$. The masses of $(N_1,~N_2,~\chi_L)$ are, in units of GeV: (a) $(200,300,150)$; (b) $(200,500,150)$; (c) $(300,200,150)$; (d) $(200,300,250)$; (e) $(200,500,250)$; (f) $(200,250,300)$.
\label{BRXH}}
\end{figure}

To prepare for the study of LHC signatures, we discuss in this subsection the decay properties of $\mathbb{Z}_3$ particles. Figure~\ref{DP} shows the decay patterns for all nine cases allowed by DM considerations, and the decay branching ratios for all $\mathbb{Z}_3$ particles are shown in Figs. \ref{BREc}--\ref{BRXH}. We assume $E$ and $N_2$ to be degenerate to reduce the number of parameters, which corresponds to a degenerate fermion doublet in the limit of no mixing. These decay patterns not only affect the DM properties discussed in the previous section, but also have a great impact on the LHC signatures. Cases AI-III correspond to fermion DM, for which we only consider the case $M_{N_1}<M_{N_2}$ due to severe constraints on the opposite mass order from direct detection \cite{Cirelli:2009uv}. Cases BI-CIII correspond to scalar DM with $M_{N_1}<M_{N_2}$ in cases BI-BIII or $M_{N_1}>M_{N_2}$ in cases CI-CIII. From these decay patterns, we see clearly that the decays of the fermion doublet are heavily dependent on the mass spectrum of the $\mathbb{Z}_3$ particles. Thus in the following studies for each $\mathbb{Z}_3$ particle, we will choose several mass spectra to illustrate such an impact.

The decay channels can be classified into three categories according to interactions via (1) the gauge coupling (e.g., $E^-\to W^-N_1$), (2) the Yukawa coupling (e.g., $E^-\to \chi_L\ell^-$), or (3) the scalar self-coupling (e.g., $\chi_H\to\chi_L h$). Decays like $E^-\to W^-N_1$ in category (1) are possible due to the mixing between the singlet and doublet neutral fermions determined by the angle $\beta$. As mentioned previously, in the case of fermion DM, $\beta$ is tightly constrained by direct detection, while in the case of scalar DM a large $\beta$ is still allowed. For simplicity, we will choose a small $\beta$ in both cases in our following discussion and other
relevant parameters, as follows:
\begin{eqnarray}
\sin\alpha=0.1,~\sin\beta=0.01,&~ &~\lambda_2=\lambda_3=0.1,\\
h^{ai}=0.01,~x^a_{L,R}=1,&~ &~\mu=10~\GeV.
\end{eqnarray}
We will take several sets of $\mathbb{Z}_3$ particle masses to illustrate different decay patterns.

We first discuss the decays of the heavy charged fermion $E^-$. There are three decay channels: \begin{equation}
E^-\to W^- N_1,~\chi_{L,H}\ell^-.
\end{equation}
The branching ratios of $E^-$ as a function of $M_E$ are presented in Fig. \ref{BREc} for three cases of $\mathbb{Z}_3$ particle spectra. Case (a) corresponds to fermion DM, while cases (b) and (c) correspond to scalar DM. In case (a), the decay channel $E^-\to W^-N_1$ is dominant in the whole mass region. BR($E^-\to \chi_L\ell^-$) reaches maximum $0.1$ around $M_E=400~\GeV$, while BR($E^-\to \chi_H\ell^-$) is a little bit smaller due to phase space suppression. In cases (b) and (c), $E^-\to \chi_L\ell^-$ is dominant before $E^-\to W^-N_1$ is kinematically opened, while $E^-\to W^-N_1$ becomes dominant quickly once allowed.

Because of the mixing between the neutral fermions, $N_2$ has more decay channels than $E$: \begin{eqnarray}
N_2\to Z N_1,~h N_1,~\chi_{L,H}^\dag \bar{N}_1,~\chi_{L,H}\nu.
\end{eqnarray}
In addition to $N_2\to Z N_1$, $N_2$ can decay into $N_1$ through emission of $h$, $\chi_L$, and $\chi_H$. More interestingly, the decay $N_2\to \chi_L\nu$ is totally invisible at colliders in the case of scalar DM, which will intensively contribute to the signature of mono-jet, -$\gamma$, and -$Z$ \cite{Abdallah:2015uba}. For the same cases of $\mathbb{Z}_3$ spectra as in the discussion of $E^-$, the branching ratios of $N_2$ as a function of $M_{N_2}$ are plotted in Fig. \ref{BRN2}. In case (a) for fermion DM, $N_2\to Z N_1$ is dominant before $N_2\to h N_1$ is opened, and $\BR(N_2\to  ZN_1)\approx\BR(N_2\to hN_1)\approx 0.5$ soon after the latter is opened. The branching rations of other decay channels are always smaller than $0.1$. With the choice of $h^{ai}=\sin\beta=0.01$, we have approximately $\BR(N_2\to \chi_{L,H}\nu)\approx\BR(N_2\to \chi_{L,H}^\dag \bar{N}_1)$, for all three sets of masses. In case (b) for scalar DM, $N_2$ decays dominantly into $\chi_L\nu$ in the low mass region below $400~\GeV$, and into $ZN_1/hN_1$ in the high mass region above $600~\GeV$. In the intermediate mass region around $500~\GeV$, the four decay channels $N_2\to\chi_L\nu$, $\chi_L^\dag \bar{N}_1$, $ZN_1$, and $hN_1$ are comparable with each other. In case (c), with $M_{\chi_H}$ lighter than $M_{N_1}$ , $\BR(N_2\to\chi_H\nu)$ could reach over $0.3$ before $ZN_1$ is open.

Although the direct production rates of $N_1$, $\chi_L$, and $\chi_H$ are small at colliders, they can be produced via the cascade decays of $E,~N_2$ and subsequent decays into lighter particles. Possible promising signatures might occur in certain cascade decay chains, thus we also present the decay channels of these singlet particles for completeness. We first discuss the decays of $N_1$, which happen only in the case of scalar DM:
\begin{eqnarray}
N_1\to W^+E^-,~ZN_2,~hN_2,~\chi_{L,H}\nu,~\chi_{L,H}^\dag \bar{N}_2.
\end{eqnarray}
In Fig. \ref{BRN1}, the branching ratios of $N_1$ is displayed as a function of $M_{N_1}$ for three cases of $\mathbb{Z}_3$ particle spectra. The decay $N_1\to\chi_L\nu$ is totally dominant in the low mass region for all the three cases. In case (b), the decay $N_1\to\chi_L^\dag \bar{N}_2$ could be dominant in the mass region $370-470~\GeV$ with a light scalar DM $M_{\chi_L}\approx M_h/2$. In case (c) where $M_{\chi_H}<M_{N_2}$, $\BR(N_1\to\chi_H\nu)$ can reach about 0.3 before $N_1\to W^+E^-$ is opened. In the high mass region where all channels are opened, the three channels $N_1\to W^+E^-,~ZN_2,~hN_2$ dominate, and have the approximate relations,
\begin{equation}
\frac{1}{2}\mbox{BR}(N_1\to W^+E^-)\approx\mbox{BR}(N_1\to ZN_2)\approx\mbox{BR}(N_1\to hN_2),
\end{equation}
due to the Goldstone nature of $W,~Z$ \cite{Wang:2015saa}.

In contrast to $N_1$, $\chi_L$ can only decay in the case of fermion DM. Being only mediated by Yukawa couplings, the decay channels of $\chi_L$ are:
\begin{eqnarray}
\chi_L\to E^-\ell^+,~N_{1,2}\nu,~\bar{N}_1\bar{N}_1,
~\bar{N}_{1}\bar{N}_{2},~\bar{N}_2\bar N_2.
\end{eqnarray}
In Fig. \ref{BRXL}, we show the branching ratios of $\chi_L$ as a function of $M_{\chi_L}$ for two sets of $\mathbb{Z}_3$ particle masses. In the low mass region, the only allowed decay is $\chi_L\to N_1\nu$. In the high mass region above $2M_{N_1}$, the decay $\chi_L\to \bar{N}_1\bar N_1$ becomes dominant. Since the mixing angle $\beta$ must be tiny in the case of fermion DM, the channels $\chi_L\to \bar N_{1}\bar N_{2},~\bar N_2\bar N_2$ are always negligible. The decay channels $\chi_L\to N_2\nu,~E^-\ell^+$ depend heavily on the mass relations between $M_{N_2}$ and $M_{N_1}$. For instance, if $M_{N_1}<M_{N_2}<2 M_{N_1}$ as in case (a), both can be the main decay channels with an approximately equal branching ratio of $\sim 0.5$ in the range between $M_{N_2}$ and $2M_{N_1}$. On the other hand, if $M_{N_2}>2 M_{N_1}$ as in case (b), neither of them dominates.

The scalar self-interactions result in a richer decay pattern for the heavier scalar $\chi_H$ than the lighter $\chi_L$:
\begin{eqnarray}
\chi_H\to \chi_L h,~\chi_L^\dag\chi_L^\dag,~E^- \ell^+,~N_{1,2}\nu,
\\
\chi_H\to \bar N_1\bar N_1,~\bar N_{1}\bar N_{2},~\bar N_2\bar N_2.
\end{eqnarray}
Among these, $\chi_H\to \bar N_{1}\bar N_{2},~\bar N_2\bar N_2$ are severely suppressed by the tiny mixing angle $\beta$. Note that these decay channels can become relatively important in the case of scalar DM, where the mixing angle $\beta$ could be larger. The branching ratios of $\chi_H$ as a function of $M_{\chi_H}$ are illustrated in Fig. \ref{BRXH} for six cases of $\mathbb{Z}_3$ particle spectra. Cases (a)-(c) correspond to scalar DM, and cases (d)-(f) to fermion DM. Similar to $\chi_L$, $\chi_H\to \bar N_1\bar N_1$ is the only dominant decay in the high mass region above $2M_{N_1}$. But in the mass region below $2M_{N_1}$, the decays of $\chi_H$ can be quite different from $\chi_L$. For scalar DM, $\chi_H$ decays into $N_1\nu$ (cases (a) and (b)) or $N_2\nu/E^-\ell^+$ (case (c)) in the low mass region, depending on which of $N_1$ and $N_2$ is lighter. In the intermediate region between $M_{\chi_L}\!+M_{h}$ and $2M_{N_1}$, the cascade decay $\chi_H\to \chi_L h$ dominates. Such decay channels play a very important role in the detection of scalar interactions at colliders. And once allowed, the branching ratio of $\chi_H\to\chi_L^\dag\chi_L^\dag$ could reach about $0.2$, which is the dominant invisible decay of $\chi_H$. Furthermore, for a large $\mu$, e.g., $\mu=100~\GeV$, the invisible decay $\chi_H\to \chi_L^\dag\chi_L^\dag$ is expected to be even larger than $\chi_H\to\chi_L h$. For fermion DM, $\chi_H$ can only decay as $\chi_L$ into $N_1\nu$ in the low mass region. Case (d) is most interesting among all three, where the four main decay channels $\chi_H\to N_1\nu,~N_2\nu/E^-\ell^+,~\chi_L h,~\bar N_1\bar N_1$ become dominant sequentially as $M_{\chi_H}$ increases. This special case requires the mass relation $M_{N_2}<M_{\chi_L}+M_h<2M_{N_1}$ to be satisfied. If not, $\chi_L h$ will be the main decay channel for a heavy $N_2$ as shown in case (e), or $N_2\nu/E^-\ell^+$ take over for a heavy $\chi_L$ as shown in (f). If both $N_2$ and $\chi_L$ are relatively heavy, $\chi_H$ will decay the same way as $\chi_L$ as shown in case (b) of Fig. \ref{BRXL}.

\subsection{LHC signatures and Constraints}
\label{SG}

After the systematic study on the decay properties of the $\mathbb{Z}_3$ particles in Sec. \ref{SecDP}, we now address their possible signatures at LHC. To a large extent, the LHC phenomenology is governed by the fermion doublet decays, since they can be pair or associated produced. The various decay channels of $N_2$ and $E^\pm$ as well as the cascade decays of other $\mathbb{Z}_3$ particles will lead to characteristic collider signatures. For instance, the final states of $\mathbb{Z}_3$ particles will always have missing transverse energy $\cancel{E}_T$ due to the existence of DM. At the same time the most interesting and easiest way to detect signatures of neutrino mass models usually involve multi-lepton final states \cite{Han:2006ip,Perez:2008ha,Bajc:2007zf,delAguila:2008cj,Deppisch:2015qwa}, and so is expected for the two-loop radiative neutrino mass model under consideration. Furthermore, with a Higgs boson $h$ \cite{Aad:2012tfa,Chatrchyan:2012xdj} in the decays of $N_2$ and $\chi_H$, it is also interesting to probe signatures with this $h$. Therefore we will explore the LHC signatures involving multi-$\ell$ ($\ell=e,\mu$) plus $\cancel{E}_T$ with or without a Higgs boson $h$. These signatures are naturally classified in terms of the number of leptons (up to four) in the final states.

\subsubsection{Signatures for $N_1$ DM}

We first highlight the signatures appearing in the case of fermion DM. Since we concentrate on the multi-lepton signatures, we will consider the leptonic decays of the gauge bosons $W$ and $Z$. The possible signatures are listed as follows.

{\bf (F1)} $0\ell 2h$ \quad This signature of no leptons and a pair of Higgs bosons \cite{Baglio:2012np,Han:2015sca} has a large $\cancel{E}_T$, which can be used to suppress the SM background. The production mechanism is
\begin{equation}
pp\to N_2\bar{N}_2\to h N_1 + h \bar{N}_1,
\end{equation}
with $h\to b\bar{b}/\gamma\gamma$. The same signature is also expected in supersymmetric (SUSY) and canonical seesaw models \cite{Kang:2015nga}. With BR($N_2\to h N_1)\approx0.5$ in our benchmark scenario, the production rate of this signature is a quarter of $\sigma(N_2\bar{N}_2)$. A search for this signature in the SUSY scenario has been performed by CMS \cite{Khachatryan:2014mma} in gauge-mediated SUSY-breaking model where the lightest superparticle (LSP) is gravitino and the next-to-lightest superparticle is higgsino. For nearly massless LSP, there is no exclusion limit for $N_2$ up to 500 GeV if one matches $N_2-N_1$ to the higgsino-gravitino system; and the sensitive mass region is $M_{N_2}>200~\GeV$ for BR($N_2\to h N_1)> 0.5$. However, for $14~\TeV$ LHC with an integrated luminosity $L=3000~\fb^{-1}$, we may have a chance to probe this signature for a small production rate down to $\sim 0.1~\fb$ or $m_{N_2}$ up to $800~\GeV$~\cite{Kang:2015nga}.

{\bf (F2)} $1\ell 1h$ \quad This signature follows from the associated production of the doublet fermions:
\begin{equation}
pp\to E^\pm N_2\to W^\pm N_1 + h N_1,
\end{equation}
with $h\to b\bar{b}/\gamma\gamma$, exactly as in the chargino-neutrolino system in SUSY models. If we further consider $h\to WW^*\to\ell^\pm\nu qq'$, this channel can also produce the like-sign dilepton signature $\ell^\pm\ell^\pm$. Searches for this signature have been carried out by CMS \cite{Khachatryan:2014mma,Khachatryan:2014qwa} and ATLAS \cite{Aad:2015jqa,Aad:2015eda}. Again, matching $E^\pm N_2$ ($N_1$) to $\tilde{\chi}^\pm_1\tilde{\chi}^0_2$ ($\tilde{\chi}^0_1$) and assuming $\BR(E^\pm\to W^\pm N_1)\approx\BR(N_2\to h N_1)\approx 100\%$, the limits on $M_{N_2}$ have been set to $200~\GeV$ by CMS \cite{Khachatryan:2014qwa} and $250~\GeV$ by ATLAS \cite{Aad:2015jqa} for massless $N_1$. But as discussed in Sec. \ref{SecDP}, $\BR(E^\pm\to W^\pm N_1)\approx100\%$ and $\BR(N_2\to h N_1)\approx50\%$ for the model considered here, one expects the limits on $M_{N_2}$ to be relaxed.

\begin{figure}[!htbp]
\begin{center}
\includegraphics[width=0.45\linewidth]{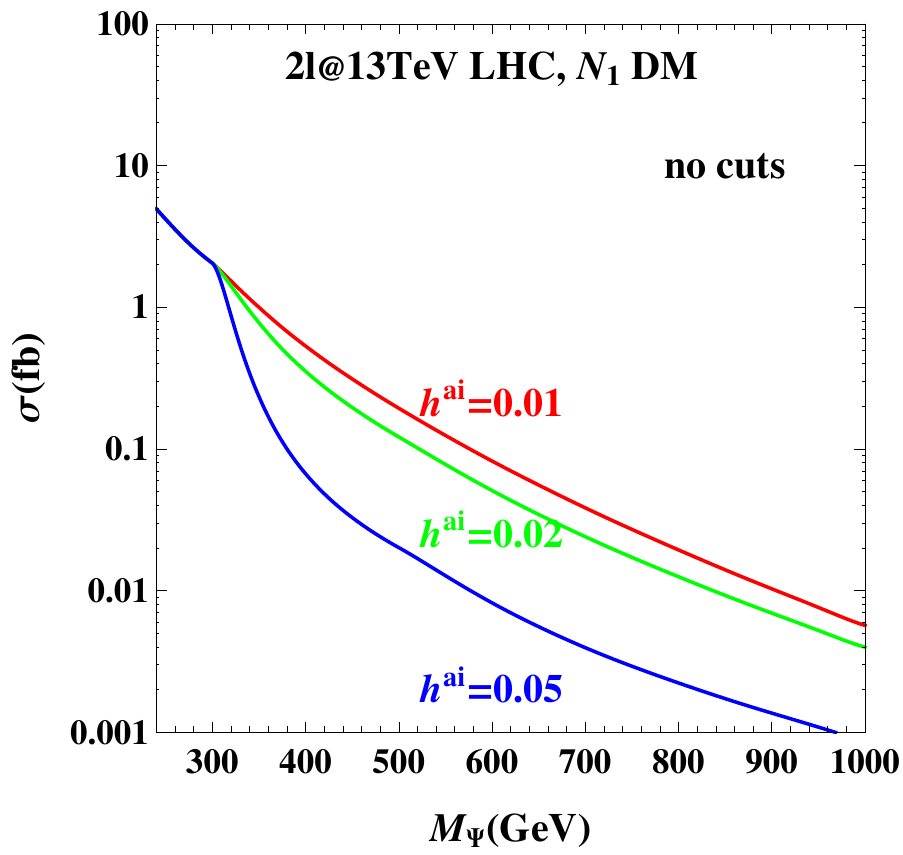}
\includegraphics[width=0.45\linewidth]{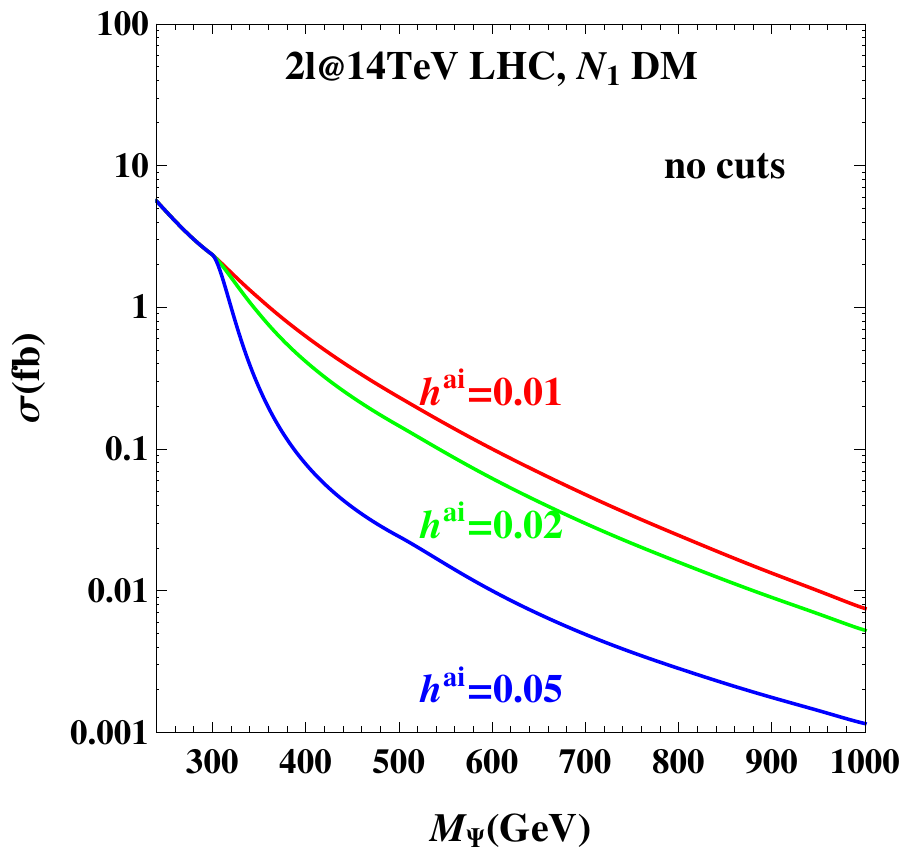}
\end{center}
\caption{Cross sections of the $2\ell+\cancel{E}_T$ signature in Eq. \ref{pro_2l_N} as a function of $M_{\Psi}$ at 13 TeV LHC (left panel) and 14 TeV LHC (right). Here, we set $\sin\beta=0.01$ and $(M_{N_1},M_{\chi_L},M_{\chi_H})=(150,300,500)~\GeV$.
\label{2l_N}}
\end{figure}

{\bf (F3)} $2\ell(\Znot)$ \quad In this signature the two opposite-sign leptons are required not to reconstruct a $Z$ boson. Such events are produced as
\begin{equation}
\label{pro_2l_N}
pp\to E^+E^- \to W^+ \bar{N}_1+ W^- N_1,
\end{equation}
with $W^\pm \to \ell^\pm\nu$. In Fig. \ref{2l_N}, the cross section of this signature at $13~(14)~\TeV$ LHC is presented. To illustrate the impact of the Yukawa couplings $h^{ai}$, we choose three typical values, $h^{ai}=0.01,~0.02,~0.05$. The cross section drops with increasing $h^{ai}$. For instance, it is approximately an order of magnitude smaller at $h^{ai}=0.05$ than at $h^{ai}=0.01$. Due to the large SM background from dibosons ($WW$) and top quarks (mainly come from $t\bar{t}$ and $Wt$), constraints on this signature are relatively loose. The current LHC limits are sensitive to this signature in the mass region $100~\GeV<M_{E}<180~\GeV$ and $M_{N_1}<30~\GeV$ which are based on $2\ell+\cancel{E}_T$ searches of direct production of electroweakinos and sleptons~\cite{Aad:2014vma}. But as discussed in Sec. \ref{DM}, such a low mass can hardly pass the constraints from DM. A brief discussion on such a signature with a much heavier $N_1$ at LHC has been performed in Ref. \cite{Bhattacharya:2015qpa}.

{\bf (F4)} $2\ell(Z)2j(Z)$ \quad In this signature a pair of opposite-sign leptons is required to reconstruct a $Z$ boson while a pair of jets is required to reconstruct a second $Z$ boson. This $ZZ$ signature comes from the decays of a neutral pair:
\begin{equation}
pp\to N_2 \bar{N}_2 \to Z N_1+ Z \bar{N}_1,
\end{equation}
with one $Z\to\ell^+\ell^-$ and the other $Z\to q\bar{q}$. There can also be fake contributions coming from $WZ$ and $hZ$ decays. Current LHC limits for this signature also come from direct searches of electroweakinos and sleptons~\cite{Khachatryan:2014mma,Khachatryan:2014qwa}. Assuming $M_{N_1}=1~\GeV$, the most sensitive mass region is $M_{N_2}>250~\GeV$ for $\BR(N_2\to ZN_1)>0.5$; and $M_{N_2}<380~\GeV$ is likely excluded by CMS for $\BR(N_2\to ZN_1)\approx 1$~\cite{Khachatryan:2014mma}. With a much heavier $N_1$ and $\BR(N_2\to ZN_1)=0.5$ in this model, the exclusion limits are considerably weakened.

{\bf (F5)} $2\ell(Z)1h$ \quad This $Zh$ signature also comes from the decays of a neutral pair $N_2\bar{N}_2$:
\begin{eqnarray}
pp\to N_2 \bar{N}_2 \to Z N_1 + h \bar{N}_1,~h N_1 + Z \bar{N}_1,
\end{eqnarray}
with $h\to b\bar{b}/\gamma\gamma$. The production rate of this $Zh$ signature is twice as large as $ZZ$ above in our benchmark scenario for case (a) of Fig. \ref{BRN2}. Analogously to previous signals, we found that most relevant LHC limits come from Ref.~\cite{Khachatryan:2014mma}. The most sensitive mass region is $160~\GeV<M_{N_2}<430~\GeV$ with $\BR(N_2\to hN_1)\in[0.45,0.85]$. For the signature dominated by the $b\bar{b}$ channel, no exclusion limits are set due to large $t\bar t$ backgrounds.

\begin{figure}[!htbp]
\begin{center}
\includegraphics[width=0.45\linewidth]{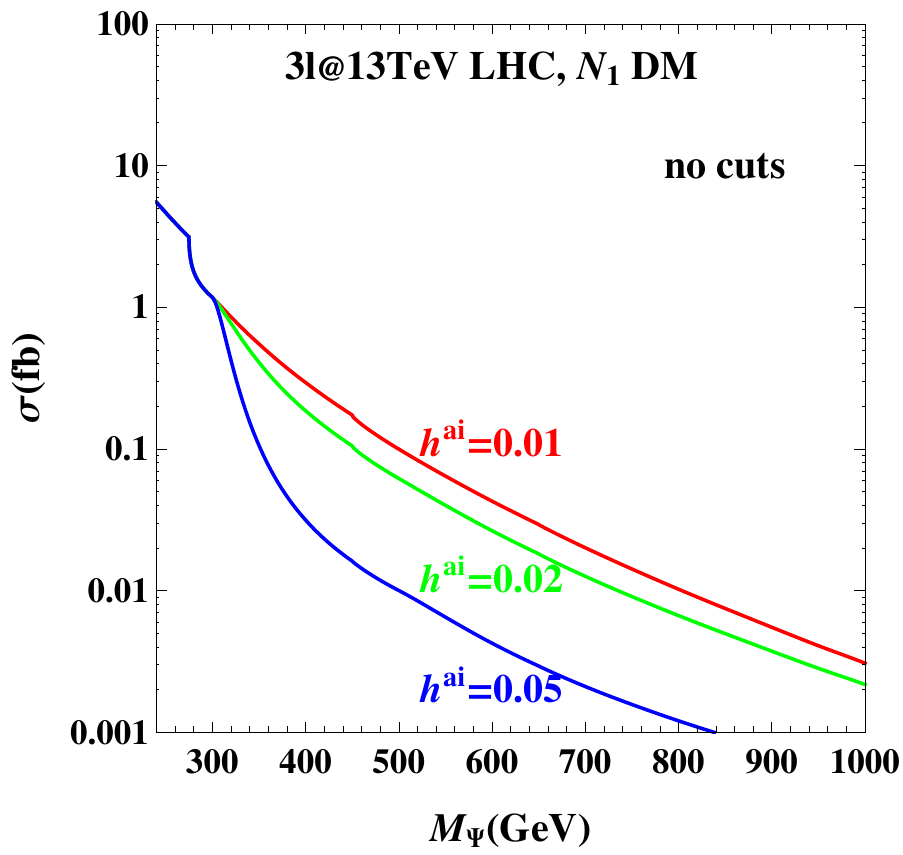}
\includegraphics[width=0.45\linewidth]{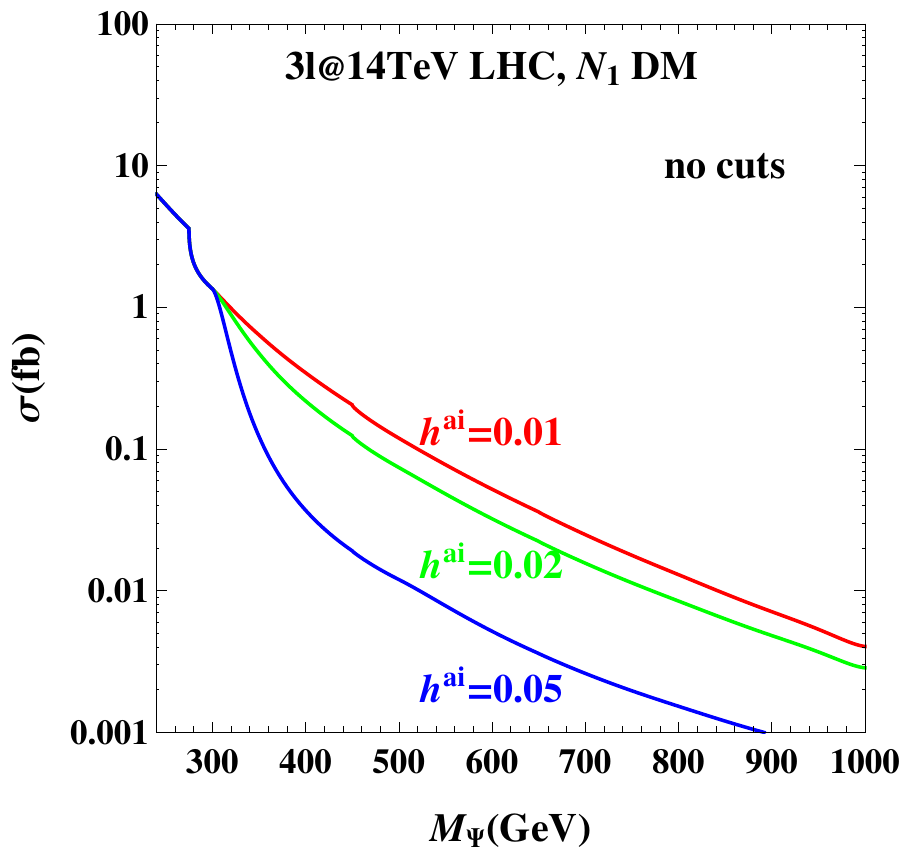}
\end{center}
\caption{Same as Fig. \ref{2l_N} but for the $3\ell+\cancel{E}_T$ signature in Eq. (\ref{pro_3l_N}).
\label{3l_N}}
\end{figure}

{\bf (F6)} $3\ell(Z)$ \quad The production mechanism for this trilepton signature is
\begin{equation}\label{pro_3l_N}
pp\to E^\pm N_2 \to W^\pm N_1 + Z N_1.
\end{equation}
The cross section for the $3\ell+\cancel{E}_T$ signature at $13~(14)~\TeV$ LHC is shown in Fig. \ref{3l_N}. It is comparable with that the di-lepton signature in Eq. (\ref{pro_2l_N}) due to a relatively large production rate of $E^\pm N_2$. But with a much cleaner background, this signature is expected to be the most promising one and to set the most stringent constraints in the mass region $M_{N_2}\lesssim 250~\GeV$. Once again, current limits for this signature have been set by ATLAS \cite{Aad:2014nua} and CMS\cite{Khachatryan:2014qwa,Chatrchyan:2014aea} from direct searches for electroweakinos. A recasting work \cite{Calibbi:2014lga} based on ATLAS limits has been performed in the gaugino-higgsino sector in MSSM with bino-like DM and decoupled sfermions. We can transfer their recasting limits to our signal. Instead of $M_{N_2}>370~\GeV$ set by Ref.~\cite{Aad:2014nua}, recasting shows that the ATLAS limits are sensitive in the mass region $M_{N_2}\lesssim 270~\GeV$ and $M_{N_1}\lesssim75~\GeV$~\cite{Calibbi:2014lga}. In addition, a combined analysis on the $2\ell$ and $3\ell$ signals by ATLAS~\cite{Aad:2014vma} shows that $M_{N_2}>425~\GeV$. However, most of current limits are based on simplified models and can be significantly relaxed with different spectra, decay chains and branching ratios.

{\bf (F7)} $4\ell(ZZ)$ \quad This signature requires two pairs of opposite-sign dilepton to reconstruct the $Z$ pair. It results from the process
\begin{equation}
pp\to N_2\bar{N}_2 \to Z N_1 + Z \bar{N}_1,
\end{equation}
with both $Z\to\ell^+\ell^-$. Although this signature is very clean, its production rate is suppressed due to the small leptonic branching ratio of $Z$. For this signal, the constraint from CMS searches \cite{Khachatryan:2014mma} is less stringent than from the $2\ell2j$ signature discussed above.

In summary, for fermion DM, the current most stringent LHC limit comes from the $3\ell$ signal resulting from $WZ$ bosons. At upcoming LHC run II, other signatures such as $4b$ from $hh$, $2\ell2b$ from $hZ$, and $2\ell2j$ from $ZZ$ are expected to have better sensitivity than this one in the high mass region. More importantly, noting the similarity of signals between fermion DM in the $\mathbb{Z}_3$ model and electroweakinos/sleptons in SUSY models, it is very interesting to recast search limits on the latter to this scenario and examine their interplay with DM constraints. For this purpose, a detailed simulation and recasting is necessary using the tools already available~\cite{Drees:2013wra,Conte:2014zja,Papucci:2014rja}. We hope to come back to this in another work.

\subsubsection{Signatures for $\chi_L$ DM}

Now we turn to the signatures related to scalar DM. A distinct decay mode of $N_2$ in this scenario is $N_2\to \nu\chi_L$, where both $\nu$ and $\chi_L$ are invisible at colliders. This results in various mono-$X$ ($X=j,~\gamma,~W,~Z,~h,~\ell$) signatures at LHC. In what follows, we first discuss these mono-$X$ signatures, and then the  signatures of multi-leptons plus $\cancel{E}_T$ with or without $h$.

{\bf (S1)} $1j$ \quad This mono-jet signature is extensively studied in DM searches at LHC. It proceeds as
\begin{equation}\label{monoj}
pp\to N_2\bar{N}_2+j\to \nu \chi_L \nu\chi_L^\dag +j,
\end{equation}
and in the low mass region $M_{\chi_L}<M_h/2$, the following signal channel should also be considered:
\begin{equation}
pp\to h+j \to \chi_L\chi_L^\dag+j.
\end{equation}
The second process depends on the coupling $\lambda_{h\chi_L}$, and according to Ref. \cite{Arhrib:2013ela}, the $14~\TeV$ LHC with $300~\fb^{-1}$ luminosity has the ability to probe $|\lambda_{h\chi_L}|<6\times10^{-3}$. The mono-jet searches by both CMS \cite{Khachatryan:2014rra} and ATLAS \cite{Aad:2015zva} are based on the effective field theory approach to weakly interacting massive particles of DM, where only the DM pair contributes to the signature $\cancel{E}_T$. Differently from this, the signature $\cancel{E}_T$ in this $\mathbb{Z}_3$ model is also contributed by the neutrino pair as shown in Eq. \ref{monoj}. Since $N_2\bar{N}_2$ can be copiously produced through the Drell-Yan process, we expect that there could be tight constraints from the mono-jet signature. Moreover, the mono-$\gamma$ \cite{Chatrchyan:2012tea,Aad:2014tda} and mono-$W/Z$ \cite{Aad:2013oja,Aad:2014vka} signatures are also possible at LHC. Although such signatures are less promising than mono-jet, they can be used as a diagnostic tool of the underlying models \cite{Abdallah:2015uba}.

{\bf (S2)} $1h$ \quad 
This is the so-called mono-$h$ signature at LHC \cite{Carpenter:2013xra,Berlin:2014cfa}, which has attracted attention since the Higgs discovery \cite{Aad:2012tfa,Chatrchyan:2012xdj}. The signature arises from
\begin{equation}
pp\to N_2\bar{N}_2 \to \nu\chi_H + \nu \chi_L^{\dag},~\nu\chi_L + \nu\chi_H^\dag,
\end{equation}
with $\chi_H^{(\dag)}\to h\chi_L^{(\dag)}$, when $\chi_{L,H}$ are both lighter than $N_2$. Searches for the signature have been recently published by ATLAS in the $h\to\gamma\gamma$ \cite{Aad:2015yga} and $h\to b\bar{b}$ \cite{Aad:2015dva} channel. The upper limit on the cross section is $0.7~\fb$ for $\gamma\gamma$ and $3.6~\fb$ for $b\bar{b}$ with $\cancel{E}_T>90~\GeV$ and $\cancel{E}_T>150~\GeV$ respectively. Similar to the mono-jet signature, this mono-$h$ also has a pair of neutrinos contributing to $\cancel{E}_T$. Since $\chi_H$ must be $125~\GeV$ heavier than $\chi_L$, $\BR(N_2\to\nu\chi_H)$ should be always smaller than $0.5$, but on the other hand $\BR(\chi_H\to h\chi_L)$ is totally dominant. Therefore, this signature is also promising.

{\bf (S3)} $2h$ \quad 
This double Higgs plus $\cancel{E}_T$ signature is also produced in the case of scalar DM
\begin{equation}
pp\to N_2\bar{N}_2 \to \nu\chi_H+\nu\chi_H^{\dag},
\end{equation}
with both $\chi_H^{(\dag)}\to h\chi_L^{(\dag)}$. The searches by CMS \cite{Khachatryan:2014mma} are also applicable here. Differently from the case of fermion DM, the $h$-pair now comes from the cascade decay of $\chi_H$ and thus their sequential decay products $b\bar{b}/\gamma\gamma$ are expected to be less energetic.

{\bf (S4)} $1\ell$ \quad 
This signature can be regarded as a mono-$\ell$ with a large $\cancel E_T$ at LHC, and it arises from
\begin{equation}
pp\to E^\pm N_2 \to \ell^\pm\chi_L^{(\dag)} + \nu \chi_L.
\end{equation}
As shown in Fig. \ref{cs}, the production rate of $E^\pm N_2$ is the largest at LHC. For both $\chi_H$ and $N_1$ heavier than $N_2$, $E^\pm\to\ell^\pm\chi_L^{(\dag)}$ and $N_2\to \nu\chi_L$ are totally dominant. The mono-$\ell$ search has been performed by CMS \cite{Khachatryan:2014tva}. With both $\nu$ and $\chi_L$ contributing to $\cancel{E}_T$, we expect severe constraints on an electroweak-scale $N_2$.

{\bf (S5)} $1\ell 1h$ \quad This signature is quite similar to the $Wh$ signature in the fermion DM case. The production mechanism is
\begin{equation}
pp\to E^\pm N_2 \to \ell^\pm \chi_L^{(\dag)} + \nu\chi_H, \ell^\pm \chi_H^{(\dag)}+\nu\chi_L,
\end{equation}
with $\chi_H^{(\dag)}\to h\chi_L^{(\dag)}$. The searches for the $Wh$ signature \cite{Khachatryan:2014mma,Khachatryan:2014qwa,Aad:2015jqa,Aad:2015eda} can be applied to set a constraint on this signature as well. But differently from the fermion DM case, the branching ratios of $E^\pm \to \ell^\pm \chi_{L,H}^{(\dag)}$ and $N_2 \to \nu \chi_{L,H}$ can be varied by tuning $M_{\chi_H}$ and the corresponding Yukawa couplings $h^{ai}$.

{\bf (S6)} $1\ell 2h$ \quad This signature can only be produced in the case of scalar DM, and thus can be used to distinguish the character of DM at LHC. It follows from the process
\begin{equation}
pp\to E^\pm N_2 \to \ell^\pm \chi_H^{(\dag)} + \nu\chi_H,
\end{equation}
with both $\chi_H^{(\dag)}\to h\chi_L^{(\dag)}$. A similar signature has been studied in the context of type-II seesaw \cite{Han:2015hba}, where the lepton comes from an off-shell $W$. The additional $\ell$ and $\cancel{E}_T$ provides more efficient cuts than the pure Higgs pair to suppress the background, hence this signature is within the reach of LHC for a light $N_2$ \cite{Han:2015hba}.

\begin{figure}[!htbp]
\begin{center}
\includegraphics[width=0.45\linewidth]{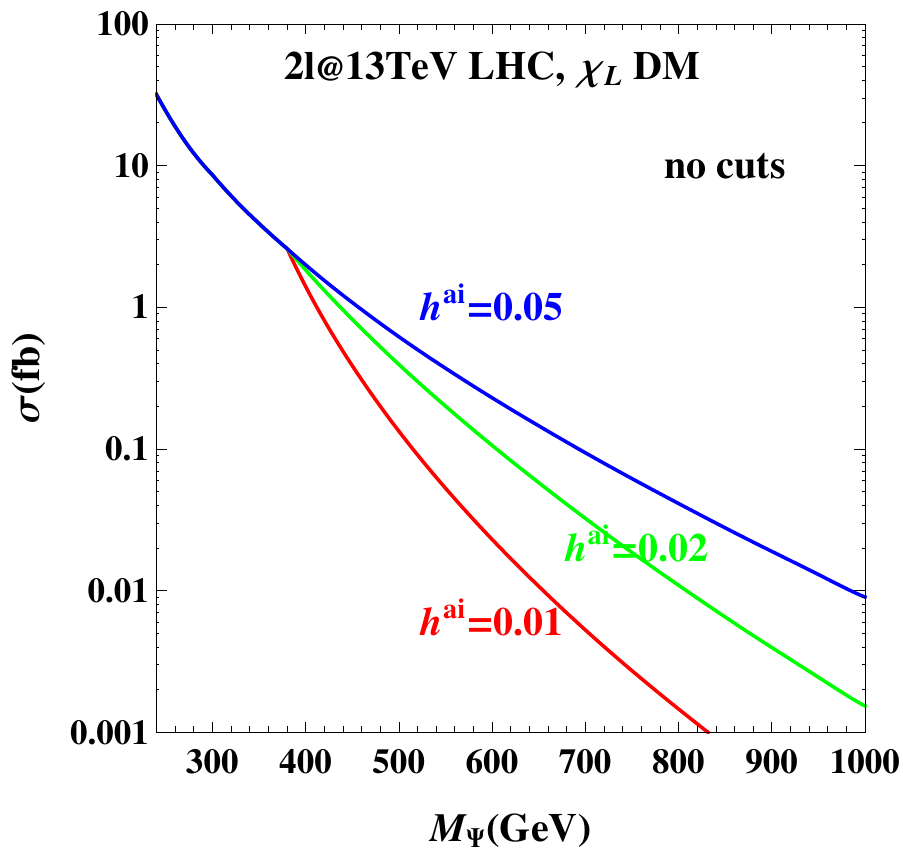}
\includegraphics[width=0.45\linewidth]{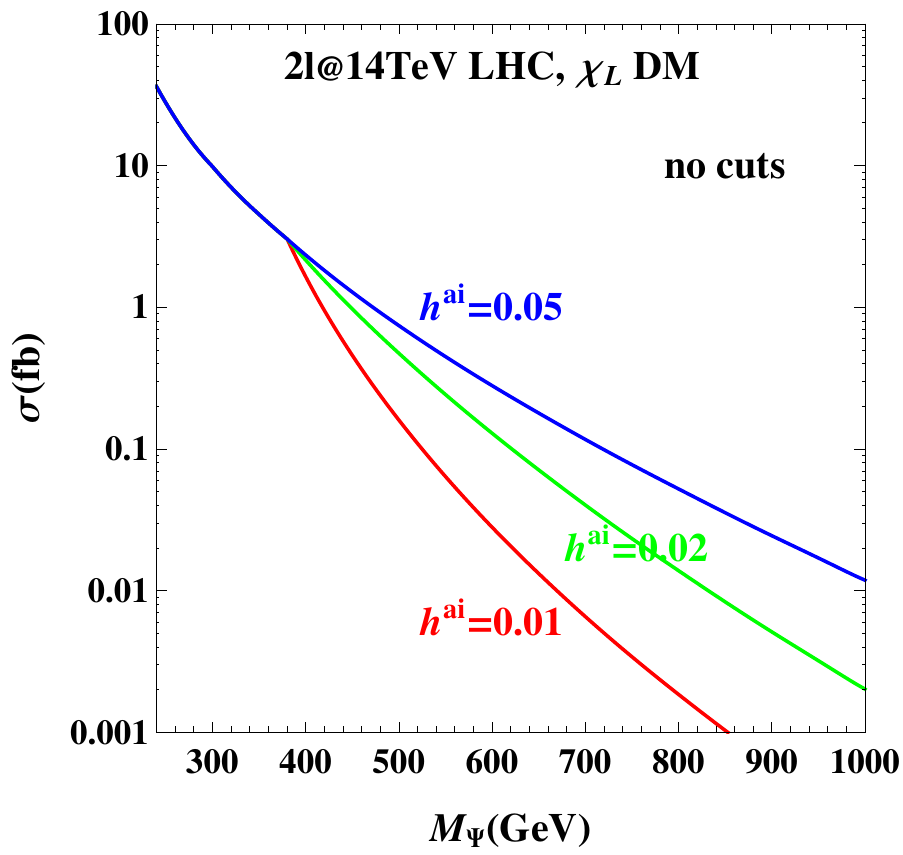}
\end{center}
\caption{Cross section of the $2\ell+\cancel{E}_T$ signature in Eq. \ref{pro_2l_X} as a function of $M_{\Psi}$ at $13~\TeV$ LHC (left panel) and $14~\TeV$ LHC (right). Here, we set $\sin\beta=0.01$ and $(M_{N_1},M_{\chi_L},M_{\chi_H})=(300,150,200)~\GeV$.
\label{2l_X}}
\end{figure}

{\bf (S7)} $2\ell~(\Znot)$ \quad Differently from the fermion DM case, the lepton pair is produced from direct decays of $E^\pm$,
\begin{eqnarray}\label{pro_2l_X}
pp &\to& E^+E^- \to \ell^+\chi_L^\dag + \ell^-\chi_L,
\end{eqnarray}
and is expected to be much more energetic for a large mass splitting between $N_2$ and $\chi_L$ than from the $W$ pair in the fermion DM case. This will lead to a more stringent constraint at colliders. The cross section at $13~(14)~\TeV$ is depicted in Fig. \ref{2l_X} for a universal Yukawa coupling $h^{ai}$, so that $\BR(E^\pm\to e^\pm\chi_{L,H})=\BR(E^\pm\to\mu^\pm\chi_{L,H})=\BR(E^\pm\to \tau^\pm\chi_{L,H})$. Contrary to the fermion DM case, the cross section now increases with $h^{ai}$. The search for this signature by ATLAS \cite{Aad:2014vma} has excluded the mass of $E^\pm$ between $160~\GeV$ and $310~\GeV$ with $M_{\chi_L}=100~\GeV$ for a simplified model.

{\bf (S8)} $2\ell(\Znot)1h$ \quad Though sharing the same final state as the $hZ$ signature in the case of fermion DM, the lepton pair is from the direct decays of $E^\pm$,
\begin{equation}
pp\to E^+E^- \to \ell^+\chi_H^{\dag} + \ell^-\chi_L, ~\ell^+\chi_L^{\dag} + \ell^-\chi_H,
\end{equation}
with $\chi_H^{(\dag)}\to h\chi_L^{(\dag)}$. As shown in case (c) of Fig.~\ref{BREc}, $\BR(E^\pm\to\ell^\pm \chi_H^{\dag})$ can reach over $0.3$, so the production rate for this signature could be promising. Since the $h\to b\bar{b}$ channel suffers from quite large background, we expect the $h\to WW/ZZ/\gamma\gamma/\tau^+\tau^-$ channels to enhance the observability.

{\bf (S9)} $2\ell(\Znot)2h$ \quad As far as we know,  the $\ell^+\ell^-hh+\cancel{E}_T$ signature has been seldom studied in previous papers. To have a pair of $h$ in the final state, we require two $\chi_H$s in the cascade decays of $E^+E^-$,
\begin{equation}
pp\to E^+E^- \to \ell^+\chi_H^{\dag} + \ell^-\chi_H,
\end{equation}
which further cascade decay as $\chi_H^{(\dag)}\to h\chi_L^{(\dag)}$. Since $\BR(E\to\ell \chi_H)\approx 0.3$ and $\BR(\chi_H\to h\chi_L)\approx 0.8-1$, the cross section of this signature is roughly one-tenth of $\sigma(E^+E^-)$. On the other hand, the backgrounds such as $ZZhh$, $WWhh$, $t\bar{t}jj$, etc., are relatively small. So this signature may also be promising at LHC.

\begin{figure}[!htbp]
\begin{center}
\includegraphics[width=0.45\linewidth]{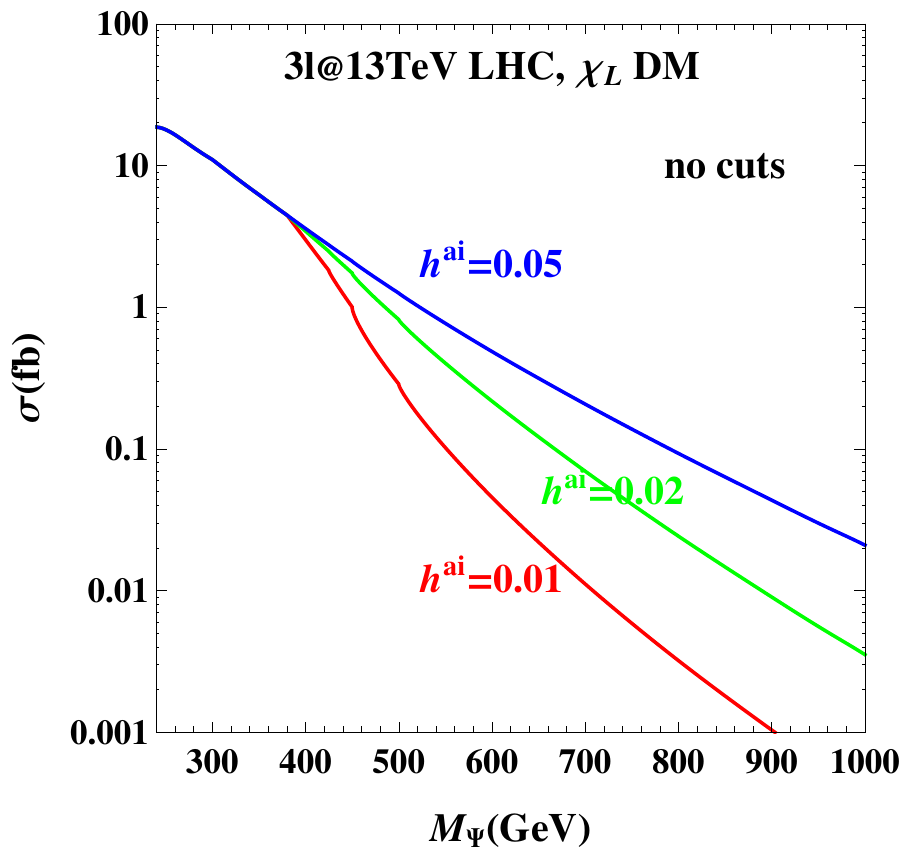}
\includegraphics[width=0.45\linewidth]{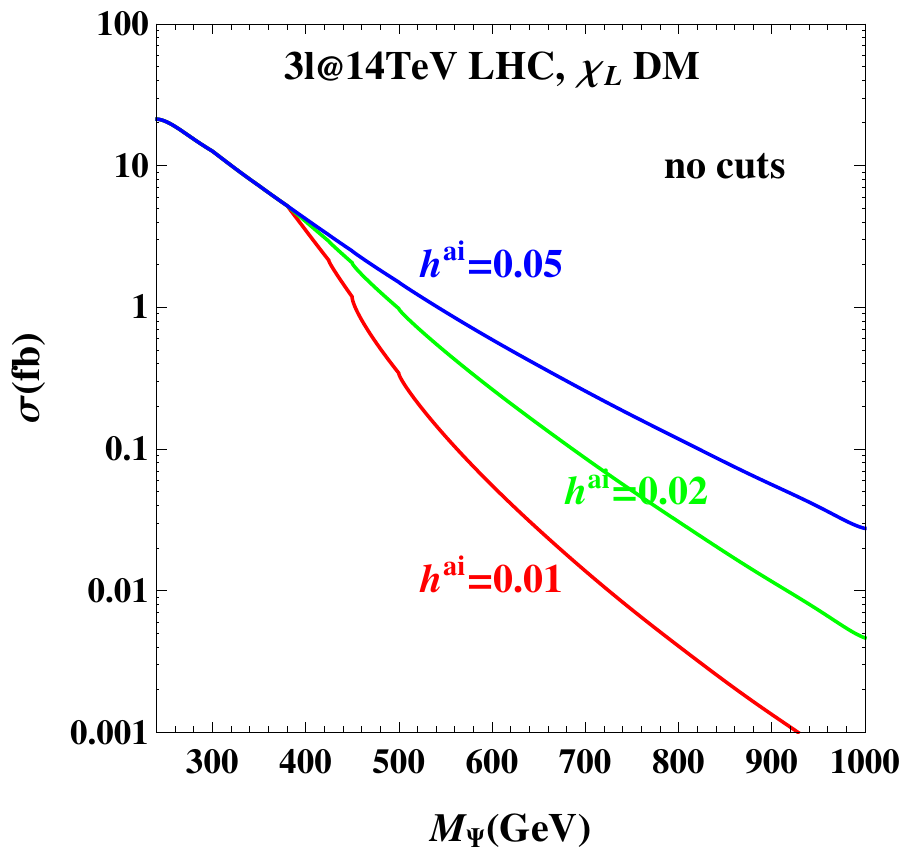}
\end{center}
\caption{Same as Fig. \ref{2l_X} but for the $3\ell+\cancel{E}_T$ signature in Eq. (\ref{pro_3l_X}).
\label{3l_X}}
\end{figure}

{\bf (S10)} $3\ell(\Znot)$ \quad The trilepton signature is also possible in the case of scalar DM following the production
\begin{eqnarray}\label{pro_3l_X}
pp\to E^\pm N_2 &\to & \ell^\pm\chi_L + \nu\chi_H,~\ell^\pm\chi_H + \nu\chi_L,
\end{eqnarray}
and decays $\chi_H^{(\dag)}\to\ell^+\ell^-\chi_L^{(\dag)}$ mediated by an off-shell $E^{\pm}$. To have a relatively large branching ratio for the decays, the mass splitting between $\chi_H$ and $\chi_L$ must be less than $M_{h}$. The theoretical cross section for the signature is plotted in Fig. \ref{3l_X}, and it can be about ten times larger than that from $WZ$ in Eq. \ref{pro_3l_N} for the fermion DM case. The same final state has been searched for by CMS \cite{Khachatryan:2014qwa} and ATLAS \cite{Aad:2014nua} for sleptons lighter than charginos and neutrolinos, with an exclusion limit on $M_{N_2}$ up to about $700~\GeV$. But these constraints cannot be applied directly to the signature here, mainly because of the softness of the dilepton from $\chi_H^{(\dag)}\to\ell^+\ell^-\chi_L^{(\dag)}$. A recasting of it on the LHC searches would reveal a more realistic constraint.

{\bf (S11)} $4\ell(\Znot)$ \quad There are two processes contributing to this signature
\begin{eqnarray}
pp\to E^+E^- &\to & \ell^+\chi_L^\dag+\ell^-\chi_H,~\ell^+\chi_H^\dag+\ell^-\chi_L, \\
pp\to N_2\bar{N}_2 & \to & \nu \chi_H+\nu\chi_H^\dag,
\end{eqnarray}
with $\chi_H^{(\dag)}\to\ell^+\ell^-\chi_L^{(\dag)}$ as well. The first process has one pair of energetic leptons from the direct decays of $E^\pm$, while all leptons in the second are expected to be soft. The search for this signature has been carried out by ATLAS based on the simplified versions of R-parity-conserving, R-parity-violating, and general gauge-mediated SUSY breaking models~\cite{Aad:2014iza}. With appropriate matching of particles and decay chains, we obtain that $M_{N_2}<600~\GeV$ with $M_{\chi_L}<100~\GeV$ has been excluded by the direct search \cite{Aad:2014iza}. For $M_{N_2}<500~\GeV$, the exclusion limit on $M_{\chi_L}$ of this 4-lepton signature is comparable with that of the trilepton signature. But for the same reason as discussed for the trilepton signature, the constraint cannot be taken for granted before a detailed simulation is performed.

To summarize the case of scalar DM, the most stringent constraint is also expected to come from the $3\ell$ signature. More interestingly, we find that various mono-$X$ ($X=j,~\gamma,~W,~Z,~h,~\ell$) signatures appear in this case, and differently from the current searches \cite{Khachatryan:2014rra,Aad:2015zva,Chatrchyan:2012tea,Aad:2014tda,Aad:2013oja,
Aad:2014vka,Aad:2015yga,Aad:2015dva},
missing transverse energy involves both scalar DM $\chi_L$ and neutrinos. A detailed simulation and recasting of these mono-$X$ and multi-$\ell$ signatures with or without $h$ is necessary to clarify the feasibility of testing the scalar DM scenario at LHC.

Before ending up this section, we briefly discuss how to distinguish between the collider signatures of $\mZ_2$ and $\mZ_3$ DM models. Based on the method developed in Refs.~\cite{Agashe:2010gt,Agashe:2010tu}, the two symmetries can be potentially discriminated by using multiple kinematical edges and cusps. The basic idea is that the cascade decay of a $\mZ_3$ particle can result in two visible particles that are separated by a DM particle. Such a decay chain involves a triple coupling of $\mZ_3$ particles which is absent in the $\mZ_2$ case. But in the minimal case with only two $\mZ_3$ scalars ($\chi_{H,L}$), the desired decay chain is hard to realize. For that purpose, we may introduce a third scalar $\chi_3$. Then a concrete example would be the decay chain, $E^-\to \ell^-\chi_3\to\ell^-\chi_L^\dag\chi_H^\dag\to\ell^-\chi_L^\dag h\chi_L^\dag$, assuming the mass hierarchy $M_E>M_{\chi_3}>M_{\chi_H}>M_{\chi_L}$ and suitable mass splitting. The charged lepton $\ell$ and Higgs boson $h$ are then separated by the DM particle $\chi_L^\dag$, which then results in a cusp in the distributions of the kinematical variables $M_{\ell h}$ (energy of the $\ell h$ system) and $M^2_{\ell h}$ (invariant mass squared) \cite{Agashe:2010gt}.

\section{Conclusion}
\label{concl}

We have made a comprehensive analysis on the phenomenology of a $\mathbb{Z}_3$ DM model that generates neutrino mass at two loops. We have examined in great detail its properties in relic density, direct detection and LHC signatures. For indirect detection, we also briefly discussed the GCE issue. To conclude, we summarize the key features separately for the scalar and fermion DM as follows.

For the scalar $\chi_L$ DM, there are three ST-A channels $\chi_L\chi_L^\dagger\to b\bar{b},~W^{+}W^{-},~hh$, and three SE-A/CO-A channels $\chi_L\chi_L\to \chi_L^\dagger h$ and $\chi_L\chi_H^\dagger,\chi_H\chi_H^\dagger\to W^{+}W^{-}$. The $\chi_L\chi_L^\dagger\to W^{+}W^{-}$ channel can satisfy both relic density and direct detection constraints in a vast mass region and thus gives the dominant contribution in the parameter space. Upon imposing the direct detection constraint, the $b\bar{b}$ channel is almost excluded while most of SE-A and CO-A processes survive. This is due to the fact that $\lambda_{h\chi_L}$ required by relic density is considerably relaxed for the SE-A/CO-A channels, thus alleviating the tension from direct detection. Concerning the LHC constraints, the $3\ell$ signal is expected to give the most stringent bound. Moreover, various mono-$X$ ($X=j,~\gamma,~W,~Z,~h,~\ell$) signatures are different from those in current LHC searches since missing transverse energy now involves both scalar DM and neutrinos. A detailed simulation and recasting of these mono-$X$ and multi-$\ell$ signatures with or without $h$ will be helpful to test the scalar DM scenario at LHC.

If the lighter of neutral fermions ($N_1$) plays the role of DM, the direct detection requires it to be an almost singlet with a mixing angle $\beta<2^{\circ}$. Compared with $\chi_L$ DM, it has more annihilation channels, including two ST-A channels $N_1\bar{N_1}\to d\bar{d},~b\bar{b}$ and eight SE-A/CO-A channels $N_1\chi_L\to \bar{N_1}h,~E^{+}W^{-},~\bar{N_1}Z$; $N_1N_1\to \chi_L^\dagger h,~E^{+}\ell^{-},~\bar{N_1}\nu$; $N_2E^{+}\to t\bar{b},~u\bar{d}$. However, only five SE-A/CO-A channels ($N_1\chi_L\to \bar{N_1}h;~N_1N_1\to E^{+}\ell^{-},~\chi_L^\dagger h;~N_2E^{+}\to u\bar{d},~t\bar{b}$) survive the LUX constraint, due to the same reason as for scalar DM. Interestingly, the LHC signatures of fermion DM are very similar to those of electroweakinos in simplified SUSY models. Currently, the $3\ell$ signal resulting from $WZ$ bosons provides the most severe bound. At upcoming LHC run II, other signatures such as $4b$ from $hh$, $2\ell2b$ from $hZ$, and $2\ell2j$ from $ZZ$ may be more promising in the high mass region. To make accurate estimation, it is necessary to recast current search limits on electroweakinos/sleptons and combine them with DM constraints.

Finally, this model can also schematically explain the GCE observed by Fermi-LAT when taking into account contributions from SE-A processes for appropriate DM mass. The corresponding annihilation channels are $\chi_L\chi_L\to \chi^\dagger_Lh$ for $\chi_L$ DM and $N_1\chi_L\to \bar N_1h$ for $N_1$ DM. A comprehensive analysis of this issue based on the MCMC method deserves a separate work.

\section*{Acknowledgement}

This work was supported in part by the Grants No. NSFC-11025525, No. NSFC-11575089 and by the CAS Center for Excellence in Particle Physics (CCEPP). The numerical analysis was done with the HPC Cluster of SKLTP/ITP-CAS.



\begin{thebibliography}{000}


\bibitem{Weinberg:1979sa}
  S.~Weinberg,
  Phys.\ Rev.\ Lett.\  {\bf 43} (1979) 1566.

\bibitem{Ma:1998dn}
  E.~Ma,
  Phys.\ Rev.\ Lett.\  {\bf 81} (1998) 1171
  [arXiv:hep-ph/9805219].

\bibitem{Ma:2006km}
  E.~Ma,
  Phys.\ Rev.\ D {\bf 73} (2006) 077301
  [hep-ph/0601225].

\bibitem{Ma:2007gq}
  E.~Ma,
  Phys.\ Lett.\ B {\bf 662}, 49 (2008)
  [arXiv:0708.3371 [hep-ph]].

\bibitem{Kanemura:2010bq}
  S.~Kanemura and T.~Ota,
  Phys.\ Lett.\ B {\bf 694} (2010) 233
  [arXiv:1009.3845 [hep-ph]].

\bibitem{Kanemura:2012rj}
  S.~Kanemura and H.~Sugiyama,
  Phys.\ Rev.\ D {\bf 86} (2012) 073006
  [arXiv:1202.5231 [hep-ph]].

\bibitem{Restrepo:2013aga}
  D.~Restrepo, O.~Zapata and C.~E.~Yaguna,
  JHEP {\bf 1311} (2013) 011
  [arXiv:1308.3655 [hep-ph]].

\bibitem{Hirsch:2013ola}
  M.~Hirsch, R.~A.~Lineros, S.~Morisi, J.~Palacio, N.~Rojas and J.~W.~F.~Valle,
  JHEP {\bf 1310} (2013) 149
  [arXiv:1307.8134 [hep-ph]].

\bibitem{Kajiyama:2013rla}
  Y.~Kajiyama, H.~Okada and T.~Toma,
  Phys.\ Rev.\ D {\bf 88} (2013) 1,  015029
  [arXiv:1303.7356].

\bibitem{Law:2013saa}
  S.~S.~C.~Law and K.~L.~McDonald,
  JHEP {\bf 1309} (2013) 092
  [arXiv:1305.6467 [hep-ph]].

\bibitem{Fabbrichesi:2014qca}
  M.~Fabbrichesi and S.~T.~Petcov,
  Eur.\ Phys.\ J.\ C {\bf 74} (2014) 2774
  [arXiv:1304.4001 [hep-ph]].

\bibitem{Ahriche:2014oda}
  A.~Ahriche, K.~L.~McDonald and S.~Nasri,
  JHEP {\bf 1410} (2014) 167
  [arXiv:1404.5917 [hep-ph]].

\bibitem{Chen:2014ska}
  C.~S.~Chen, K.~L.~McDonald and S.~Nasri,
  Phys.\ Lett.\ B {\bf 734} (2014) 388
  [arXiv:1404.6033 [hep-ph]].

\bibitem{Kanemura:2014rpa}
  S.~Kanemura, T.~Matsui and H.~Sugiyama,
  Phys.\ Rev.\ D {\bf 90} (2014) 1,  013001
  [arXiv:1405.1935 [hep-ph]].

\bibitem{Okada:2014qsa}
  H.~Okada, T.~Toma and K.~Yagyu,
  Phys.\ Rev.\ D {\bf 90} (2014) 9,  095005
  [arXiv:1408.0961 [hep-ph]].


\bibitem{Krauss:1988zc}
  L.~M.~Krauss and F.~Wilczek,
  Phys.\ Rev.\ Lett.\  {\bf 62}, 1221 (1989).
  doi:10.1103/PhysRevLett.62.1221

\bibitem{D'Eramo:2010ep}
  F.~D'Eramo and J.~Thaler,
  JHEP {\bf 1006}, 109 (2010)
  doi:10.1007/JHEP06(2010)109
  [arXiv:1003.5912 [hep-ph]].

\bibitem{Batell:2010bp}
  B.~Batell,
  Phys.\ Rev.\ D {\bf 83}, 035006 (2011)
  doi:10.1103/PhysRevD.83.035006
  [arXiv:1007.0045 [hep-ph]].

\bibitem{Belanger:2012vp}
  G.~Belanger, K.~Kannike, A.~Pukhov and M.~Raidal,
  JCAP {\bf 1204}, 010 (2012)
  doi:10.1088/1475-7516/2012/04/010
  [arXiv:1202.2962 [hep-ph]].

\bibitem{Aoki:2012ub}
  M.~Aoki, M.~Duerr, J.~Kubo and H.~Takano,
  Phys.\ Rev.\ D {\bf 86}, 076015 (2012)
  doi:10.1103/PhysRevD.86.076015
  [arXiv:1207.3318 [hep-ph]].

\bibitem{Belanger:2012zr}
  G.~Belanger, K.~Kannike, A.~Pukhov and M.~Raidal,
  JCAP {\bf 1301}, 022 (2013)
  [arXiv:1211.1014 [hep-ph]].

\bibitem{Belanger:2014bga}
  G.~B¨¦langer, K.~Kannike, A.~Pukhov and M.~Raidal,
  JCAP {\bf 1406}, 021 (2014)
  doi:10.1088/1475-7516/2014/06/021
  [arXiv:1403.4960 [hep-ph]].

\bibitem{Agashe:2010gt}
  K.~Agashe, D.~Kim, M.~Toharia and D.~G.~E.~Walker,
  Phys.\ Rev.\ D {\bf 82}, 015007 (2010)
  [arXiv:1003.0899 [hep-ph]].

\bibitem{Agashe:2010tu}
  K.~Agashe, D.~Kim, D.~G.~E.~Walker and L.~Zhu,
  Phys.\ Rev.\ D {\bf 84}, 055020 (2011)
  [arXiv:1012.4460 [hep-ph]].

\bibitem{Belanger:2011ww}
  G.~Belanger and J.~C.~Park,
  JCAP {\bf 1203}, 038 (2012)
  doi:10.1088/1475-7516/2012/03/038
  [arXiv:1112.4491 [hep-ph]].



\bibitem{Aoki:2014cja}
  M.~Aoki and T.~Toma,
  JCAP {\bf 1409}, 016 (2014)
  [arXiv:1405.5870 [hep-ph]].

\bibitem{Lerner:2009xg}
  R.~N.~Lerner and J.~McDonald,
  Phys.\ Rev.\ D {\bf 80}, 123507 (2009)
  [arXiv:0909.0520 [hep-ph]].

\bibitem{Liao:2009fm}
  Y.~Liao and J.~Y.~Liu,
  Phys.\ Rev.\ D {\bf 81}, 013004 (2010)
  [arXiv:0911.3711 [hep-ph]].

\bibitem{Ding:2014nga}
  R.~Ding, Z.~L.~Han, Y.~Liao, H.~J.~Liu and J.~Y.~Liu,
  Phys.\ Rev.\ D {\bf 89}, no. 11, 115024 (2014)
  [arXiv:1403.2040 [hep-ph]].

\bibitem{Adam:2013mnn}
  J.~Adam {\it et al.} [MEG Collaboration],
  Phys.\ Rev.\ Lett.\  {\bf 110}, 201801 (2013)
  [arXiv:1303.0754 [hep-ex]].

\bibitem{Aubert:2009ag}
  B.~Aubert {\it et al.} [BaBar Collaboration],
  Phys.\ Rev.\ Lett.\  {\bf 104}, 021802 (2010)
  [arXiv:0908.2381 [hep-ex]].

\bibitem{Vicente:2014wga}
  A.~Vicente and C.~E.~Yaguna,
  JHEP {\bf 1502}, 144 (2015)
  [arXiv:1412.2545 [hep-ph]].

\bibitem{Cynolter:2008ea}
  G.~Cynolter and E.~Lendvai,
  Eur.\ Phys.\ J.\ C {\bf 58}, 463 (2008)
  [arXiv:0804.4080 [hep-ph]].

\bibitem{Lavoura:1992np}
  L.~Lavoura and J.~P.~Silva,
  Phys.\ Rev.\ D {\bf 47}, 2046 (1993).

\bibitem{Baak:2014ora}
  M.~Baak {\it et al.} [Gfitter Group Collaboration],
  Eur.\ Phys.\ J.\ C {\bf 74}, 3046 (2014)
  [arXiv:1407.3792 [hep-ph]].

\bibitem{Bhattacharya:2015qpa}
  S.~Bhattacharya, N.~Sahoo and N.~Sahu,
  arXiv:1510.02760 [hep-ph].

\bibitem{Belyaev:2012qa}
  A.~Belyaev, N.~D.~Christensen and A.~Pukhov,
  Comput.\ Phys.\ Commun.\  {\bf 184}, 1729 (2013)
  doi:10.1016/j.cpc.2013.01.014
  [arXiv:1207.6082 [hep-ph]].

  \bibitem{feynrules}
  N.~D.~Christensen and C.~Duhr,
  Comput.\ Phys.\ Commun.\  {\bf 180}, 1614 (2009)
  [arXiv:0806.4194 [hep-ph]];
  A.~Alloul, N.~D.~Christensen, C.~Degrande, C.~Duhr and B.~Fuks,
  Comput.\ Phys.\ Commun.\  {\bf 185}, 2250 (2014)
  [arXiv:1310.1921 [hep-ph]].

\bibitem{Belanger:2014vza}
  G.~B¨¦langer, F.~Boudjema, A.~Pukhov and A.~Semenov,
  Comput.\ Phys.\ Commun.\  {\bf 192}, 322 (2015)
  doi:10.1016/j.cpc.2015.03.003
  [arXiv:1407.6129 [hep-ph]].

\bibitem{Ade:2013zuv}
  P.~A.~R.~Ade {\it et al.} [Planck Collaboration],
  Astron.\ Astrophys.\  {\bf 571}, A16 (2014)
  [arXiv:1303.5076 [astro-ph.CO]].

\bibitem{Akerib:2013tjd}
  D.~S.~Akerib {\it et al.} [LUX Collaboration],
  Phys.\ Rev.\ Lett.\  {\bf 112}, 091303 (2014)
  [arXiv:1310.8214 [astro-ph.CO]].


\bibitem{Aad:2015txa}
  G.~Aad {\it et al.} [ATLAS Collaboration],
  arXiv:1508.07869 [hep-ex].

\bibitem{Chatrchyan:2014tja}
  S.~Chatrchyan {\it et al.} [CMS Collaboration],
  Eur.\ Phys.\ J.\ C {\bf 74}, 2980 (2014)
  [arXiv:1404.1344 [hep-ex]].

\bibitem{Aad:2014iia}
  G.~Aad {\it et al.} [ATLAS Collaboration],
  Phys.\ Rev.\ Lett.\  {\bf 112}, 201802 (2014)
  [arXiv:1402.3244 [hep-ex]].

\bibitem{Bechtle:2014ewa}
  P.~Bechtle, S.~Heinemeyer, O.~St\aa l, T.~Stefaniak and G.~Weiglein,
  JHEP {\bf 1411}, 039 (2014)
  [arXiv:1403.1582 [hep-ph]].

\bibitem{Corbett:2015ksa}
  T.~Corbett, O.~J.~P.~Eboli, D.~Goncalves, J.~Gonzalez-Fraile, T.~Plehn and M.~Rauch,
  JHEP {\bf 1508}, 156 (2015)
  [arXiv:1505.05516 [hep-ph]].

\bibitem{Aad:2015pla}
  G.~Aad {\it et al.} [ATLAS Collaboration],
  arXiv:1509.00672 [hep-ex].

\bibitem{ALEPH:2005ab}
  S.~Schael {\it et al.} [ALEPH and DELPHI and L3 and OPAL and SLD and LEP Electroweak Working Group and SLD Electroweak Group and SLD Heavy Flavour Group Collaborations],
  Phys.\ Rept.\  {\bf 427}, 257 (2006)
  [hep-ex/0509008].


\bibitem{Daylan:2014rsa}
  T.~Daylan, D.~P.~Finkbeiner, D.~Hooper, T.~Linden, S.~K.~N.~Portillo, N.~L.~Rodd and T.~R.~Slatyer,
  arXiv:1402.6703 [astro-ph.HE].

\bibitem{Calore:2014xka}
  F.~Calore, I.~Cholis and C.~Weniger,
  JCAP {\bf 1503}, 038 (2015)
  doi:10.1088/1475-7516/2015/03/038
  [arXiv:1409.0042 [astro-ph.CO]].

\bibitem{Goodenough:2009gk}
  L.~Goodenough and D.~Hooper,
  arXiv:0910.2998 [hep-ph].

\bibitem{Hooper:2010mq}
  D.~Hooper and L.~Goodenough,
  Phys.\ Lett.\ B {\bf 697}, 412 (2011)
  doi:10.1016/j.physletb.2011.02.029
  [arXiv:1010.2752 [hep-ph]].

\bibitem{Boyarsky:2010dr}
  A.~Boyarsky, D.~Malyshev and O.~Ruchayskiy,
  Phys.\ Lett.\ B {\bf 705}, 165 (2011)
  doi:10.1016/j.physletb.2011.10.014
  [arXiv:1012.5839 [hep-ph]].

\bibitem{Hooper:2011ti}
  D.~Hooper and T.~Linden,
  Phys.\ Rev.\ D {\bf 84}, 123005 (2011)
  doi:10.1103/PhysRevD.84.123005
  [arXiv:1110.0006 [astro-ph.HE]].

\bibitem{Abazajian:2012pn}
  K.~N.~Abazajian and M.~Kaplinghat,
  Phys.\ Rev.\ D {\bf 86}, 083511 (2012)
  [Phys.\ Rev.\ D {\bf 87}, 129902 (2013)]
  doi:10.1103/PhysRevD.86.083511, 10.1103/PhysRevD.87.129902
  [arXiv:1207.6047 [astro-ph.HE]].

\bibitem{TheFermi-LAT:2015kwa}
  M.~Ajello {\it et al.} [Fermi-LAT Collaboration],
  arXiv:1511.02938 [astro-ph.HE].

\bibitem{Agrawal:2014oha}
  P.~Agrawal, B.~Batell, P.~J.~Fox and R.~Harnik,
  JCAP {\bf 1505}, 011 (2015)
  doi:10.1088/1475-7516/2015/05/011
  [arXiv:1411.2592 [hep-ph]].

\bibitem{Cline:2015qha}
  J.~M.~Cline, G.~Dupuis, Z.~Liu and W.~Xue,
  Phys.\ Rev.\ D {\bf 91}, no. 11, 115010 (2015)
  doi:10.1103/PhysRevD.91.115010
  [arXiv:1503.08213 [hep-ph]].

\bibitem{Elor:2015tva}
  G.~Elor, N.~L.~Rodd and T.~R.~Slatyer,
  Phys.\ Rev.\ D {\bf 91}, 103531 (2015)
  doi:10.1103/PhysRevD.91.103531
  [arXiv:1503.01773 [hep-ph]].

\bibitem{Calore:2014nla}
  F.~Calore, I.~Cholis, C.~McCabe and C.~Weniger,
  Phys.\ Rev.\ D {\bf 91}, no. 6, 063003 (2015)
  doi:10.1103/PhysRevD.91.063003
  [arXiv:1411.4647 [hep-ph]].

\bibitem{Cirelli:2014lwa}
  M.~Cirelli, D.~Gaggero, G.~Giesen, M.~Taoso and A.~Urbano,
  JCAP {\bf 1412}, no. 12, 045 (2014)
  doi:10.1088/1475-7516/2014/12/045
  [arXiv:1407.2173 [hep-ph]].

\bibitem{Ade:2015xua}
  P.~A.~R.~Ade {\it et al.} [Planck Collaboration],
  arXiv:1502.01589 [astro-ph.CO].

\bibitem{Slatyer:2015jla}
  T.~R.~Slatyer,
  arXiv:1506.03811 [hep-ph].

\bibitem{Gordon:2013vta}
  C.~Gordon and O.~Macias,
  Phys.\ Rev.\ D {\bf 88}, no. 8, 083521 (2013)
  [Phys.\ Rev.\ D {\bf 89}, no. 4, 049901 (2014)]
  doi:10.1103/PhysRevD.88.083521, 10.1103/PhysRevD.89.049901
  [arXiv:1306.5725 [astro-ph.HE]].

\bibitem{Abazajian:2014fta}
  K.~N.~Abazajian, N.~Canac, S.~Horiuchi and M.~Kaplinghat,
  Phys.\ Rev.\ D {\bf 90}, no. 2, 023526 (2014)
  doi:10.1103/PhysRevD.90.023526
  [arXiv:1402.4090 [astro-ph.HE]].

\bibitem{Yuan:2014rca}
  Q.~Yuan and B.~Zhang,
  JHEAp {\bf 3-4}, 1 (2014)
  doi:10.1016/j.jheap.2014.06.001
  [arXiv:1404.2318 [astro-ph.HE]].

\bibitem{Bartels:2015aea}
  R.~Bartels, S.~Krishnamurthy and C.~Weniger,
  arXiv:1506.05104 [astro-ph.HE].

\bibitem{Lee:2015fea}
  S.~K.~Lee, M.~Lisanti, B.~R.~Safdi, T.~R.~Slatyer and W.~Xue,
  arXiv:1506.05124 [astro-ph.HE].


\bibitem{Berlin:2014tja}
  A.~Berlin, D.~Hooper and S.~D.~McDermott,
  Phys.\ Rev.\ D {\bf 89}, no. 11, 115022 (2014)
  doi:10.1103/PhysRevD.89.115022
  [arXiv:1404.0022 [hep-ph]].

\bibitem{Alves:2014yha}
  A.~Alves, S.~Profumo, F.~S.~Queiroz and W.~Shepherd,
  Phys.\ Rev.\ D {\bf 90}, no. 11, 115003 (2014)
  doi:10.1103/PhysRevD.90.115003
  [arXiv:1403.5027 [hep-ph]].

\bibitem{Agrawal:2014una}
  P.~Agrawal, B.~Batell, D.~Hooper and T.~Lin,
  Phys.\ Rev.\ D {\bf 90}, no. 6, 063512 (2014)
  doi:10.1103/PhysRevD.90.063512
  [arXiv:1404.1373 [hep-ph]].

\bibitem{Abdullah:2014lla}
  M.~Abdullah, A.~DiFranzo, A.~Rajaraman, T.~M.~P.~Tait, P.~Tanedo and A.~M.~Wijangco,
  Phys.\ Rev.\ D {\bf 90}, 035004 (2014)
  doi:10.1103/PhysRevD.90.035004
  [arXiv:1404.6528 [hep-ph]].

\bibitem{Martin:2014sxa}
  A.~Martin, J.~Shelton and J.~Unwin,
  Phys.\ Rev.\ D {\bf 90}, no. 10, 103513 (2014)
  doi:10.1103/PhysRevD.90.103513
  [arXiv:1405.0272 [hep-ph]].

\bibitem{Berlin:2014pya}
  A.~Berlin, P.~Gratia, D.~Hooper and S.~D.~McDermott,
  Phys.\ Rev.\ D {\bf 90}, no. 1, 015032 (2014)
  doi:10.1103/PhysRevD.90.015032
  [arXiv:1405.5204 [hep-ph]].

\bibitem{Basak:2014sza}
  T.~Mondal and T.~Basak,
  Phys.\ Lett.\ B {\bf 744}, 208 (2015)
  doi:10.1016/j.physletb.2015.03.055
  [arXiv:1405.4877 [hep-ph]].

\bibitem{Cline:2014dwa}
  J.~M.~Cline, G.~Dupuis, Z.~Liu and W.~Xue,
  JHEP {\bf 1408}, 131 (2014)
  doi:10.1007/JHEP08(2014)131
  [arXiv:1405.7691 [hep-ph]].

\bibitem{Wang:2014elb}
  L.~Wang and X.~F.~Han,
  Phys.\ Lett.\ B {\bf 739}, 416 (2014)
  doi:10.1016/j.physletb.2014.11.016
  [arXiv:1406.3598 [hep-ph]].

\bibitem{Cheung:2014lqa}
  C.~Cheung, M.~Papucci, D.~Sanford, N.~R.~Shah and K.~M.~Zurek,
  Phys.\ Rev.\ D {\bf 90}, no. 7, 075011 (2014)
  doi:10.1103/PhysRevD.90.075011
  [arXiv:1406.6372 [hep-ph]].

\bibitem{Ko:2014loa}
  P.~Ko and Y.~Tang,
  JCAP {\bf 1501}, 023 (2015)
  doi:10.1088/1475-7516/2015/01/023
  [arXiv:1407.5492 [hep-ph]].

\bibitem{Bell:2014xta}
  N.~F.~Bell, S.~Horiuchi and I.~M.~Shoemaker,
  Phys.\ Rev.\ D {\bf 91}, no. 2, 023505 (2015)
  doi:10.1103/PhysRevD.91.023505
  [arXiv:1408.5142 [hep-ph]].

\bibitem{Okada:2014usa}
  N.~Okada and O.~Seto,
  Phys.\ Rev.\ D {\bf 90}, no. 8, 083523 (2014)
  doi:10.1103/PhysRevD.90.083523
  [arXiv:1408.2583 [hep-ph]].

\bibitem{Borah:2014ska}
  D.~Borah and A.~Dasgupta,
  Phys.\ Lett.\ B {\bf 741}, 103 (2015)
  doi:10.1016/j.physletb.2014.12.023
  [arXiv:1409.1406 [hep-ph]].

\bibitem{Cahill-Rowley:2014ora}
  M.~Cahill-Rowley, J.~Gainer, J.~Hewett and T.~Rizzo,
  JHEP {\bf 1502}, 057 (2015)
  doi:10.1007/JHEP02(2015)057
  [arXiv:1409.1573 [hep-ph]].

\bibitem{Guo:2014gra}
  J.~Guo, J.~Li, T.~Li and A.~G.~Williams,
  Phys.\ Rev.\ D {\bf 91}, no. 9, 095003 (2015)
  doi:10.1103/PhysRevD.91.095003
  [arXiv:1409.7864 [hep-ph]].

\bibitem{Cao:2014efa}
  J.~Cao, L.~Shang, P.~Wu, J.~M.~Yang and Y.~Zhang,
  Phys.\ Rev.\ D {\bf 91}, no. 5, 055005 (2015)
  doi:10.1103/PhysRevD.91.055005
  [arXiv:1410.3239 [hep-ph]].

\bibitem{Freytsis:2014sua}
  M.~Freytsis, D.~J.~Robinson and Y.~Tsai,
  Phys.\ Rev.\ D {\bf 91}, no. 3, 035028 (2015)
  doi:10.1103/PhysRevD.91.035028
  [arXiv:1410.3818 [hep-ph]].

\bibitem{Buckley:2014fba}
  M.~R.~Buckley, D.~Feld and D.~Goncalves,
  Phys.\ Rev.\ D {\bf 91}, 015017 (2015)
  doi:10.1103/PhysRevD.91.015017
  [arXiv:1410.6497 [hep-ph]].

\bibitem{Hooper:2014fda}
  D.~Hooper,
  Phys.\ Rev.\ D {\bf 91}, 035025 (2015)
  doi:10.1103/PhysRevD.91.035025
  [arXiv:1411.4079 [hep-ph]].

\bibitem{Dolan:2014ska}
  M.~J.~Dolan, F.~Kahlhoefer, C.~McCabe and K.~Schmidt-Hoberg,
  JHEP {\bf 1503}, 171 (2015)
  [JHEP {\bf 1507}, 103 (2015)]
  doi:10.1007/JHEP07(2015)103, 10.1007/JHEP03(2015)171
  [arXiv:1412.5174 [hep-ph]].

\bibitem{Cerdeno:2015ega}
  D.~G.~Cerdeno, M.~Peiro and S.~Robles,
  Phys.\ Rev.\ D {\bf 91}, no. 12, 123530 (2015)
  doi:10.1103/PhysRevD.91.123530
  [arXiv:1501.01296 [hep-ph]].

\bibitem{Alves:2015pea}
  A.~Alves, A.~Berlin, S.~Profumo and F.~S.~Queiroz,
  Phys.\ Rev.\ D {\bf 92}, no. 8, 083004 (2015)
  doi:10.1103/PhysRevD.92.083004
  [arXiv:1501.03490 [hep-ph]].

\bibitem{Kaplinghat:2015gha}
  M.~Kaplinghat, T.~Linden and H.~B.~Yu,
  Phys.\ Rev.\ Lett.\  {\bf 114}, no. 21, 211303 (2015)
  doi:10.1103/PhysRevLett.114.211303
  [arXiv:1501.03507 [hep-ph]].

\bibitem{Chen:2015nea}
  C.~H.~Chen and T.~Nomura,
  Phys.\ Lett.\ B {\bf 746}, 351 (2015)
  doi:10.1016/j.physletb.2015.05.027
  [arXiv:1501.07413 [hep-ph]].

\bibitem{Modak:2015uda}
  K.~P.~Modak and D.~Majumdar,
  Astrophys.\ J.\ Suppl.\  {\bf 219}, no. 2, 37 (2015)
  doi:10.1088/0067-0049/219/2/37
  [arXiv:1502.05682 [hep-ph]].

\bibitem{Gherghetta:2015ysa}
  T.~Gherghetta, B.~von Harling, A.~D.~Medina, M.~A.~Schmidt and T.~Trott,
  Phys.\ Rev.\ D {\bf 91}, 105004 (2015)
  doi:10.1103/PhysRevD.91.105004
  [arXiv:1502.07173 [hep-ph]].

\bibitem{Rajaraman:2015xka}
  A.~Rajaraman, J.~Smolinsky and P.~Tanedo,
  arXiv:1503.05919 [hep-ph].

\bibitem{Mondal:2015rba}
  T.~Mondal and T.~Basak,
  arXiv:1507.01793 [hep-ph].

\bibitem{Butter:2015fqa}
  A.~Butter, T.~Plehn, M.~Rauch, D.~Zerwas, S.~Henrot-Versill¨¦ and R.~Lafaye,
  arXiv:1507.02288 [hep-ph].

\bibitem{Buckley:2015cia}
  M.~R.~Buckley and D.~Feld,
  Phys.\ Rev.\ D {\bf 92}, no. 7, 075024 (2015)
  doi:10.1103/PhysRevD.92.075024
  [arXiv:1508.00908 [hep-ph]].

\bibitem{Freese:2015ysa}
  K.~Freese, A.~Lopez, N.~R.~Shah and B.~Shakya,
  arXiv:1509.05076 [hep-ph].

\bibitem{Williams:2015bfa}
  A.~J.~Williams,
  arXiv:1510.00714 [hep-ph].

\bibitem{Duerr:2015bea}
  M.~Duerr, P.~Fileviez Perez and J.~Smirnov,
  arXiv:1510.07562 [hep-ph].

\bibitem{Cai:2015zza}
  Y.~Cai and A.~P.~Spray,
  arXiv:1509.08481 [hep-ph].

\bibitem{Cai:2015tam}
  Y.~Cai and A.~P.~Spray,
  arXiv:1511.09247 [hep-ph].

\bibitem{Ko:2014nha}
  P.~Ko and Y.~Tang,
  JCAP {\bf 1405}, 047 (2014)
  doi:10.1088/1475-7516/2014/05/047
  [arXiv:1402.6449 [hep-ph], arXiv:1402.6449].

\bibitem{Choi:2015bya}
  S.~M.~Choi and H.~M.~Lee,
  JHEP {\bf 1509}, 063 (2015)
  doi:10.1007/JHEP09(2015)063
  [arXiv:1505.00960 [hep-ph]].

  \bibitem{MG5}
  J.~Alwall, M.~Herquet, F.~Maltoni, O.~Mattelaer and T.~Stelzer,
  JHEP {\bf 1106}, 128 (2011)
  [arXiv:1106.0522 [hep-ph]];
  J.~Alwall, R.~Frederix, S.~Frixione, V.~Hirschi, F.~Maltoni, O.~Mattelaer, H.-S.~Shao and T.~Stelzer {\it et al.},
  JHEP {\bf 1407}, 079 (2014)
  [arXiv:1405.0301 [hep-ph]].

\bibitem{Nadolsky:2008zw}
  P.~M.~Nadolsky, H.~-L.~Lai, Q.~-H.~Cao, J.~Huston, J.~Pumplin, D.~Stump, W.~-K.~Tung and C.~-P.~Yuan,
  Phys.\ Rev.\ D {\bf 78} (2008) 013004
  [arXiv:0802.0007 [hep-ph]].

\bibitem{Ruiz:2015zca}
  R.~Ruiz,
  arXiv:1509.05416 [hep-ph].

\bibitem{Han:2013kza}
  T.~Han, S.~Padhi and S.~Su,
  Phys.\ Rev.\ D {\bf 88}, no. 11, 115010 (2013)
  [arXiv:1309.5966 [hep-ph]].

\bibitem{Cirelli:2009uv}
  M.~Cirelli and A.~Strumia,
  New J.\ Phys.\  {\bf 11}, 105005 (2009)
  [arXiv:0903.3381 [hep-ph]].

\bibitem{Abdallah:2015uba}
  W.~Abdallah, J.~Fiaschi, S.~Khalil and S.~Moretti,
  arXiv:1510.06475 [hep-ph].

\bibitem{Wang:2015saa}
  W.~Wang and Z.~L.~Han,
  Phys.\ Rev.\ D {\bf 92}, 095001 (2015)
  [arXiv:1508.00706 [hep-ph]].

\bibitem{Han:2006ip}
  T.~Han and B.~Zhang,
  Phys.\ Rev.\ Lett.\  {\bf 97}, 171804 (2006)
  [hep-ph/0604064].

\bibitem{Perez:2008ha}
  P.~Fileviez Perez, T.~Han, G.~y.~Huang, T.~Li and K.~Wang,
  Phys.\ Rev.\ D {\bf 78}, 015018 (2008)
  [arXiv:0805.3536 [hep-ph]].

\bibitem{Bajc:2007zf}
  B.~Bajc, M.~Nemevsek and G.~Senjanovic,
  Phys.\ Rev.\ D {\bf 76}, 055011 (2007)
  [hep-ph/0703080].

\bibitem{delAguila:2008cj}
  F.~del Aguila and J.~A.~Aguilar-Saavedra,
  Nucl.\ Phys.\ B {\bf 813}, 22 (2009)
  [arXiv:0808.2468 [hep-ph]].

\bibitem{Deppisch:2015qwa}
  F.~F.~Deppisch, P.~S.~Bhupal Dev and A.~Pilaftsis,
  New J.\ Phys.\  {\bf 17}, no. 7, 075019 (2015)
  [arXiv:1502.06541 [hep-ph]].

\bibitem{Aad:2012tfa}
  G.~Aad {\it et al.} [ATLAS Collaboration],
  Phys.\ Lett.\ B {\bf 716}, 1 (2012)
  [arXiv:1207.7214 [hep-ex]].

\bibitem{Chatrchyan:2012xdj}
  S.~Chatrchyan {\it et al.} [CMS Collaboration],
  Phys.\ Lett.\ B {\bf 716}, 30 (2012)
  [arXiv:1207.7235 [hep-ex]].

\bibitem{Baglio:2012np}
  J.~Baglio, A.~Djouadi, R.~Gr\"{o}ber, M.~M.~M\"{u}hlleitner, J.~Quevillon and M.~Spira,
  JHEP {\bf 1304}, 151 (2013)
  [arXiv:1212.5581 [hep-ph]].

\bibitem{Han:2015sca}
  Z.~L.~Han, R.~Ding and Y.~Liao,
  Phys.\ Rev.\ D {\bf 92}, no. 3, 033014 (2015)
  [arXiv:1506.08996 [hep-ph]].

\bibitem{Kang:2015nga}
  Z.~Kang, P.~Ko and J.~Li,
  arXiv:1504.04128 [hep-ph].

\bibitem{Khachatryan:2014mma}
  V.~Khachatryan {\it et al.} [CMS Collaboration],
  Phys.\ Rev.\ D {\bf 90}, no. 9, 092007 (2014)
  [arXiv:1409.3168 [hep-ex]].

\bibitem{Khachatryan:2014qwa}
  V.~Khachatryan {\it et al.} [CMS Collaboration],
  Eur.\ Phys.\ J.\ C {\bf 74}, no. 9, 3036 (2014)
  [arXiv:1405.7570 [hep-ex]].

\bibitem{Aad:2015jqa}
  G.~Aad {\it et al.} [ATLAS Collaboration],
  Eur.\ Phys.\ J.\ C {\bf 75}, no. 5, 208 (2015)
  [arXiv:1501.07110 [hep-ex]].

\bibitem{Aad:2015eda}
  G.~Aad {\it et al.} [ATLAS Collaboration],
  arXiv:1509.07152 [hep-ex].

\bibitem{Aad:2014vma}
  G.~Aad {\it et al.} [ATLAS Collaboration],
  JHEP {\bf 1405}, 071 (2014)
  [arXiv:1403.5294 [hep-ex]].

\bibitem{Aad:2014nua}
  G.~Aad {\it et al.} [ATLAS Collaboration],
  JHEP {\bf 1404}, 169 (2014)
  [arXiv:1402.7029 [hep-ex]].

\bibitem{Chatrchyan:2014aea}
  S.~Chatrchyan {\it et al.} [CMS Collaboration],
  Phys.\ Rev.\ D {\bf 90}, 032006 (2014)
  doi:10.1103/PhysRevD.90.032006
  [arXiv:1404.5801 [hep-ex]].

\bibitem{Calibbi:2014lga}
  L.~Calibbi, J.~M.~Lindert, T.~Ota and Y.~Takanishi,
  JHEP {\bf 1411}, 106 (2014)
  [arXiv:1410.5730 [hep-ph]].

\bibitem{Drees:2013wra}
  M.~Drees, H.~Dreiner, D.~Schmeier, J.~Tattersall and J.~S.~Kim,
  Comput.\ Phys.\ Commun.\  {\bf 187}, 227 (2014)
  doi:10.1016/j.cpc.2014.10.018
  [arXiv:1312.2591 [hep-ph]].

\bibitem{Conte:2014zja}
  E.~Conte, B.~Dumont, B.~Fuks and C.~Wymant,
  Eur.\ Phys.\ J.\ C {\bf 74}, no. 10, 3103 (2014)
  doi:10.1140/epjc/s10052-014-3103-0
  [arXiv:1405.3982 [hep-ph]].

\bibitem{Papucci:2014rja}
  M.~Papucci, K.~Sakurai, A.~Weiler and L.~Zeune,
  Eur.\ Phys.\ J.\ C {\bf 74}, no. 11, 3163 (2014)
  doi:10.1140/epjc/s10052-014-3163-1
  [arXiv:1402.0492 [hep-ph]].

\bibitem{Arhrib:2013ela}
  A.~Arhrib, Y.~L.~S.~Tsai, Q.~Yuan and T.~C.~Yuan,
  JCAP {\bf 1406}, 030 (2014)
  [arXiv:1310.0358 [hep-ph]].

\bibitem{Khachatryan:2014rra}
  V.~Khachatryan {\it et al.} [CMS Collaboration],
  Eur.\ Phys.\ J.\ C {\bf 75}, no. 5, 235 (2015)
  [arXiv:1408.3583 [hep-ex]].

\bibitem{Aad:2015zva}
  G.~Aad {\it et al.} [ATLAS Collaboration],
  Eur.\ Phys.\ J.\ C {\bf 75}, no. 7, 299 (2015)
  [Eur.\ Phys.\ J.\ C {\bf 75}, no. 9, 408 (2015)]
  [arXiv:1502.01518 [hep-ex]].

\bibitem{Chatrchyan:2012tea}
  S.~Chatrchyan {\it et al.} [CMS Collaboration],
  Phys.\ Rev.\ Lett.\  {\bf 108}, 261803 (2012)
  [arXiv:1204.0821 [hep-ex]].

\bibitem{Aad:2014tda}
  G.~Aad {\it et al.} [ATLAS Collaboration],
  Phys.\ Rev.\ D {\bf 91}, no. 1, 012008 (2015)
  [Phys.\ Rev.\ D {\bf 92}, no. 5, 059903 (2015)]
  [arXiv:1411.1559 [hep-ex]].

\bibitem{Aad:2013oja}
  G.~Aad {\it et al.} [ATLAS Collaboration],
  Phys.\ Rev.\ Lett.\  {\bf 112}, no. 4, 041802 (2014)
  [arXiv:1309.4017 [hep-ex]].

\bibitem{Aad:2014vka}
  G.~Aad {\it et al.} [ATLAS Collaboration],
  Phys.\ Rev.\ D {\bf 90}, no. 1, 012004 (2014)
  [arXiv:1404.0051 [hep-ex]].

\bibitem{Carpenter:2013xra}
  L.~Carpenter, A.~DiFranzo, M.~Mulhearn, C.~Shimmin, S.~Tulin and D.~Whiteson,
  Phys.\ Rev.\ D {\bf 89}, no. 7, 075017 (2014)
  [arXiv:1312.2592 [hep-ph]].

\bibitem{Berlin:2014cfa}
  A.~Berlin, T.~Lin and L.~T.~Wang,
  JHEP {\bf 1406}, 078 (2014)
  [arXiv:1402.7074 [hep-ph]].

\bibitem{Aad:2015yga}
  G.~Aad {\it et al.} [ATLAS Collaboration],
  Phys.\ Rev.\ Lett.\  {\bf 115}, no. 13, 131801 (2015)
  [arXiv:1506.01081 [hep-ex]].

\bibitem{Aad:2015dva}
  G.~Aad {\it et al.} [ATLAS Collaboration],
  arXiv:1510.06218 [hep-ex].

\bibitem{Khachatryan:2014tva}
  V.~Khachatryan {\it et al.} [CMS Collaboration],
  Phys.\ Rev.\ D {\bf 91}, no. 9, 092005 (2015)
  [arXiv:1408.2745 [hep-ex]].

\bibitem{Han:2015hba}
  Z.~L.~Han, R.~Ding and Y.~Liao,
  Phys.\ Rev.\ D {\bf 91}, 093006 (2015)
  [arXiv:1502.05242 [hep-ph]].

\bibitem{Aad:2014iza}
  G.~Aad {\it et al.} [ATLAS Collaboration],
  Phys.\ Rev.\ D {\bf 90}, no. 5, 052001 (2014)
  [arXiv:1405.5086 [hep-ex]].

\end{thebibliography}
\end{document}